\documentclass[useAMS,usenatbib]{mn2e}

\usepackage{psfig} 
\usepackage{times}
\usepackage{bm}
\usepackage{amsmath}
\usepackage{url}
\usepackage{amssymb}
\usepackage{dblfnote}
\usepackage{aas_macros}
\usepackage{hyperref}
\usepackage{natbib}
\usepackage{ifthen}
\usepackage{color}
\usepackage[T1]{fontenc}
\font\bolditalics = cmmib10
\def\bx#1{\leavevmode\thinspace\hbox{vrule\vtop{\vbox{\hrule\kern1pt
        \hbox{\vphantom{\tt/}\thinspace{\bf#1}\thinspace}}
      \kern1pt\hrule}\vrule}\thinspace}
\def \vc #1{{\textfont1=\bolditalics \hbox{$\bf#1$}}}

\def\sg{\bmath{s}}

\def\kg{\bmath{k}}

\def\gg{{\vc \gamma}}

\def\be{\begin{equation}}
\def\ee{\end{equation}}

\def\ba{\begin{eqnarray}}
\def\ea{\end{eqnarray}}

\def \vc #1{{\textfont1=\bolditalics \hbox{$\bf#1$}}}{\catcode`\@=11

\def\ave#1{\left\langle #1 \right\rangle}

\def\d{{\rm d}}

\def\ltsima{$\; \buildrel < \over \sim \;$}
\def\lsim{\lower.5ex\hbox{\ltsima}}
\def\gtsima{$\; \buildrel > \over \sim \;$}
\def\gsim{\lower.5ex\hbox{\gtsima}}

\renewcommand{\vec}[1]{{\boldsymbol{#1}}}
\newcommand{\e}[0]{{\rm e}}
\renewcommand{\i}[0]{{\rm i}}
\newcommand{\Ref}[1]{(\ref{#1})}

\newcommand{\mat}[1]{\mathrm{#1}}

\newcommand{\lensfit}[0]{\emph{lens}fit}

\title[CFHTLenS: non-Gaussian analysis of 3pt shear
correlations]{CFHTLenS: A Gaussian likelihood is a sufficient
  approximation for a cosmological analysis of third-order cosmic
  shear statistics}

\author[Simon et al.]{P. Simon$^{1}$\thanks{psimon@astro.uni-bonn.de},
  E. Semboloni$^{2,3}$, L. van Waerbeke$^{2}$, H. Hoekstra$^{3}$,
  T. Erben$^{1}$, L. Fu$^{4}$, \newauthor J. Harnois-D\'eraps$^{2,5}$,
  C. Heymans$^6$, H. Hildebrandt$^1$, M. Kilbinger$^{7,8}$,
  T.D. Kitching$^9$, \newauthor L. Miller$^{10}$, T. Schrabback$^1$\\
  $^1$ Argelander Institut f\"ur Astronomie, Auf dem H\"ugel 71, D-53121, Bonn, Germany\\
  $^2$ University of British Columbia, Department of Physics \&
  Astronomy, 6224 Agricultural Road, Vancouver, B.C. V6T 1Z1, Canada \\
  $^3$ Leiden Observatory, Leiden University, NL-2333 CA Leiden, The
  Netherlands\\
  $^4$ Shanghai Key Lab for Astrophysics, Shanghai Normal University,
  100
  Guilin Road, 200234, Shanghai, China\\
  $^5$ Canadian Institute for Theoretical Astrophysics, University of
  Toronto, M5S 3H8, On., Canada\\
  $^6$ Scottish Universities Physics Alliance, Institute for Astronomy, University of Edinburgh, Royal Observatory, Blackford Hill, Edinburgh, EH9 3HJ, UK\\
  $^7$ Institut d'Astrophysique de Paris, UMR7095 CNRS, Universit\'{e}
  Pierre \& Marie Curie, 98 bis boulevard Arago, F-75014 Paris, France\\
  $^8$ CEA/Irfu/SAp Saclay, Laboratoire AIM, F-91191 Gif-sur-Yvette,
  France\\
  $^9$ Mullard Space Science Laboratory, University College London,
  Holmbury St Mary, Dorking, Surrey RH5 6NT, UK\\
  $^{10}$Department of Physics, Oxford University, Keble Road, Oxford
  OX1 3RH, UK.}

\def\LaTeX{L\kern-.36em\raise.3ex\hbox{a}\kern-.15em
    T\kern-.1667em\lower.7ex\hbox{E}\kern-.125emX}

\begin{document}
\maketitle
\begin{abstract}

  We study the correlations of the shear signal between triplets of
  sources in the Canada-France-Hawaii Lensing Survey (CFHTLenS) to
  probe cosmological parameters via the matter bispectrum.  In
  contrast to previous studies, we adopted a non-Gaussian model of the
  data likelihood which is supported by our simulations of the
  survey. We find that for state-of-the-art surveys, similar to
  CFHTLenS, a Gaussian likelihood analysis is a reasonable
  approximation, albeit small differences in the parameter constraints
  are already visible. For future surveys we expect that a Gaussian
  model becomes inaccurate.  Our algorithm for a refined non-Gaussian
  analysis and data compression is then of great utility especially
  because it is not much more elaborate if simulated data are
  available.  Applying this algorithm to the third-order correlations
  of shear alone in a blind analysis, we find a good agreement with
  the standard cosmological model:
  \mbox{$\Sigma_8=\sigma_8(\Omega_{\rm
      m}/0.27)^{0.64}=0.79^{+0.08}_{-0.11}$} for a flat $\Lambda\rm
  CDM$ cosmology with \mbox{$h=0.7\pm0.04$} ($68\%$ credible
  interval).  Nevertheless our models provide only moderately good
  fits as indicated by \mbox{$\chi^2/{\rm dof}=2.9$}, including a
  $20\%$ r.m.s. uncertainty in the predicted signal amplitude. The
  models cannot explain a signal drop on scales around 15 arcmin,
  which may be caused by systematics. It is unclear whether the
  discrepancy can be fully explained by residual PSF systematics of
  which we find evidence at least on scales of a few arcmin. Therefore
  we need a better understanding of higher-order correlations of
  cosmic shear and their systematics to confidently apply them as
  cosmological probes.

\end{abstract}


\section{Introduction}

The statistics of the distribution of matter on large cosmological
scales, when combined with other cosmological probes, is a powerful
tool to discriminate between different cosmological models
\citep[e.g.][]{2003moco.book.....D,redbook}. Gravitational lensing is
a technique to assess the mass distribution in the Universe in a way
that is independent of the exact nature of dark matter and its
dynamical state \citep[][for an extensive review]{BaSc01}. One of the
consequences of gravitational lensing is cosmic shear, which we
statistically infer from correlations between shapes of distant
galaxies \citep[][for a recent review on weak gravitational
lensing]{2006glsw.conf..269S,2014arXiv1411.0115K}. The correlations
between shapes of galaxy pairs give a measurement of the projected
matter density power spectrum, which in turn constrains the geometry
of the Universe and the growth of structure.

The most recent cosmological constraints from cosmic shear are
reported by \citet{2014MNRAS.441.2725F}, F14 hereafter,
\citet{2013MNRAS.430.2200K}, \citet{2014MNRAS.442.1326K},
\citet{2013MNRAS.431.1547B}, and \citet{Heetal13} where the authors
analyse the latest data release by the Canada France Hawaii Telescope
Lensing Survey team
(CFHTLenS\footnote{\url{http://www.cfhtlens.org}}). This lensing
catalogue builds upon the Canada France Hawaii Telescope Legacy Survey
(CFHTLS), which represents together with the Red Cluster Sequence
Lensing Survey\footnote{\url{http://www.rcslens.org}} the
state-of-the-art of gravitational lensing surveys from the ground. A
preliminary weak lensing analysis of the CFHTLS has been presented
earlier in
\citet{Hoekstraetal06,Sembolonietal06,Benjamin07,Fuetal08}. Since
then, however, the CFHTLenS data has significantly improved in terms
of the characterisation of the residual systematics and the estimation
of galaxy redshifts, making the full scientific potential of weak
lensing a reality
\citep{Hildebrandtetal12,Erbenetal12,Heymans12,2013MNRAS.433.3373V,2013MNRAS.431.1439G,2013MNRAS.430.2476S,2013MNRAS.429.2249S,
  2014MNRAS.437.2111V}.

While current studies mainly focus on the two-point correlations in
the cosmic shear field, higher-order statistics contain more
information, and this can improve constraints on cosmological models
\citep{Beetal97,1999A&A...342...15V,2003A&A...397..405B,TaJa03b,ScLo03,KiSc05,ScKiLo05,Beetal09,Vaetal10,Kayoetal13}. For
the ongoing wide field surveys, such as the Kilo Degree Survey
(KiDS\footnote{\url{http://kids.strw.leidenuniv.nl}}), the Dark Energy
Survey (DES\footnote{\url{http://www.darkenergysurvey.org}}), the
Hyper Suprime-Cam survey
(HSC\footnote{\url{http://www.subarutelescope.org/Projects/HSC}}), and
future surveys such as the Large Synoptic Survey Telescope
(LSST\footnote{\url{http://www.lsst.org}}) and
Euclid\footnote{\url{http://sci.esa.int/euclid}}, the statistical
power of the three-point shear statistics alone is comparable to that
of two-point shear statistics \citep{Vaetal10,Kayoetal13}.  Therefore,
prospects on obtaining cosmological information from third-order shear
statistics are high.

On the observational side, early attempts to measure the three-point
shear statistics were carried out by \citet{Peetal03} and
\citet{JaBeJa04}. These two studies were performed using small data
sets and shape measurement algorithms that are not as robust as
algorithms today. While in both studies a signal was detected, the
results were strongly affected by residual point spread function (PSF)
systematics.  More recently, \citet{Seetal11a} used high-quality
space-based data, the HST/COSMOS dataset, to perform a measurement of
three-point shear statistics that did not show evidence of residual
systematics; however the analysis was limited to $1.6$ deg$^2$ of the
COSMOS data. The latest successful measurements of third-order shear
statistics have been performed by \citet{2014MNRAS.441.2725F} and
\citet{2013MNRAS.433.3373V} based on the CFHTLenS data set. These two
different approaches, shear correlation functions and moments in the
reconstructed lensing mass map, are complementary and are sensitive to
different residual systematics.

The interpretation of these statistics is still plagued and possibly
limited by theoretical uncertainties. A correct interpretation of this
signal can only be performed by accurately modelling the evolution of
the matter bispectrum in the non-linear regime. Analytical fitting
formulae such as \citet{ScCo01} are only accurate at the 10-20\% level
\citep{Waetal01,Seetal11a,HDetal12}. Alternative approaches based on
the halo model also have a limited accuracy \citep{VaNi11a,VaNi11b,
  Kayoetal13}.

Moreover, other phenomena are expected to affect the
measured signal, such as baryonic physics in the non-linear regime,
intrinsic alignments of source galaxies and source-lens
clustering. These phenomena have not yet been extensively studied and
are uncertain
\citep{Hamanaetal02,Seetal08,Seetal12,2014arXiv1407.4301H,2014MNRAS.441.2725F}.

Improvements of both the theoretical cosmological models and the
reduction of systematics in observational data are only two pillars of
a successful exploitation of the plentiful cosmological information in
the higher-order shear statistics. The success will also depend on
realistic models of statistical uncertainties in the shear
estimators. For this, a Gaussian likelihood is typically used in the
statistical analysis, such as in F14, whereas at least for
second-order cosmic shear statistics there is evidence in favour of
more complex models
\citep{2009A&A...504..689H,2011A&A...534A..76K,2011PhRvD..83b3501S,2013A&A...556A..70W}. For
this paper, we hypothesise that a Gaussian model for the data
likelihood of third-order shear correlations possibly yields biased
results for cosmological parameters. We motivate this hypothesis by
our observation in Sect. \ref{sec:3ptresults} that the distribution of
the estimates in simulations of the CFHTLenS data exhibits a
non-Gaussian distribution on angular scales of around 10-30 arcmin,
violating the assumption of Gaussian noise. To test our hypothesis for
CFHTLenS data we compare the cosmological constraints obtained from
measurements of the third-moment of the aperture mass when based on
Gaussian versus non-Gaussian likelihoods \citep{Scetal98,ScKiLo05}.
The CFHTLenS data is briefly summarised in Sect. \ref{sec:data}. Our
first analysis uses a commonly used Gaussian likelihood as in F14,
whereas the second analysis uses a non-Gaussian model.  Our estimator
of the third-order shear statistics is detailed in
Sect. \ref{sec:3ptstat}. For the cosmology, we assume a flat
$\Lambda\rm CDM$ model with the matter density parameter $\Omega_{\rm
  m}$ and the amplitude of fluctuations in the matter density field
$\sigma_8$ as free parameters; a flat $\Lambda\rm CDM$ model is
strongly supported by recent constraints from the cosmic microwave
background \citep{WMAP5,2013arXiv1303.5076P}.  Based on our new
technique in Sect.  \ref{sec:likelihood}, we construct the
non-Gaussian likelihood from a set of simulated measurements.

We present the results in Sect. \ref{sec:results} and discuss them in
Sect. \ref{sec:discussion}. In comparison to the two-point systematics
analysis of \citet{Heymans12}, H12 hereafter, we perform new tests for
third-order shear systematics of CFHTLenS that we present in
Sect. \ref{sect:cfhtlensmeasure} and in Appendix
\ref{sec:systematics}.

\section{Data}\label{sec:data}

\subsection{CFHTLenS}
\label{sec:cfhtlens}

The CFHTLS-Wide survey area is divided into four independent fields
(W1, W2, W3, W4), with a total area of $154\, {\rm deg^2}$, observed
in the five optical bands
$u^\ast,g^\prime,r^\prime,i^\prime,z^\prime$. Each field is a mosaic
of several MEGACAM fields, called pointings. More details about the
data set itself are given in \citet{Erbenetal12}.  The procedure for
the shape measurements using \lensfit~can be found in
\citet{Milleretal12}, \citet{Milleretal07},
\citet{2008MNRAS.390..149K}, and the photometric redshifts are
described in \citet{Hildebrandtetal12}. 

\newpage\noindent
A description of the CFHTLenS
shear catalogue, and the residual systematics based on the shear
two-points correlation function, is given in H12.

For the measurement of the three-point shear statistics presented in
this paper, we use galaxies from the 129 pointings selected by H12,
with \mbox{$0.2<z_{\rm phot}<1.3$} and \mbox{$i^\prime<24.7$}. The
mosaic for each field has been constructed by merging the single
pointings, so that overlaps are eliminated, and each galaxy appears
only once. Each field is projected on the tangential plane centred in
the middle using a gnomonic projection (see for example
\citealt{CaGi00}).  Three-point shear correlation functions are
measured for each of the four fields, i.e., not for individual
pointings, using for each galaxy the final Cartesian coordinates
$(x,y)$ (flat sky approximation), the ellipticity
($\epsilon_1,\epsilon_2$) and weights $w$ provided by \lensfit.

\begin{figure}
  \psfig{figure=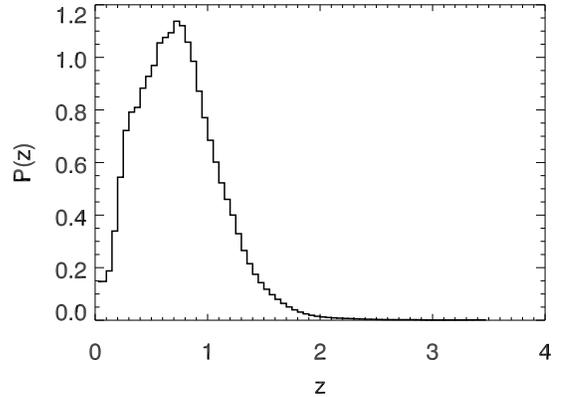,width=0.5\textwidth}
  \caption{\label{fig:redshift} Source redshift distribution $p_z(z)$
    obtained by stacking the probability densities of all galaxies
    with \mbox{$0.2<z_{\rm phot}<1.3$}, \mbox{$i^\prime<24.7$},
    weighted by the \lensfit~weight $w$. This distribution is used
    here to construct a forecast of the cosmological shear signal for
    a WMAP5 cosmology.}
\end{figure}

In order to interpret the shear signal we need to know the redshift
distribution of the sources. The redshift probability distribution
function (PDF) of each galaxy in the CFHTLenS catalogues is sampled in
70 redshift bins of width $0.05$ between $0$ and $3$. We obtain the
source redshift distribution of the full dataset by stacking the
distributions of all galaxies.  However, since each galaxy in our
sample is weighted according to $w$ when we compute the shear signal,
we need to weight the PDF of each galaxy to obtain the effective
redshift distribution of the sources, $p_z(z)$. This technique is
explained and tested in \citet{2013MNRAS.431.1547B}.  The final
redshift distribution, shown in Figure \ref{fig:redshift}, has a mean
redshift of \mbox{${\bar z}_{\rm phot}=0.74$}, and it is sampled in 30
steps between redshift zero and 3.

\subsection{Clone simulations of CFHTLenS}

The CFHTLenS {\tt clone} is a mock survey in which the lensing signal
obtained from N-body simulations is known, and the observational
properties, such as galaxy position, ellipticity, magnitude, weight,
are included such that the {\tt clone}'s are consistent with the
data. In this paper, we use the CFHTLenS {\tt clone} for various
purposes. A thorough description of the dark-matter-only simulations
can be found in \citet{HDetal12}.  These simulations have been
constructed using the \mbox{WMAP5+SN+BAO} cosmology:
$\{\Omega_m,\Omega_\Lambda,
\Omega_b,n,h,\sigma_8\}=\{0.279,0.721,0.046,0.96,0.701,0.817\}$
\citep{WMAP5}.  Starting at an initial redshift of $200$, the mass
density in the simulation box is sampled at $26$ different redshifts
between \mbox{$z=3$} and zero. The density fields are collapsed along
one of the Cartesian axes, and the resulting series of planes are used
to generate shear, convergence, and mass maps inside the light
cone. The 184 independent line-of-sights cover an area of $12.84~{\rm
  deg}^2$ each and have been populated with sources using the same
redshift distribution and galaxy density as in the CFHTLenS
observations.

The {\tt clone} does not include density fluctuations larger than the
simulation box, and this is known to affect the covariance estimated
from these simulated maps.  These missing super-survey modes propagate
in many ways in the shear covariance \citep{2014PhRvD..89h3519L}. Our
measurement, however, is very weakly impacted by these for two
reasons. First, our shear aperture statistics uses only aperture
scales up to 30 arcmin, which is well below the maximum usable scale
of 70 arcmin. This makes the {\it Finite Support effect} described in
\cite{2014arXiv1406.0543H} at most a 5 percent deficit on the error
bar of the largest angles. Second, we must also examine the
contribution from the {\it Halo Sampling Variance}, {\it Beat
  Coupling} and {\it Dilation}, which causes the small scale
clustering variance to be under-estimated
\citep{2006MNRAS.371.1188H,2006MNRAS.371.1205R}. As summarised in
\cite{2014arXiv1406.0543H}, simulation boxes of $500 \,h^{-1}\rm Mpc$
miss about 90 percent of the non-Gaussian part of the variance at
\mbox{$z=0$}. We can expect that the {\tt clone} misses even more due
to the smaller size of the box. However, weak lensing projects many
scales onto the same angular measurement, which dramatically decreases
the non-Gaussian contribution to the error bar.  In this case, the
most important contribution to the super-sample covariance comes from
the \emph{Halo Sampling Variance}, which peaks at small scales. As
argued in \cite{2013MNRAS.430.2200K}, this causes the small scale
covariance to be under-estimated by less than 10 percent, hence we do
not explicitly correct for super-survey modes in the {\tt clone}.

\section{Three-point shear statistics}
\label{sec:3ptstat}

\subsection{Definition of statistics}
\label{sec:method}

In this section, we briefly outline the relation between cosmic shear
and the statistics of the matter density fluctuation. The matter
density field at comoving position $\boldsymbol{\chi}$ and at redshift
$z$ is $\rho(\boldsymbol{\chi};z)\equiv \bar
\rho(z)[1+\delta(\boldsymbol{\chi};z)]$, where $\delta$ is the density
contrast and $\bar \rho(z)$ the average density at redshift $z$. The
power spectrum $P(k;z)$ and the bispectrum $B(k_1,k_2,k_3;z)$ are
defined by the correlators
\begin{equation}
  \label{eq:spectrum}
  \ave{\tilde\delta(\kg_1;z)\tilde\delta(\kg_2;z)}\equiv(2\pi)^3\delta_D(\kg_1+\kg_2)
  P(k_1;z)
\end{equation}
and
\begin{eqnarray}
  \label{eq:bispectrum}  
  \lefteqn{\ave{\tilde\delta(\kg_1;z)\tilde\delta(\kg_2;z)\tilde\delta(\kg_3;z)}}\\
  &&\nonumber
  \equiv(2\pi)^3\delta_D(\kg_1+\kg_2+\kg_3)B(k_1,k_2,k_3;z)\;,
\end{eqnarray}
of the Fourier transform $\tilde\delta(\kg_i;z)$ of the density
contrast $\delta$ for the 3D wave number $\kg_i$ at redshift $z$.

We apply a flat-sky approximation in the following.  Let
$\boldsymbol{\chi}_\perp$ be the two dimensional vector that we obtain
by projecting $\boldsymbol{\chi}$ onto the tangential plane on the
celestial sphere that is defined by the line-of-sight direction; the
$x$- and $y$-coordinates of $\boldsymbol{\chi}_\perp$ are $x,y$
respectively.  The complex shear $\gg=\gamma_1+\i\gamma_2$ and the
convergence $\kappa$ at angle $\vec{s}$ are both functions of the
second-order derivatives of the gravitational potential
$\phi(\boldsymbol{\chi}_\perp;\chi)$ at
\mbox{$\boldsymbol{\chi}_\perp:=\sg f_{\rm K}(\chi)$} with $f_{\rm
  K}(\chi)$ being the comoving angular diameter distance at radial
distance \mbox{$\chi=|\boldsymbol{\chi}|$}. Following \cite{Scetal98}
we find at angle $\sg$ in the tangential plane:
\begin{eqnarray}
  \label{eq:kappa}
  \lefteqn{\kappa(\sg)}\\
  &&\nonumber
  =\frac{3 \Omega_{\rm m} H_0^2}{2
    c^2}\int_0^{\chi_{\rm h}}\!\!\!\d\chi\frac{g(\chi)f_{\rm
      K}(\chi)}{a(\chi)}(\partial^2_x+\partial^2_y)\phi\Big(\boldsymbol{\chi}_\perp;\chi\Big)\;,
  \\
  \lefteqn{\gamma_1(\sg)}\\
  &&\nonumber
  =\frac{3 \Omega_{\rm m} H_0^2}{2 c^2}\int_0^{\chi_{\rm h}}\!\!\!\d\chi
  \frac{g(\chi)f_{\rm
      K}(\chi)}{a(\chi)}(\partial^2_x-\partial^2_y)\phi\Big(\boldsymbol{\chi}_\perp;\chi\Big)\;,\\
  \lefteqn{\gamma_2(\sg)}\\
  &&\nonumber
  =\frac{3 \Omega_{\rm m} H_0^2}{2c^2} \int_0^{\chi_{\rm h}}\!\!\!\d\chi
  \frac{g(\chi)f_{\rm K}(\chi)}{a(\chi)}2\partial_x\partial_y
  \phi\Big(\boldsymbol{\chi}_\perp;\chi\Big)\;,
\end{eqnarray}
where $H_0$ is the Hubble constant; $\Omega_{\rm m}$ is the matter
density parameter; $c$ is the vacuum speed of light; $a$ is the
cosmological scale factor; $\chi_{\rm h}$ is the size of the Hubble
horizon; and $p_\chi(\chi)\d \chi$ in
\begin{equation}
  \label{eq:redshift} 
  g(\chi):=\int_\chi^{\chi_{\rm h}}\!\!\!\d\chi^\prime 
  p_\chi(\chi^\prime)\frac{f_{\rm K}(\chi^\prime -\chi)}{f_{\rm K}(\chi^\prime)}
\end{equation}
describes the distribution of sources per distance interval $\d\chi$,
or per redshift interval in the case of $p_z(z)\d z$.

In the weak lensing regime of cosmic shear, we find
\mbox{$\kappa\ll1$} and \mbox{$|\gg|\ll1$} such that the observable
galaxy ellipticity \mbox{$\vec{\epsilon}=\epsilon_1+\i\epsilon_2$}
becomes an unbiased estimator of the galaxy shear to first
approximation, i.e.,
\mbox{$\ave{\boldsymbol{\epsilon}}=\boldsymbol{\gamma}$}. In addition,
by assuming that galaxies are randomly oriented intrinsically, the
Fourier transform of the angular correlation
\mbox{$\ave{\vec{\epsilon}_i\vec{\epsilon}_j^\ast}=\ave{\vec{\gamma}_i\vec{\gamma}_j^\ast}=\ave{\kappa_i\kappa_j}$}
between the ellipticities of pairs of galaxies $i,j$ can be
interpreted as a direct measure of the matter power spectrum in
projection on the sky,
\begin{equation}
  P_\kappa(\ell)=P_\gamma(\ell)=
  \frac{9\Omega_{\rm m}^2H_0^4}{4c^4}
  \int_0^{\chi_{\rm h}}\d\chi\frac{g^2(\chi)}{a^2(\chi)}
  P\left(\frac{\ell}{f_{\rm K}(\chi)};z(\chi)\right)\;.
\end{equation}
By $\vec{\gamma}^\ast$ we denote the complex conjugate of
$\vec{\gamma}$. The angular power spectrum $P_\kappa(\ell)$ of
$\kappa$ is defined by
\begin{equation}
  \ave{\tilde\kappa(\vec{\ell}_1)\tilde\kappa(\vec{\ell}_2)}
  =(2\pi)^2\delta_D(\vec{\ell}_1+\vec{\ell}_2)
  P_\kappa(\ell_1)
\end{equation}
for the angular wave number $\ell$.

A similar relation exists between the angular bispectrum $B_\kappa$ of
$\kappa$,
\begin{equation}
  \label{eq:Bkappa0}
  \ave{\tilde\kappa(\vec{\ell}_1)\tilde\kappa(\vec{\ell}_2)\tilde\kappa(\vec{\ell}_3)}
  =
  (2\pi)^2\delta_D(\vec{\ell}_1+\vec{\ell}_2+\vec{\ell}_3)
  B_\kappa(\ell_1,\ell_2,\ell_3)\;,
\end{equation}
attainable through the correlation of three galaxy ellipticities, and
the projected matter bispectrum
\begin{eqnarray}
  \label{eq:Bkappa}
  B_\kappa(\ell_1,\ell_2,\ell_3)
  &=&
  \frac{27H_0^6\Omega_{\rm m}^3}{8c^6}
  \int\d\chi\,\frac{g^3(\chi)}{f_{\rm K}(\chi)a^3(\chi)}\\
  &&\nonumber
  ~~~B\left(\frac{\ell_1}{f_{\rm K}(\chi)},
    \frac{\ell_2}{f_{\rm K}(\chi)},
    \frac{\ell_3}{f_{\rm K}(\chi)};z(\chi)\right)
\end{eqnarray}
\citep{Beetal97,Scetal98}. In contrast to second-order statistics,
four independent correlation functions of the third-order statistics
can be defined.  We utilise the representation of the correlator in
terms of its natural components as advocated by \citet{ScLo03}. Let
${\bf r}$, ${\bf r+s}$, and ${\bf r+t}$ be the angular positions of
three galaxies, while $\alpha$ denotes the angle between the $x$-axis
and ${\bf s}$. The natural components are then given by:
\begin{eqnarray}
  {\bf \Gamma}_0(s,{\bf t^\prime})&=&\ave{{\vc \gamma}({\bf r}) {\vc
      \gamma}({\bf r+s}) {\vc \gamma}({\bf r+t})
    e^{-6i\alpha}}\label{eq:natural_components_0}\,,\\
  {\bf \Gamma}_1(s,{\bf t^\prime})&=&\ave{{\vc \gamma}^\ast({\bf r})
    {\vc \gamma}({\bf r+s}) {\vc \gamma}({\bf r+t})
    e^{-2i\alpha}}\,,\\
  {\bf \Gamma}_2(s,{\bf t^\prime})&=&\ave{{\vc \gamma}({\bf r}) {\vc
      \gamma}^\ast({\bf r+s}) {\vc \gamma}({\bf r+t})
    e^{-2i\alpha}}\,,\\
  {\bf \Gamma}_3(s,{\bf t^\prime})&=&\ave{{\vc \gamma}({\bf r}) {\vc
      \gamma}({\bf r+s}) {\vc \gamma}^\ast({\bf r+t})
    e^{-2i\alpha}}\label{eq:natural_components_3}\;,
\end{eqnarray}
where \mbox{$s=\sqrt{\vec{s}\vec{s}^\ast}$} of ${\bf s}$ and
\mbox{${\bf t^\prime}={\bf t} {\bf s}^\ast/s$}. In equations
\Ref{eq:natural_components_0}--\Ref{eq:natural_components_3}, the
ensemble averages are performed over triangles invariant under
translation and rotation. These triangles can hence be characterised
by three real-valued variables $s$ and $\vec{t}^\prime=t^\prime_1+\i
t^\prime_2$. The component $\vec{\Gamma}_0$ is invariant under a
parity transformation, whereas $\vec{\Gamma}_{1,2,3}$ are not
\citep{ScLo03}.

For our analysis, we integrate the natural components to obtain an
alternative third-order statistic of shear, the third moment of the
aperture mass \citep{Scetal98},
\begin{equation}
  \label{eq:apmass}
  M_{\rm ap}(\theta)=
  \int d^2r\,U_{\theta}(|\vec{r}|)\kappa(\vec{r})\;,
\end{equation}
for an aperture filter $U_\theta(r)$ and a triplet
$(\theta_1,\theta_2,\theta_3)$ of aperture radii,
\begin{eqnarray}
  \label{eq:map3}
  \lefteqn{\ave{M_{\rm ap}^3(\theta_1,\theta_2,\theta_3)}}\\
  &&\nonumber
  \hspace{-0.4cm}=
  \int\left\{\prod_{j=1}^3\frac{\d^2r_j\d^2\ell_j U_{\theta_j}(|\vec{r}_j|)
    e^{i {\bf r_j}\cdot\vec{\ell}_j}}{(2\pi)^2}\right\}
  \ave{\tilde\kappa(\vec{\ell}_1)\tilde\kappa(\vec{\ell}_2)
    \tilde\kappa(\vec{\ell}_3)}
\end{eqnarray}
\citep{Peetal03,JaBeJa04,ScKiLo05}.  Roughly speaking, $M_{\rm
  ap}(\theta)$ is the $U_\theta$-smoothed convergence field at
\mbox{$\vec{r}=0$}, and $\ave{M_{\rm
    ap}^3(\theta_1,\theta_2,\theta_3)}$ is the $U_\theta$-smoothed
version of the correlator
$\ave{\tilde\kappa(\vec{\ell}_1)\tilde\kappa(\vec{\ell}_2)
  \tilde\kappa(\vec{\ell}_3)}$, which is a measure of the projected
matter bispectrum in Eq. \Ref{eq:Bkappa}. In particular, by choosing
an exponential aperture filter as in \cite{Wae98},
\begin{equation}
  \label{eq:filter} 
  U_\theta(r)=\frac{1}{2\pi
    \theta^2}\Big(1-\frac{r^2}{2 \theta^2} \Big)
  e^{-\frac{1}{2}(\frac{ r}{\theta})^2}\;,
\end{equation}
the relation between three-point correlation functions and the third
moment of $\ave{M_{\rm ap}^3}$ is relatively simple. Note that our
filter has been rescaled by $2\sqrt{2}\theta$ in comparison to
\citet{Wae98}. Our filter peaks at
\mbox{$\ell=\sqrt{2}/\theta\approx4862\,(\theta/1^\prime)^{-1}$} in
wave number space.

\subsection{Model of the matter bispectrum}
\label{sec:model}

The equations \Ref{eq:Bkappa0}, \Ref{eq:Bkappa}, and \Ref{eq:map3}
establish an explicit relation between $B(k_1,k_2,k_3;z)$ and the
third moment $\ave{M_{\rm ap}^3}$. Thus, in order to predict
$\ave{M_{\rm ap}^3}$ we need to model the bispectrum of matter
fluctuations. It is important that the bispectrum model captures the
mode coupling beyond perturbation theory, since our CFHTLenS
measurements are probing the non-linear regime of
\mbox{$k\sim0.1-10\,h\rm Mpc^{-1}$}.  To this end, we employ the
bispectrum fit of \citet[][SC01 hereafter]{ScCo01}.  The work of SC01
produced an analytical fit to the non-linear evolution of the
bispectrum based on a suite of cold dark matter N-body simulations
that was available at that time. The accuracy of this fit is limited
by the accuracy of the N-body simulations in their study and by the
fact that the effect of baryons on the small-scale clustering of
matter is not included. In short, the fit of SC01 consists of a
refinement lowest-order perturbation theory,
\begin{eqnarray}
  \label{eq:approx_from} 
  \lefteqn{B(k_1,k_2,k_3;z)}\\
  &&\nonumber
  =2 F(\kg_1,\kg_2;z) P(k_1;z)P(k_2;z)+ cycl. 
\end{eqnarray}
where $cycl.$ indicates a cyclic permutation of the indices. SC01
express the non-linear extension of the mode coupling factors
$F(\kg_1,\kg_2;z)$ as
\begin{eqnarray}
  F(\kg_1,\kg_2,z)&=&\frac{5}{7}a(n,k_1;z)a(n,k_2;z)\\
  \nonumber
  &+&\frac{1}{2}\frac{\kg_1\cdot\kg_2}{k_1k_2}\Big(\frac{k_1}{k_2}+
  \frac{k_2}{k_1}\Big)b(n,k_1;z)b(n,k_2;z)\\
  \nonumber
  &+&\frac{2}{7}\Big(\frac{\kg_1\cdot\kg_2}{k_1k_2}\Big)^2c(n,k_1;z)c(n,k_2;z)
\end{eqnarray}
where the coefficients 
\begin{eqnarray}
  a(n,k;z)&=&\frac{1+\sigma_8(z)^{-0.2}[0.7
    Q_3(n)]^{1/2}(q[z]/4)^{n+3.5}}{1+(q[z]/4)^{n+3.5}}\,,\\
  b(n,k;z)&=&\frac{1+0.4(n+0.3)q[z]^{n+3}}{1+q[z]^{n+3.5}}\,,\\
  c(n,k;z)&=&\frac{1+4.5/[1.5+(n+3)^4](2q[z])^{n+3}}{1+(2q[z])^{n+3.5}}\,,
\end{eqnarray}
have been fitted to the N-body simulations with
\begin{equation}
  \label{eq:approx_to}
  Q_3(n)=\frac{4-2^n}{1+2^{n+1}}\,.
\end{equation}
Here $\sigma_8(z)$ is the standard deviation of matter density
fluctuations within a sphere of radius $8h^{-1}{\rm Mpc}$ linearly
devolved from zero to redshift $z$, and $n$ is the spectral index of
the primordial power spectrum. The time dependence of
$F(\kg_1,\kg_2;z)$ is given by both the evolution of $\sigma_8(z)$ and
the function \mbox{$q[z]=k/k_{\rm nl}[z]$} where \mbox{$4 \pi k_{\rm
    nl}[z]^3 P_{\rm lin}(k_{\rm nl}[z];z)=1$} defines the wave number
$k_{\rm nl}[z]$ of the non-linear regime at redshift $z$; $P_{\rm
  lin}(k;z)$ denotes the linear matter power spectrum.

SC01 showed that this approximation is accurate to within $15\%$ up to
$k$ of a few $h{\rm Mpc}^{-1}$. \citet{Waetal01} compared the
third-order moments of the projected density field measured directly
on simulated $\kappa$ maps with predictions obtained using the fitting
formula. They found a similar accuracy. In agreement with these
previous results, \citet{Seetal11a} found that this approximation
systematically underestimates $\ave{M_{\rm ap}^3}$ on small angular
scales. A different approach to compute the bispectrum has been
recently suggested by \citet{VaNi11a,VaNi11b}. It uses a combination
of perturbation theory and the halo model. This approach is promising
but its performance depends on the accuracy of the halo-model which is
in general still limited.  Moreover, none of these approximations
accounts for the potentially large effects from baryonic physics
\citep{Seetal12}. Overall, the accuracy of bispectrum predictions is
therefore still an open issue. Current models cannot claim an accuracy
better than \mbox{$\sim 20\%$} which we include in our cosmological
analysis.  Consequently further improvements in modeling are necessary
in the future, which is beyond the scope of this work. 

We note here that the coefficients of SC01 have recently been updated
by \citet{2012GILMAR}. Our analysis does not include this update as
the corrections are smaller than the 20\% model error that we include
in our analysis, and as such this update would not impact our
results. Moreover, as non-linear power spectrum $P(k;z)$ in
Eq. \Ref{eq:approx_from} we use the model of \citet{Smetal03} with the
transfer function of \cite{EiHu98}. As recently reported in
\citet{2014arXiv1407.4301H}, this model lacks power on small scales in
comparison to N-body simulations. As shown in
  Sect. \ref{sect:clonetest}, however, this bias is negligible on the
  angular scales that we exploit for our analysis.
        
\subsection{Estimators of the natural components}
\label{sec:measure}

The measurement of the third-order moment of the aperture mass
statistics, $\ave{M^3_{\rm ap} (\theta_1,\theta_2, \theta_3)}$, is
performed using the same procedure described in \citet{Seetal11a} and
in the recent F14 analysis (see their Sect. 2.3). This procedure
consists of reconstructing $\ave{M^3_{\rm ap}}$ by numerical
integration of estimates of the natural components $\vec{\Gamma}_i$.

For the estimators of the natural components equations
\Ref{eq:natural_components_0}--\Ref{eq:natural_components_3}, we bin
the products of three source ellipticities of similar triangle
configurations. One possible choice is to bin the triangles according
to their values of $(s,t^\prime_1,t^\prime_2)$. However,
\citet{JaBeJa04} suggested a more suitable binning scheme. Given three
sides $s,t,d_1$ with \mbox{$s<t<d_1$}, the following variables are
defined:
\begin{eqnarray}
  d=&s   & ,\, d_{\rm min}<d <d_{\rm max}\,,\\
  u=&s/t & ,\,0<u<1\,, \\
  v=&\pm \frac{d_1 -t}{s} & ,\,-1<v<1\,.
\end{eqnarray}
Assigning a sign to $v$ keeps track of the triangle orientation: if
\mbox{$t^\prime_x=t_x$} (that is \mbox{$s_x>0$}, i.e., the triangle is
clockwise oriented) then \mbox{$v>0$}, otherwise \mbox{$v<0$}. The
limits $d_{\rm min}$ and $d_{\rm max}$ are the minimal and maximal
lengths for the smallest triangle side $s$ and define the range on
which the natural components are sampled.  For the CFHTLenS fields,
$d_{\rm min}$ is set to $9\, {\rm arcsec}$ in order to avoid bias from
close galaxies pairs with overlapping isophotes.  The maximum
separation is set to \mbox{$d_{\rm max}=400\, {\rm arcmin}$} which
means that $\ave{M^3_{\rm ap}}$ can be reconstructed up to angular
scales \mbox{$\lesssim100\,{\rm arcmin}$}. We compute the correlation
function of the galaxy triplets by a tree-code approach, similar to
the one suggested by \citet{ZhPe05}. In our implementation of the tree
code, we require \mbox{$(S_1+S_2)/D<0.1$} as criterion for stopping a
deeper search into the tree; $S_1$ and $S_2$ are the sizes of any pair
of tree nodes belonging to a node triplet, and $D$ is the distance
between the centres of the nodes.

In the case of the CFHTLenS catalogue, it is necessary to account for
a multiplicative correction factor $m(\nu_{\rm SN},r_{\rm gal})$
assigned to each galaxy (H12, \citealt{Milleretal12}). The correction
factor depends on the galaxy signal-to-noise $\nu_{\rm SN}$ and size
$r_{\rm gal}$. According to \citet{Milleretal12}, for an average shear
$\ave{\gamma_i}$ the corrected estimate of the $i$-component is given
by
\begin{equation}
  \ave{\gamma_i}_{\rm cal}=\frac{\ave{\gamma_i}}{\ave{1+m}}\,, {\rm
    with }\; i=1,2 
\end{equation}
where $\ave{\ldots}$ indicates ensemble weighted averages for $n_{\rm
  gal}$ galaxies defined using the \lensfit~weights $w$, namely
\begin{eqnarray}
  \ave{\gamma_i}&=&
  \left(\sum \limits_{\rm
      a=1}\limits^{n_{\rm gal}} w_{\rm a}\right)^{-1}
  \sum \limits_{\rm a=1} \limits^{n_{\rm gal}}
  \epsilon_{i,{\rm a} } w_{\rm a}\,, \\
  \ave{1+m}&=&
  \left(\sum\limits_{\rm a=1}
    \limits^{n_{\rm gal}} w_{\rm a}\right)^{-1}
  \sum \limits_{\rm a=1} \limits^{n_{\rm gal}}w_{\rm
    a}(1+m_{\rm a}(\nu_{\rm SN},r_{\rm gal}))\,. 
\end{eqnarray}
The extension of this calibration scheme to the three-point shear
statistics is straightforward: 
\begin{equation}
  \label{eq:correction}
  \ave{\gamma_i \gamma_j
    \gamma_k}_{\rm cal}=\frac{\sum\limits_{\widehat {\rm abc}}
    \epsilon_{i,{\rm a}} \epsilon_{j,{\rm b}} \epsilon_{k,{\rm c}} w_{\rm a}
    w_{\rm b} w_{\rm c}}{\sum\limits_{\widehat{\rm abc}} w_{\rm a}
    w_{\rm b} w_{\rm c} (1+m_{\rm a})(1+m_{\rm b})(1+m_{\rm
      c})}\,,
\end{equation}
where the sum is over all triplets $\widehat{\rm a,b,c}$ belonging to
a bin and $i,j,k,=1,2$ denote the shear components. This correction is
applied to our practical estimators of the natural components,
\begin{equation}
  \vec{\Gamma}_0(s,\vec{t}^\prime)=
  \frac{\sum_{\widehat{\rm a,b,c}}w_{\rm a}w_{\rm b} w_{\rm
      c}\epsilon_{\rm a}\epsilon_{\rm b}\epsilon_{\rm
      c}\e^{-6\i\alpha_{\rm b}}}
  {\sum_{\widehat{\rm a,b,c}}w_{\rm a}w_{\rm b} w_{\rm
      c}(1+m_{\rm a})(1+m_{\rm b})(1+m_{\rm
      c})}
\end{equation}
and
\begin{equation}
  \vec{\Gamma}_1(s,\vec{t}^\prime)=
  \frac{\sum_{\widehat{\rm a,b,c}}w_{\rm a}w_{\rm b} w_{\rm
      c}\epsilon^\ast_{\rm a}\epsilon_{\rm b}\epsilon_{\rm
      c}\e^{-2\i\alpha_{\rm b}}}
  {\sum_{\widehat{\rm a,b,c}}w_{\rm a}w_{\rm b} w_{\rm
      c}(1+m_{\rm a})(1+m_{\rm b})(1+m_{\rm
      c})}\;.
\end{equation}
We obtain the remaining two other estimators
$\vec{\Gamma}_{2,3}(s,\vec{t}^\prime)$ by cyclic permutations of the
triangle parameters \citep{ScLo03}.

\subsection{Estimators of modes of the aperture statistics}

In practice, we encounter four distinct modes of the aperture
statistics due to the possible presence of B-modes in the shear field.
Known sources of B-modes are systematics due to residuals in the PSF
correction or intrinsic alignments between the sources
\citep[e.g.,][]{2000MNRAS.319..649H,2012MNRAS.423.3163K}. Even without
these systematics higher-order effects such as lens coupling and
corrections beyond the Born approximation give rise to B-modes,
however much smaller than the amplitude of the E-modes
\citep{Co&Hu02,HiSe03,Scetal98}.

For the aperture mass of an aperture centred on the origin
$\vec{r}=0$, the formal separation between E- and B-modes is given by
\begin{eqnarray}
  M(\theta)&=&
  M_{\rm ap}(\theta)+\i M_\perp(\theta)\\
  \nonumber
  &=&-\int d^2r\,Q_\theta(|\vec{r}|) {\gg}(\vec{r}) e^{-2\i\phi}\;,
\end{eqnarray}
where the imaginary part $M_\perp(\theta)$ is the B-mode of the
aperture mass, the real part $M_{\rm ap}(\theta)$ is the E-mode of
Eq. \Ref{eq:apmass}, $\phi$ is the polar angle of $\vec{r}$, and
\begin{equation}
  Q_\theta(x)=
  \left(\frac{2}{x^2}
    \int_0^{x}\d ssU_\theta(s)\right)-U_\theta(x)
\end{equation}
is the filter of the shear field that corresponds to $U_\theta$. We
denote by $M^\ast(\theta)$ the complex conjugate of $M(\theta)$. To
lowest order in $\delta\phi/c^2$, gravitational lensing can only
produce E-modes, hence \mbox{$M_\perp=0$} in this regime. In this
case, the third-order moment $\ave{M_{\rm
    ap}^3(\theta_1,\theta_2,\theta_3)}$ in Eq. \Ref{eq:map3} is given
by the average $\ave{M(\theta_1)M(\theta_2)M(\theta_3)}$ over all
available aperture positions.  Generally, however, we have E/B mixing
inside the correlator, giving rise to different modes of the
statistics: the EEE, EEB, EBB, and BBB mode,
\begin{eqnarray}
  \lefteqn{{\rm EEE:~}\ave{M_{\rm ap}^3(\theta_1,\theta_2,\theta_3)}}\\
  &&\nonumber
  =\frac{1}{4}\operatorname{Re}\left(
    \ave{M^3(\theta_1,\theta_2,\theta_3)}+
    \ave{M^2M^\ast(\theta_1,\theta_2,\theta_3)}\right.\\
  &&\nonumber
  \left.+\ave{M^2M^\ast(\theta_3,\theta_1,\theta_2)}+
    \ave{M^2M^\ast(\theta_2,\theta_3,\theta_1)}\right)\\
  \nonumber\\
  \lefteqn{{\rm EEB:~}\ave{M_{\rm ap}^2M_\perp(\theta_1,\theta_2,\theta_3)}}\\
  &&\nonumber
  =\frac{1}{4}\operatorname{Im}\left(
    \ave{M^3(\theta_1,\theta_2,\theta_3)}+
    \ave{M^2M^\ast(\theta_3,\theta_1,\theta_2)}\right.\\
  &&\nonumber
  \left.+\ave{M^2M^\ast(\theta_2,\theta_3,\theta_1)}-
    \ave{M^2M^\ast(\theta_1,\theta_2,\theta_3)}\right)\;,\\
  \nonumber\\
  \lefteqn{{\rm EBB:~}\ave{M_{\rm ap}M^2_\perp(\theta_1,\theta_2,\theta_3)}}\\
  &&\nonumber
  =\frac{1}{4}\operatorname{Re}\left(
    -\ave{M^3(\theta_1,\theta_2,\theta_3)}+
    \ave{M^2M^\ast(\theta_1,\theta_2,\theta_3)}\right.\\
  &&\nonumber
  \left.+\ave{M^2M^\ast(\theta_3,\theta_1,\theta_2)}-
    \ave{M^2M^\ast(\theta_2,\theta_3,\theta_1)}\right)\;,\\
  \nonumber\\
  \lefteqn{{\rm BBB:~}\ave{M^3_\perp(\theta_1,\theta_2,\theta_3)}}\\
  &&\nonumber
  =\frac{1}{4}\operatorname{Im}\left(
    \ave{M^3(\theta_1,\theta_2,\theta_3)}+
    \ave{M^2M^\ast(\theta_1,\theta_2,\theta_3)}\right.\\
  &&\nonumber
  \left.+\ave{M^2M^\ast(\theta_3,\theta_1,\theta_2)}+
    \ave{M^2M^\ast(\theta_2,\theta_3,\theta_1)}\right)\;.
\end{eqnarray}
Note the order of the arguments $(\theta_1,\theta_2,\theta_3)$ in the
previous equations.  The actual cosmological signal EEE is the focus
of this study. The modes EEB and BBB, or parity modes, are only
present for parity violation \citep{2003A&A...408..829S}, whereas the
B-mode EBB may be the product of intrinsic alignments of the galaxy
ellipticities or an indicator of PSF systematics. To obtain the
different modes, we estimate the two statistics $\ave{M^3}$ and
$\ave{M^2M^\ast}$ from the measured natural components as in F14
(their equations 16 and 17).

Ideally, all four modes can be separated perfectly if the natural
components are measured with infinite resolution and over the entire
sky. However, the observed shear fields are finite, incomplete, and
they are only sampled at the positions of sources. For second-order
statistics, specially designed filters can be used to do this job; see
for example \citet{ScEiKr10} and references therein. For the
third-order statistics in this study, on the other hand, this
separation is currently not perfect. However, all these effects can be
quantified and are much smaller than the expected signal. In addition,
given the large errors in our measurements these inaccuracies are not
important for this study.  Therefore separating the signal into E-,
B-, and parity components is nevertheless a very effective way to
assess the cosmological origin of measured shear statistics.

\section{Simulated data versus measurements}
\label{sec:3ptresults}

\subsection{Clone Simulation}
\label{sect:clonetest}

\begin{figure}
  \begin{center}
    \psfig{figure=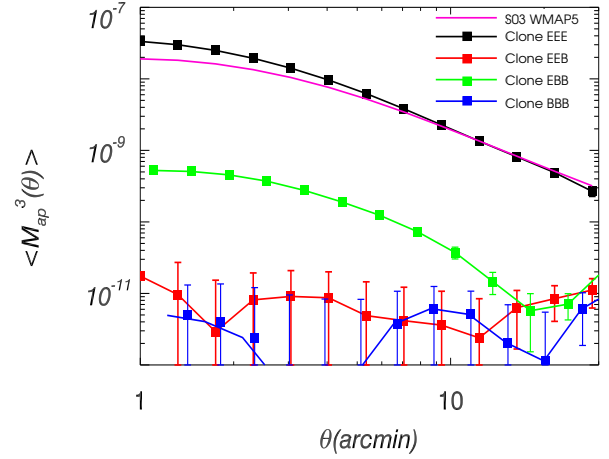,width=0.51\textwidth}
    \vspace{-0.8cm}
  \end{center}
  \caption{\label{fig:3pt_clone} Measurement of the equilateral
    $\ave{M^3_{\rm ap}}$ for the \emph{noise-free} data of the
    CFHTLenS {\tt clone}. The signal has been divided in EEE (black
    solid squares), absolute values of the two parity modes BBB (blue)
    and EEB (red), and the B-mode EBB (green). Error-bars represent
    the error on the average computed using the full suite of {\tt
      clone} mock catalogues.  For comparison, we show WMAP5
    predictions done using \citet{Smetal03} and the fitting formula
    SC01 in Sect. \ref{sec:method}.}
\end{figure}

We measure the three-point shear signal on the 184 simulated $12.84\,
{\rm deg}^2$ lines-of-sights from the CFHTLenS {\tt clone}.  For this
purpose, we perform two separate measurements.  Both sets of
measurements are used for different purposes. First, for the
\emph{noise-free} sample, we analyse shear catalogues that assume
intrinsically round sources. Second, for the \emph{noisy} sample, we
add ellipticity noise with an amplitude similar to the CFHTLenS
data. The \emph{noisy} sample is attained by adding a Gaussian random
value with an average of zero and a 1D dispersion of
\mbox{$\sigma_\epsilon=0.28$} to a noise-free sample where this value
is the measured ellipticity dispersion of the CFHTLenS including both
intrinsic ellipticity dispersion and measurement noise; we do not
include intrinsic alignments.  For each line-of sight, we measure the
natural components ${\bf \Gamma}_i$ using the same binning and the
same criteria we use on the CFHTLenS data. From the natural
components, we then compute the third-order moments of the aperture
mass by a numerical integration.

Figure \ref{fig:3pt_clone} verifies our method of computational
analysis, supporting previous findings that the SC01 matter bispectrum
is a reasonable fit to a dark-matter only simulation of the standard
cosmological model.  Using the \emph{noise-free} sample the figure
displays the mean and standard error, due to cosmic variance, of the
equilateral $\ave{M^3_{\rm ap}
  (\theta_1=\theta_2=\theta_3)}\equiv\ave{M^3_{\rm ap}(\theta)}$
obtained by averaging the 184 {\tt clone} fields. The signal has been
separated into E-, B- and two parity modes. As mentioned before, due
to E/B mixing of insufficiently sampled fields we expect a small level
of B-modes in the estimators even for the pure cosmic shear fields as
the ones in the simulation. We find that the BBB and EEB modes are
consistent with zero, whereas the EBB mode exhibits a positive signal,
albeit almost two orders of magnitude lower than the cosmological
signal EEE. A possible explanation for this (negligible) EBB
contamination, apart from the mixing, might be a numerical residual
due to our numerical transformation from natural components to the
aperture statistics.  For reference, in the figure we also show the
prediction of $\ave{M_{\rm ap}^3}$ for a WMAP5 cosmology and our SC01
model of the matter bispectrum. Moreover, we include the redshift
distribution $p_z(z)$ of sources as shown in Sect.
\ref{sec:cfhtlens}. 

The agreement of the prediction and the measurement for the EEE signal
is good overall. There is a discrepancy of about $10\%$ at around $5$
arcmin increasing to $30\%$ at 1 arcmin. The discrepancy may be
explained by the limited resolution of the simulations and the limited
accuracy of our model of the non-linear matter power spectrum
\citep{HDetal12,2014arXiv1407.4301H}. In addition, there is a
comparable inaccuracy in the non-linear regime due to baryonic physics
that is not accounted for in the {\tt clone}. We therefore take a
conservative stand in our analysis and utilise only measurements on
scales larger than 5 arcmin.

\begin{figure}
  \begin{center}
    \psfig{figure=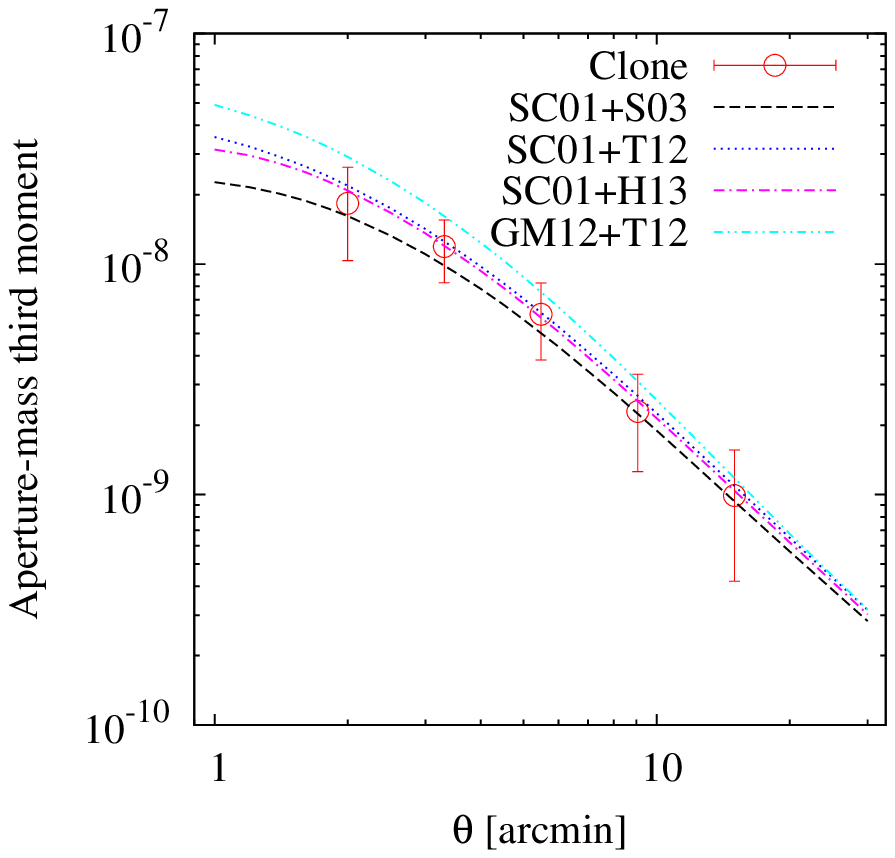,width=0.55\textwidth}
    \vspace{-0.5cm}
  \end{center}
  \caption{\label{fig:3pt_clone2} Data points of the
      equilateral $\ave{M^3_{\rm ap}}$ in the {\tt clone} (open
      circles) in comparison to different analytical descriptions of
      the matter bispectrum (lines). The models are combinations of a
      fitting formula for the bispectrum with different prescriptions
      of the non-linear matter power spectrum; SC01: \citet{ScCo01};
      GM12: \citet{2012GILMAR}; H13: \citet{2014ApJ...780..111H}; T12:
      \citet{2012ApJ...761..152T}; S03: \citet{Smetal03}. We use
      SC01+S03 for the scope of this study (bottom black line). The
      figure is a reproduction of Fig. A1 in F14.}
\end{figure}

In contrast to our relatively good agreement between simulation and
analytical model, \citet{2012A&A...541A.162V} find that the SC01 model
underestimates the power of a simulation by roughly a factor of two at
5 arcmin of $\ave{M_{\rm ap}^3}$ (their Fig. 2; lower left
panel). However, the authors in this paper employ a polynomial filter
that peaks at \mbox{$\theta_{\rm poly}\approx5/\ell$}, whereas our
Gaussian aperture filter peaks at \mbox{$\theta\approx\sqrt{2}/\ell$}
\citep{Scetal98,2002ApJ...568...20C}. This means that the
corresponding scales of the polynomial filter are \mbox{$\theta_{\rm
    poly}\gtrsim17.6$} arcmin for \mbox{$\theta\gtrsim5$} arcmin,
where the agreement of SC01 is much better compared to the
simulation. All the same, it is still possible that our good agreement
between simulation and model is a coincidence that results from an
offset of the {\tt clone} power due to the limited resolution and
volume of the simulation. In order to further investigate this, we
reproduce the Fig. A1 of our companion paper F14, see our
Fig. \ref{fig:3pt_clone2}. In this figure, we compare the {\tt clone}
measurements to a suit of recent, more accurate models of the matter
bispectrum discussed in the literature (see figure caption). All these
models agree with the {\tt clone} within \mbox{$\sim10\%$} for angular
scales larger than or equal 5 arcmin (Gaussian aperture filter). In
addition, our model SC01+S03 is below the amplitude of all recent
models by about 10\%-20\% for scales larger than 5 arcmin (bottom
black line). We therefore conclude that (i) analytical models of the
matter bispectrum on the scales considered here root-mean-square vary
within \mbox{$\sim20\%$}, and (ii) the bispectrum power in the {\tt
  clone} is sufficiently well described by SC01 on the angular scales
that we consider by a cosmological analysis.

\subsection{CFHTLenS measurement}
\label{sect:cfhtlensmeasure}

\begin{figure*} 
  \begin{tabular}{|@{}l@{}|@{}l@{}|@{}l@{}|}
    \psfig{figure=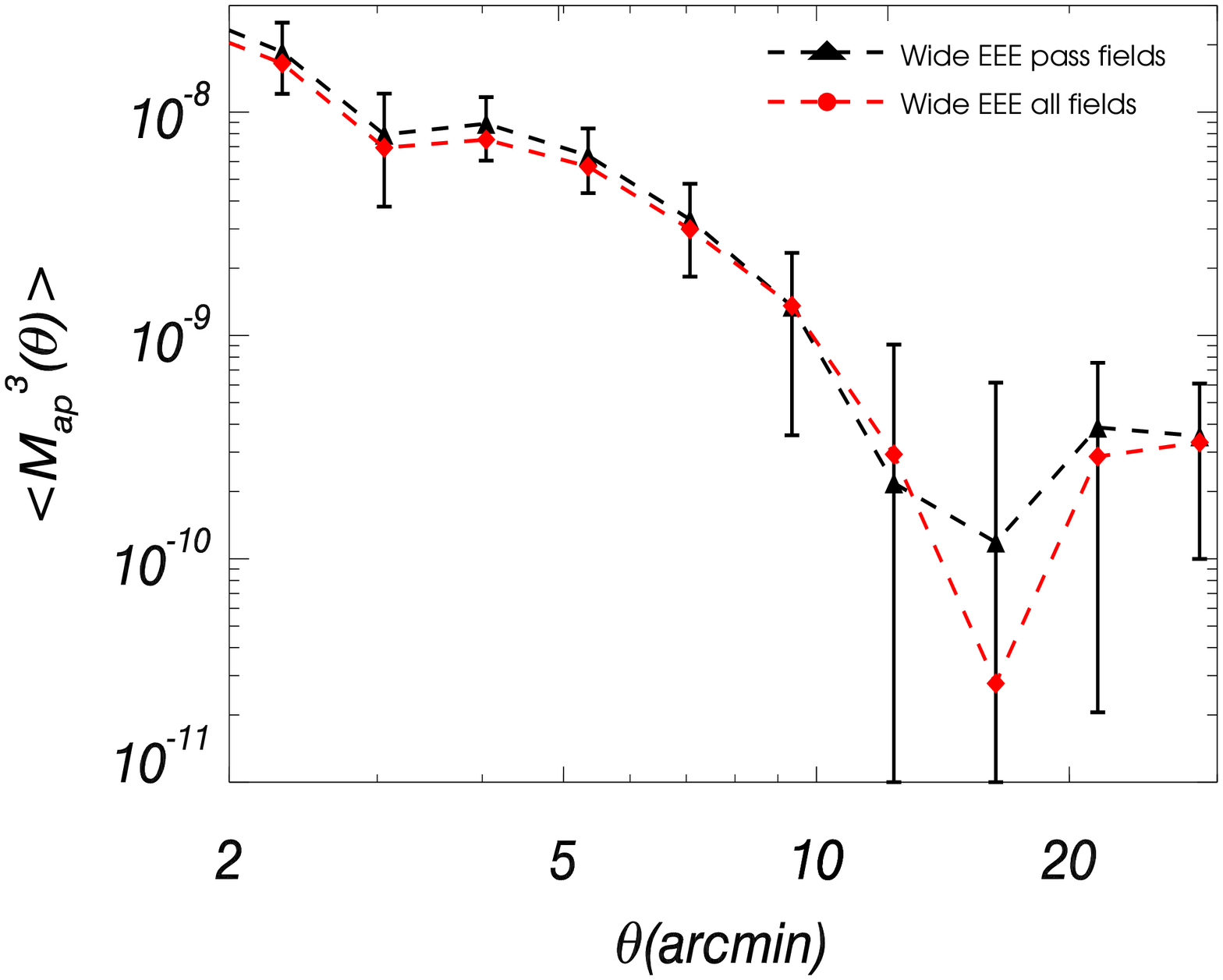,width=0.51\textwidth}
    &\psfig{figure=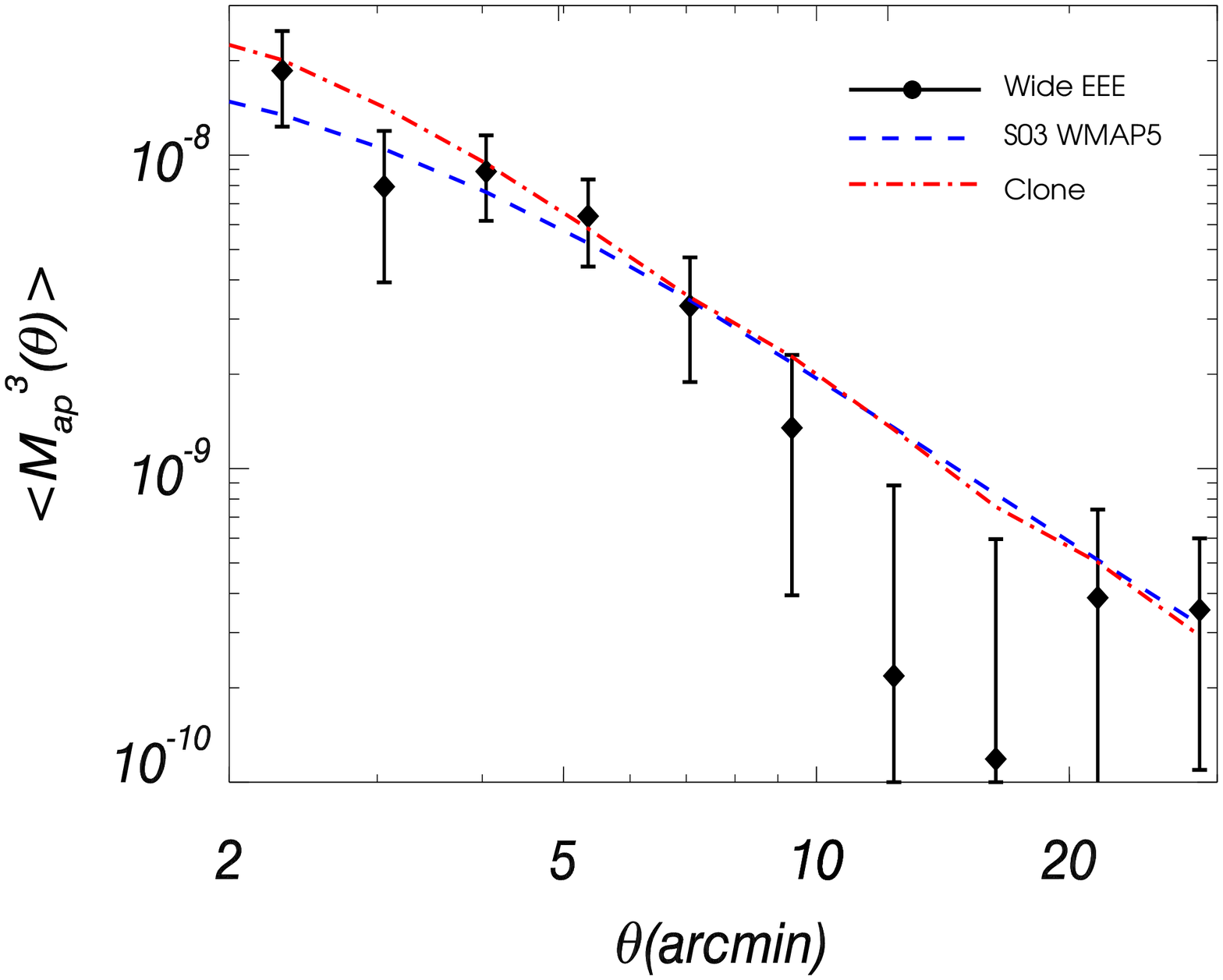,width=0.51\textwidth} 
  \end{tabular} 
  \caption{\label{fig:3pt_data_selected} \emph{Left panel}: the black
    solid triangles show the equilateral $\ave{M^3_{\rm ap}(\theta)}$
    computed using all the 120 CFHTLenS fields passing the systematic
    tests by H12 and our additional test in the Appendix
    \ref{sec:systematics}. The red squares show the same signal
    measured using all 129 fields that passed H12 tests only.  The
    error bars have been computed using the {\tt clone}.  \emph{Right
      panel}: display of the measured $\ave{M^3_{\rm ap}(\theta)}$
    from the 120 \emph{pass fields} together with the WMAP5 prediction
    and the average signal from the {\tt clone}.}
\end{figure*}

In this section, we present our measurements of $\ave{M_{\rm ap}^3}$
in the CFHTLenS data.  To start, in the left panel of Fig.
\ref{fig:3pt_data_selected} we evaluate the difference between
$\ave{M_{\rm ap}^3}$ as measured using all the 129 pointings to the
same quantity measured excluding 9 fields that fail our new residual
systematics test in Appendix \ref{sec:systematics}. Our systematics
test refines the tests done in H12 for third-order shear data. In the
following, we refer to the 120 remaining fields as \emph{pass
  fields}. These sample is utilised to constrain cosmological
parameters.  For Fig. \ref{fig:3pt_data_selected} we calculate the
error bars by rescaling the standard error of the mean in the 184
\emph{noisy} {\tt clone} simulation by the factor $\sqrt{A_{\rm
    ratio}}$, where \mbox{$A_{\rm ratio}=0.12$} is the area ratio
between the \emph{pass fields} and the simulation. The amplitude
difference between the two measurements of $\ave{M_{\rm ap}^3}$ with
and without the rejected 9 fields are smaller than the statistical
error. Nevertheless the two measurements are performed essentially
from the same sample, therefore the change in amplitude at aperture
radii $\gtrsim15\,\rm arcmin$ may be indicative of residual PSF
systematics on these scales. The right panel of
Fig. \ref{fig:3pt_data_selected} shows the CFHTLenS measurement of
$\ave{M_{\rm ap}^3(\theta)}$ for the \emph{pass fields} in comparison
to the predicted signal for a WMAP5 cosmology, and again the
measurement from the {\tt clone}.

\begin{figure*}
  \begin{tabular}{|@{}l@{}|@{}l@{}|@{}l@{}|}
    \psfig{figure=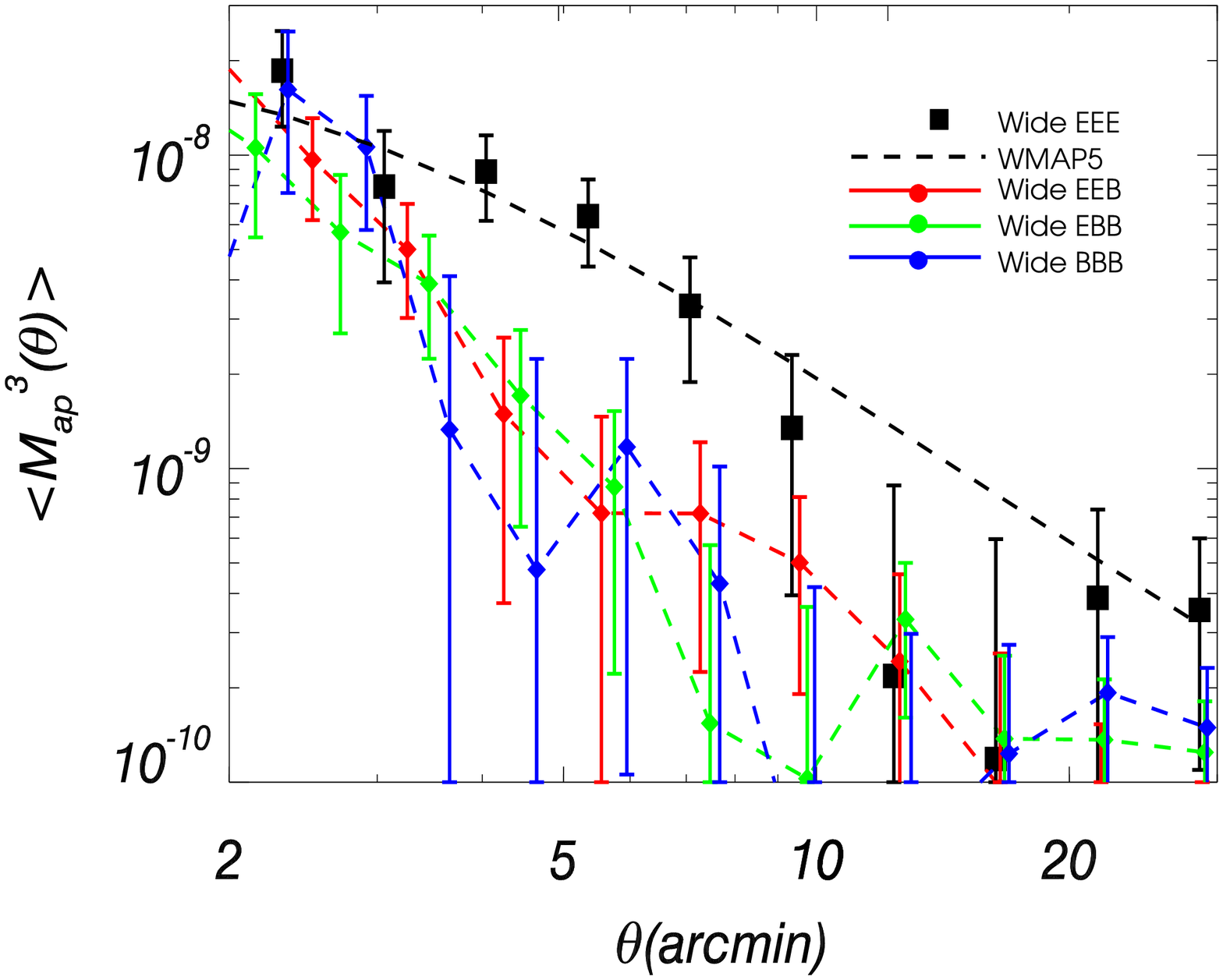,width=0.5\textwidth}&\psfig{figure=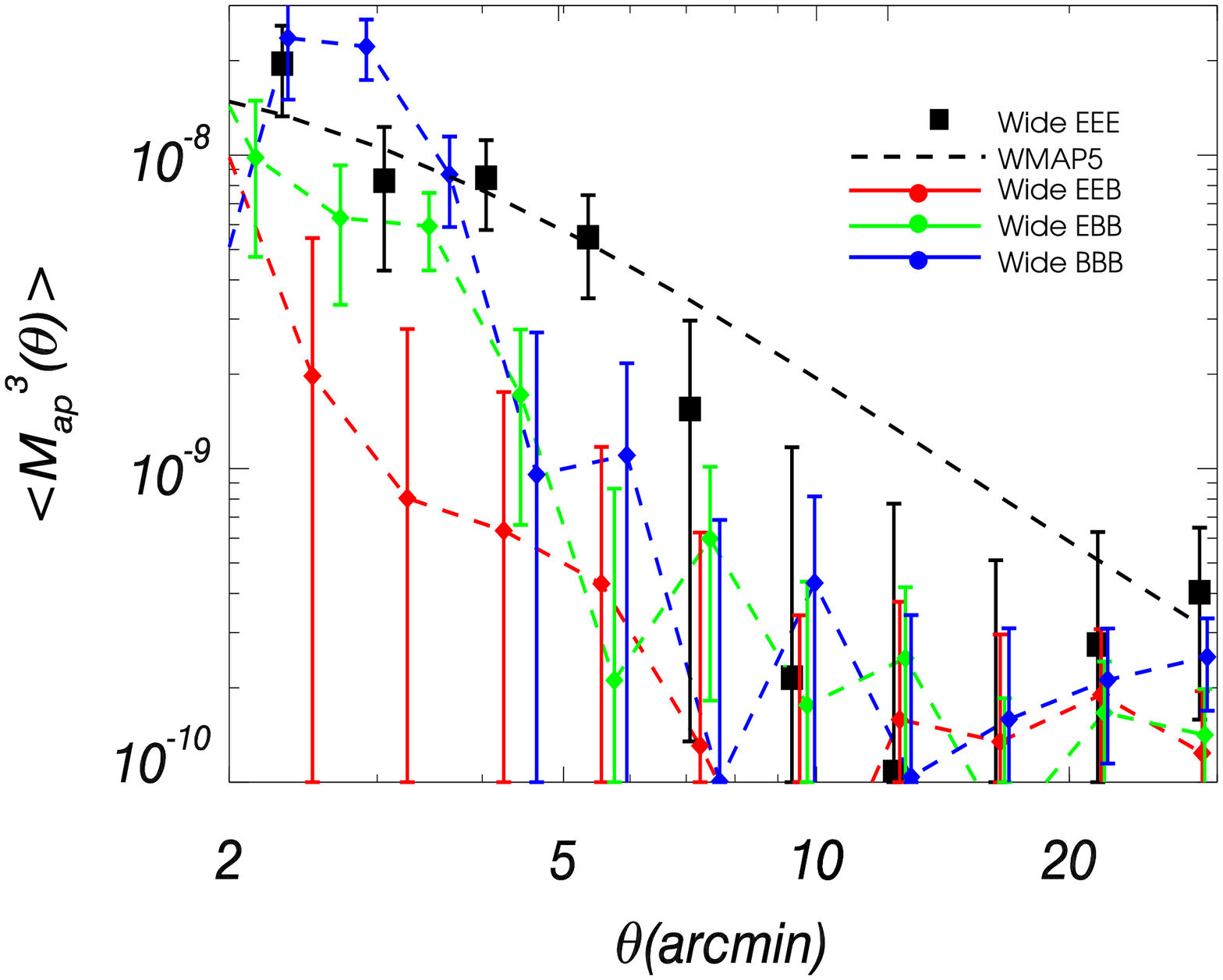,width=0.5\textwidth}
  \end{tabular}
  \caption{\label{fig:3pt_data} \emph{Left panel}: comparison of the
    CFHTLenS EEE mode of $\ave{M_{\rm ap}^3(\theta)}$ (black data
    points) to the absolute values of EEB, EBB, and BBB modes (red,
    green, and blue lines). The error bars have been computed using
    the {\tt clone}. The dashed black line is a WMAP5 prediction.
    \emph{Right panel}: same quantities as in the left panel after
    excluding elliptical galaxies, i.e., sources with BPZ
    \mbox{$T_B\le2$}.}
\end{figure*}

In Fig. \ref{fig:3pt_data}, we compare our measurement of the EEE mode
to the other modes of $\ave{M_{\rm ap}^3(\theta)}$, and we possibly
find more evidence for PSF systematics on scales of a few arcmin and
below. For scales \mbox{$\theta\gtrsim3$} arcmin, the EEE signal is
roughly one order of magnitude larger than the B- and parity modes
whenever it is not consistent with zero. Meanwhile, the amplitude of
the B- and parity modes are for \mbox{$\theta\lesssim5$} arcmin
clearly larger than those measured on the idealistic data in the {\tt
  clone}, Fig. \ref{fig:3pt_clone}, where typical amplitudes are
always smaller than $10^{-9}$.  Possible origins of the increase of
the systematics indicators can be intrinsic alignments (IA) and PSF
residuals, both of which are not accounted for in the {\tt clone}. In
principle, IA can generate B-modes, but the correlations in the source
ellipticities should remain parity invariant. The presence of both
parity modes EEB and BBB signals hence indicates that IA alone cannot
explain the systematics in the data. This hints to an imperfect PSF
correction in the shape measurement that may be relevant for
$\ave{M^3_{\rm ap}}$. This further supports our decision to reject
measurements below 5 arcmin in the cosmological analysis.

Nevertheless IA could also affect the measurement of our cosmological
shear signal
\citep{2014MNRAS.445.2918M,2014A&A...561A..53V,Seetal08}. For this
reason, we try to quantify the impact of IA.  Using N-body
simulations, \citet{Seetal08} find that mostly the IA of early type
galaxies contaminates three-point shear statistics.  In their model
intrinsic ellipticities of galaxies are given by the ellipticity of
the parent matter halo. Early type galaxies are perfectly aligned to
the halo, whereas spiral galaxies have a random misalignment to their
parent halo. In surveys comparable to CFHTLenS wide, the IA amplitude
of the EEE signal due to intrinsic-intrinsic-intrinsic (III)
correlations becomes strong compared to the cosmic shear signal for
aperture radii below a few arcmin and grows to a comparable amplitude
at $\sim1$ arcmin. By considering tidal torque theory
\citet{2014MNRAS.445.2918M} find that third-order correlations between
the intrinsic shapes of sources and shear (GGI) are negligible with
respect to the III correlations for our angular scales of interest;
the absolute amplitude of III and GGI correlations is highly uncertain
in their model though.  Using the assumption of a linear relation
between intrinsic shapes and the local matter density as well as
hierarchical clustering, \citet{2014A&A...561A..53V} finds that IA
contamination is of the level of typically about 10\% for different
source redshifts which is ubiquitous in our data. This figure includes
a bias due to lens-source clustering which arises because of a
correlation of source number density and the matter density (treated
as linear in the model). In summary, these pieces of research imply
that IA alignments should make only a small contribution to the total
EEE signal for \mbox{$\theta\gtrsim5$} arcmin but may be substantial
for $\theta\sim1$ arcmin.

Using the conclusions from \citet{Seetal08}, it is possible to
evaluate the impact of III in our analysis by measuring the
three-point signal with and without elliptical galaxies. The effective
removal of ellipticals is achieved by selecting galaxies with
classification \mbox{$T_B>2$} from the Bayesian Photometric Redshift
Estimation \citep[BPZ;][]{2000ApJ...536..571B,Heetal13}. This
decreases the effective number of sources by about 20 percent. With
this selection the change in the EEE signal is within the error bars,
as we show in the right panel of Figure \ref{fig:3pt_data}. The EBB,
EEB and BBB components qualitatively retain their behaviour compared
to the full galaxy sample: their amplitude is large at small aperture
radii and falls off quickly. The amplitude itself, however, changes in
a way that is difficult to reconcile with IA: the BBB signal increases
significantly at about 3 arcmin rather than decreasing as anticipated.
Our result therefore suggests that (i) the systematics are associated
to residual PSF systematics which affect late-type galaxies more than
early-type galaxies and (ii) that the precision of the measurement is
still not sufficient to set constraints on the IA from the
three-points statistics alone.  For reason (i), we decide not to apply
a morphological cut for our cosmological analysis. Hence our final
catalogue is composed of 120 \emph{pass fields} with the original
selection cuts previously described in Section \ref{sec:data}. The
final selection of pointings correspond to a total effective area of
roughly 100 square degrees, with masked regions taken into account.

\begin{figure*}
  \begin{center}
    \psfig{figure=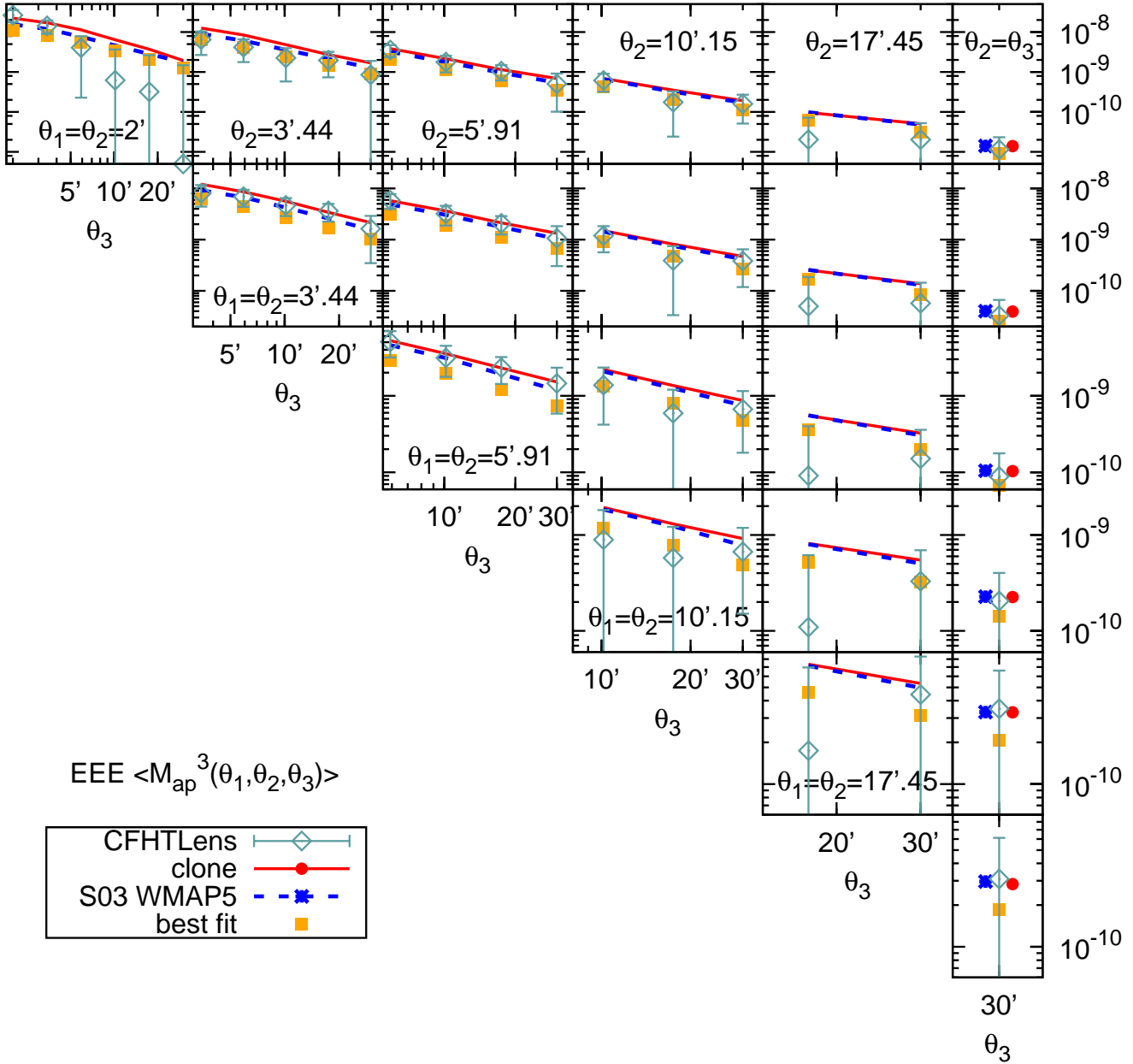,width=1.0\textwidth}
    \vspace{-1.5cm}
  \end{center}
  \caption{\label{fig:3pt_multi} Measurements of the EEE
    $\ave{M^3_{\rm ap} (\theta_1, \theta_2,\theta_3)}$ in the 120
    CFHTLenS \emph{pass fields} (dark green diamonds). The aperture
    radii $\theta_i$ are quoted in units of arcmin. The error bars
    have been computed using 184 sight lines of the {\tt clone}. The
    average signal in the simulations is shown in red (solid lines or
    filled circles), and the analytical WMAP5 predictions with the
    SC01 bispectrum are shown in blue (dashed lines or stars).  Each
    panel show the EEE signal as a function of
    \mbox{$\theta_3\ge\theta_2$} for a fixed value of $\theta_1$ and
    $\theta_2$.  Panels on the diagonal correspond to equilateral
    configurations with, increasing size from top to bottom.  In each
    row, from left to right we increase the size of $\theta_2$,
    keeping $\theta_1$ fixed. The filled orange boxes are the
      best-fit \mbox{$(\Omega_{\rm m},\sigma_8)=(0.32,0.7)$} of the
      SC01 model to the CFHTLenS data for \mbox{$\theta_i\ge5$
        arcmin. For details see Sect. \ref{sec:results}.}}
\end{figure*}

Figure \ref{fig:3pt_multi} shows, for this final catalogue, the
measurement of the \emph{full} $\ave{M^3_{\rm ap} (\theta_1,
  \theta_2,\theta_3)}$ in comparison to the WMAP5 predictions and the
{\tt clone} measurement. As before with the equilateral $\ave{M_{\rm
    ap}^3(\theta)}$ we find a reasonable agreement also between the
full statistics of the measurement, WMAP5 prediction, and the {\tt
  clone}. For \mbox{$\theta_i\lesssim5\,\rm arcmin$}, the SC01 model
falls slightly below the {\tt clone}, on larger scales the CFHTLenS
measurement appears to be below the prediction; the statistical errors
are, however, relatively large in this regime. In the next section, we
use this vector of data points for $\theta_i\ge5\,\rm arcmin$ to infer
cosmological parameters through a likelihood analysis.

\subsection{Evidence for a non-Gaussian likelihood}
\label{sec:evidence}

The matter density field obeys non-Gaussian statistics for smoothing
scales below the typical size of large galaxy clusters, and it
asymptotically approaches Gaussian statistics towards larger scales
and higher redshifts for a Gaussian primordial density field. Because
of this non-Gaussian nature on small physical scales we expect that
measurements of the three-point statistics of cosmic shear will
exhibit a distinct non-Gaussian distribution on small angular scales,
if these scales are not dominated by the shape noise of the
sources. In order to investigate whether the estimator of
$\ave{M^3_{\rm ap}}$ shows any signs of non-Gaussianity, we study its
distribution in 184 realisations of the \texttt{clone} for both the
\emph{noise-free} and the \emph{noisy} samples. For this purpose, we
compute estimates of $\ave{M^3_{\rm ap}}$ for the same combination of
aperture radii as for the CFHTLenS data. The resulting 184 simulated
data vectors represent the likelihood of obtaining a value of
$\ave{M_{\rm ap}^3(\theta)}$ on a $12.84\,{\rm deg^2}$ survey
(corresponding to the field of view of each simulation), given the
particular set of cosmological parameters in the {\tt clone}.

\begin{figure}
   \begin{tabular}{|@{}l@{}|@{}l@{}|@{}l@{}|}
     \psfig{figure=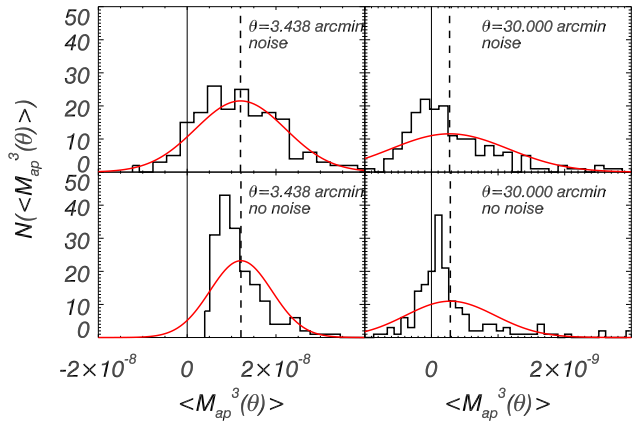,width=0.5\textwidth}
   \end{tabular}
   \caption{\label{fig:distribution} \emph{Top panels}: frequency of
     values of the equilateral $\ave{M_{\rm ap}^3(\theta)}$ from 184
     lines-of-sight from \emph{noisy} version of the {\tt clone}
     ($12.84\,\rm deg^2)$. In the left panels, we show the result for
     $\theta=3.438\, {\rm arcmin}$, while in the right panels we show
     the distribution for $\theta=30\, {\rm arcmin}$. Black dashed
     lines show the average values, while the solid red lines,
     indicate the best-fitting Gaussian. The black solid line
     indicates the zero value. \emph{Bottom panel}: the same as the
     top panels for the \emph{noise-free} version of the {\tt clone}.}
\end{figure}

\begin{figure}
  \begin{center}
    \psfig{figure=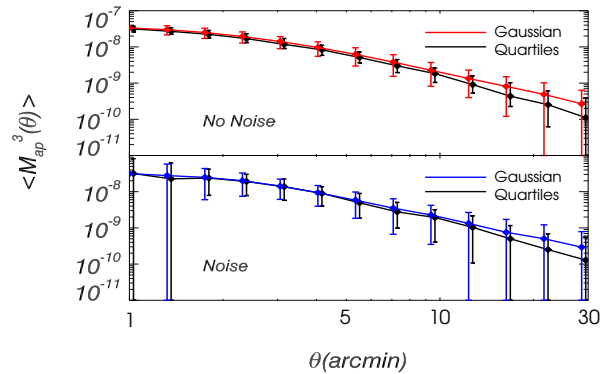,width=0.51\textwidth}
  \end{center}
  \caption{\label{fig:quartiles}\emph{Top
      panel}: median signal and the $\pm25\%$ quartiles obtained from
    the 184 \emph{noise-free} clone realisations (black line;
    \emph{Quartiles}). We compare this signal with the average signal
    and its error which is obtained multiplying the standard deviation
    by $0.67$ (red line; \emph{Gaussian}). In the case of a Gaussian
    distribution, the curves should coincide and the error bars should
    be the same. \emph{Bottom panel}: same results as top panel but
    using the \emph{noisy} version of the {\tt clone}.}
\end{figure}

The distributions are shown in Figure \ref{fig:distribution}, for
angular scales \mbox{$\theta=3.4,30.0$} arcmin of the equilateral
$\ave{M^3_{\rm ap}}$. The functional form of these distributions
depends on the sampling and the presence of shape noise
(\emph{noisy}).  In order to better illustrate the non-Gaussian
feature of the likelihoods we plot their best-fitting Gaussian inside
the panel; the Gaussian fits have the same mean and variance as our
observed values. In this figure, we find clear deviations from a
Gaussian model in the absence of shape noise, the observed
distributions are skewed towards large values of $\ave{M^3_{\rm ap}}$
(bottom panels). The skewness becomes weaker in the presence of shape
noise but deviations from the Gaussian description are still
discernible, especially for \mbox{$\theta=30$ arcmin} (top right
panel).

For the range of angular scales considered in this paper, the
non-Gaussian features are most prominent for
\mbox{$\theta\gtrsim10\,\rm arcmin$} as highlighted by
Fig. \ref{fig:quartiles}. In this figure, we plot for \emph{Gaussian}
the average and for \emph{Quartiles} the median of $\ave{M^3_{\rm
    ap}(\theta)}$ as function of $\theta$ using the simulations. In
addition, we indicate as error bars the standard deviation times
$0.67$ (\emph{Gaussian}) and the $\pm25$ percent quartiles
(\emph{Quartiles}). In the case of a Gaussian distribution of
$\ave{M^3_{\rm ap}}$, the data points and error bars of
\emph{Gaussian} and \emph{Quartiles} should be identical, whereas
differences indicate deviations from a Gaussian distribution for a
given $\theta$. Such deviations become most obvious for
$\theta\gtrsim10\,\rm arcmin$, which could indicate that we have to
find a non-Gaussian model of the likelihood for a cosmological
analysis of CFHTLenS. To test this hypothesis in the following, we
utilise the {\tt clone} distribution of $\ave{M^3_{\rm ap}}$ of the
\emph{noisy} sample to construct a numerically sampled non-Gaussian
likelihood and compare its constraints on \mbox{$(\Omega_{\rm
    m},\sigma_8)$} to the constraints from a traditional analytical
Gaussian model.

\section{Cosmological analysis}
\label{sec:likelihood}

In this section, we describe the details of the cosmological analysis
of $\ave{M_{\rm ap}^3}$. For the analysis we use \mbox{$N_{\rm d}=20$}
combinations of aperture radii between \mbox{$5\le\theta_i\le30$}
arcmin in $\ave{M^3_{\rm ap}(\theta_1,\theta_2,\theta_3)}$, shown in
Figure \ref{fig:3pt_multi}. We combine all $N_{\rm d}$ data points
into one data vector $\vec{d}$.  The key features of our analysis are
a non-Gaussian likelihood function that we estimate from the {\tt
  clone} and a data compression to suppress the numerical noise in the
final result.  We compare the results of this analysis to the results
with a standard Gaussian likelihood. Moreover, our new approach allows
us to factor in model uncertainties as well as measurement noise.  We
include errors due to shape noise, cosmic variance, measurement errors
of source ellipticities, uncertainties in the Hubble parameter, and a
multiplicative error in the overall amplitude of the predicted matter
bispectrum. We embed everything into a Monte-Carlo scheme for which no
analytical form of the likelihood has to be specified; only
realisations of statistical errors have to be provided.

\subsection{Model parameters and priors}
 
For the comparison of our non-Gaussian likelihood to a standard
Gaussian likelihood, we jointly constrain the two cosmological
parameters of a flat $\Lambda\rm CDM$ cosmology with a cosmological
constant \mbox{$\Omega_\Lambda=1-\Omega_{\rm m}$}: the matter density
$\Omega_{\rm m}$ and the r.m.s. dispersion $\sigma_8$ of matter
fluctuations on a scale of $8h^{-1}\rm Mpc$ linearly evolved to the
Universe of today. Moreover, we assume for the Hubble parameter
\mbox{$H_0=h\,100\,\rm km\,s^{-1}$} a Gaussian prior with mean
\mbox{$h=0.7$} and relative variance of $5\%$, compatible with the
CMB-only constraints of WMAP5 for a flat universe. We compile the
model parameters into the parameter vector \mbox{$\vec{p}=(\Omega_{\rm
    m},\sigma_8,h)$}.  All other cosmological parameters are fixed to
their WMAP5 best-fit values to be consistent with the {\tt clone}; see
Sect. \ref{sec:method} and \ref{sec:model} for all relevant model
parameters.  We denote all priors by the probability density function
(PDF) $P_p(\vec{p})$.

In addition, we assume that the overall amplitude of $\ave{M_{\rm
    ap}^3}$ can only be predicted up to multiplicative factor $1+f$
with a Gaussian prior $P_f(f)$ of r.m.s. 20\% and mean
$\ave{f}=0$. This directly accounts for the spread in predictions for
the matter bispectrum by competing models (see
Fig. \ref{fig:3pt_clone2}).

\subsection{Likelihood}

Let $\vec{d}$ be a vector of observables in an experiment, such as our
measurements of $\ave{M^3_{\rm ap}}$. For the statistical analysis of
these data $\vec{d}$, we can distinguish between two categories of
errors.

Firstly, errors originating from the theory side, or \emph{theory
  errors}, which depend on the model parameters $\vec{p}$ in
general. Theory errors are present if a model $\vec{m}(\vec{p})$
cannot perfectly fit $\vec{d}$ in the total absence of measurement
noise.  In our analysis, three kinds of theory error are accounted
for: cosmic variance, intrinsic shapes of the sources, and model
uncertainties in the (cosmic average) matter bispectrum. Cosmic
variance arises because our model predicts the cosmic average of
$\ave{M^3_{\rm ap}}$ but not the statistics for a particular volume of
the Universe. Intrinsic shapes of sources are not observable although
they can practically be inferred with reasonable assumptions. Hence
intrinsic galaxy shapes are nuisance parameters in our model that we
marginalise over.  And, as reflected by Fig. \ref{fig:3pt_clone2},
there is a set of theoretical models for the matter bispectrum
currently available that vary among themselves by about 20\% in
amplitude for the range of angular scales that we study here. We
parametrise this particular theory uncertainty by the nuisance
parameter $f$.

Secondly, even if a model fits the data we expect residuals of the fit
due to errors in the measurement process: the \emph{measurement
  noise}. Measurement noise is by definition only related to the
observables and hence unrelated to $\vec{p}$. Our main source of
measurement errors, and the only one we account for here, are
uncertainties in the galaxy ellipticities $\vec{\epsilon}_i$. Other
conceivable sources such as source positions or their redshifts are
neglected here.

In our model of $\vec{d}$, we include theory and measurement errors by
the Ansatz
\begin{equation}
  \vec{d}=
  (1+f)\vec{m}(\vec{p})+\Delta\vec{d}(\vec{p})
  =: m_f(\vec{p})+\Delta\vec{d}(\vec{p})\;,
\end{equation}
where $f$ is the multiplicative error in the model amplitude
$\vec{m}(\vec{p})$, while $\Delta\vec{d}(\vec{p})$ comprises all
remaining theory and measurement errors combined. In more complex
applications, theory errors and measurement errors could be
separated and simulated independently from each other. On average we
have $\ave{\Delta\vec{d}(\vec{p})}=0$ and $\ave{f}=0$. Our model of
the likelihood function is then as follows.  We denote by $P_{\Delta
  d}(\Delta\vec{d}|\vec{p})$ the probability density of an error
$\Delta\vec{d}$ given the parameters $\vec{p}$. To include the error
of $f$, we then write the likelihood ${\cal L}(\vec{d}|\vec{p})$ of
$\vec{d}$ given $\vec{p}$ as
\begin{equation}
  \label{eq:likelihood}
  {\cal L}(\vec{d}|\vec{p})
  =
  \!\!\int\d f\,P_f(f)
  P_{\Delta d}\Big(\vec{d}-\vec{m}_f(\vec{p})\Big|\vec{p}\Big)\;.
\end{equation}
Up to a normalisation constant $E(\vec{d})$ the posterior PDF of
$(\Omega_{\rm m},\sigma_8)$ given $\vec{d}$ is therefore
\begin{equation}
  \label{eq:posterior}
  P_p(\Omega_{\rm m},\sigma_8|\vec{d})=
  E^{-1}(\vec{d})\int\d h\,P_p(\vec{p}){\cal L}(\vec{d}|\vec{p})\;.
\end{equation}
The exact value of the so-called evidence $E(\vec{d})$ is irrelevant
for this analysis and hence set to unity. The Hubble parameter $h$ is
weakly constrained by the data. Therefore, we marginalise the
posterior over $h$.

A reasonable and common approximation of the likelihood ${\cal
  L}(\vec{d}|\vec{p})$ would be a multivariate Gaussian PDF as, for
instance, applied in F14. As shown in Sect. \ref{sec:evidence} by our
simulation of the CFHTLenS data set, however, the $\ave{M^3_{\rm ap}}$
measurement exhibits deviations from Gaussian statistics. This
motivates us to apply a non-Gaussian model of the likelihood. Towards
this goal, we outline in the following an algorithm that estimates
${\cal L}(\vec{d}|\vec{p})$ based on a discrete set of Monte-Carlo
realisations of $\Delta\vec{d}(\vec{p})$ for the posterior
$P_p(\Omega_{\rm m},\sigma_8|\vec{d})$.

\subsection{Monte-Carlo sampling of likelihood}

An opportunity for an approximation of ${\cal
  L}(\vec{d}|\vec{p})$ by a Monte-Carlo process can be seen by
re--writing the PDF $P_{\Delta d}$ as
\begin{eqnarray}
  P_{\Delta d}(\Delta\vec{d}|\vec{p})
  &=&
  \int\d\Delta\vec{x}\,
  P_{\Delta
    d}(\Delta\vec{x}|\vec{p})
  \delta_{\rm D}\Big(\Delta\vec{d}-\Delta\vec{x}\Big)
  \\
  &\approx&
  \frac{1}{N_{\Delta d}}\sum_{k=1}^{N_{\Delta d}}
  \delta_{\rm D}\Big(\Delta\vec{d}-\Delta\vec{x}_k\Big)\;
\end{eqnarray}
with $\delta_{\rm D}(\vec{x})$ being the Dirac delta function.  The
sum in last line approximates the integral in the first line by a
numerical Monte-Carlo integration with $N_{\Delta d}$ points
$\Delta\vec{x}_k$ \citep{1992nrca.book.....P}. For this approximation,
we produce, for a given $\vec{p}$, $N_{\Delta d}$ random realisations
of $\Delta\vec{x}$ from $P_{\Delta d}(\Delta\vec{x}|\vec{p})$ by means
of the {\tt clone}. We denote this process simply by
\mbox{$\Delta\vec{x}_k\curvearrowleft P_{\Delta
    d}(\Delta\vec{x}|\vec{p})$} in the following.  In the limit
\mbox{$N_{\Delta d}\to\infty$}, the number density of the points
$\Delta\vec{x}_k$ at $\Delta\vec{d}$ converges to $P_{\Delta
  d}(\Delta\vec{d}|\vec{p})$ up to a normalisation constant. Even for
finite $N_{\Delta d}$, the number density of points still provides an
useful estimator for the likelihood. This is the basis of our
Monte-Carlo scheme.  For this scheme, we address the infinite noise in
the previous estimator by smoothing the number density of sampling
points.

\subsection{ICA-based interpolation}
\label{sec:ica}

Let \mbox{$\Delta\vec{x}_k\curvearrowleft P_{\Delta
    d}(\Delta\vec{x}|\vec{p})$} be a sample of $N_{\Delta d}$
simulated discrete points with average
\mbox{$\ave{\Delta\vec{x}_k}=0$}. In addition, let the matrix
\mbox{$\mat{N}=\sum_k\Delta\vec{x}_k\Delta\vec{x}_k^{\rm T}/N_{\Delta
    d}$} be an estimator of the covariance of $\Delta\vec{x}_k$. Our
aim is to use the smoothed local number density of the set of points
$\{\Delta\vec{x}_k\}$ at $\Delta\vec{d}$ as approximation of
$P_{\Delta d}(\Delta\vec{d}|\vec{p})$.  To this end, we exploit the
independent component analysis (ICA) to factorise the sparsely sampled
higher-dimensional likelihood $P_{\Delta d}(\Delta\vec{d}|\vec{p})$
into a product of one dimensional (1D) histograms as in
\citet{2009A&A...504..689H}. We refer the reader to this paper for a
general discussion of the ICA and only briefly summarise the practical
steps here.

The ICA defines a linear transformation $\mat{M}$ and new coordinates
\mbox{$\vec{c}:=\mat{M}\mat{N}^{-1/2}\Delta\vec{d}$} that allow us to
express the $N_{\rm d}$-dimensional PDF
\begin{equation}
  P_{\Delta d}(\Delta\vec{d}|\vec{p})\approx
  \det{\mat{M}\mat{N}^{-1/2}}\times
  \prod_{i=1}^{N_{\rm d}}
  P_c^{(i)}\big([\mat{M}\mat{N}^{-1/2}\Delta\vec{d}]_i\big|\vec{p}\big)
\end{equation}
as a product of the 1D probability densities
$P_c^{(i)}(c_i|\vec{p})$. The constant pre-factor on the right-hand
side is the Jacobian of the mapping
$\mat{M}\mat{N}^{-1/2}$. Therefore, by means of the ICA all components
$c_i$ of $\vec{c}$ are made statistically independent.  Note that for
a Gaussian $P_{\Delta d}(\Delta\vec{d}|\vec{p})$ the ICA is
essentially a principal component analysis. We estimate the
transformation matrix $\mat{M}$ from $\{\Delta\vec{x}_k\}$ by
application of the \texttt{fastICA} algorithm \citep{761722}. The
matrix $\mat{M}$ is a \mbox{$N_{\rm d}\times N_{\rm d}$} square matrix
in our case, i.e., the number of components of $\vec{c}$ and
$\Delta\vec{d}$ are the same.

For the sample $\{\Delta\vec{x}_k\}$, the ICA transformation then
gives us a new set of $N_{\Delta d}$ sampling points
\mbox{$\vec{c}_k:=\mat{M}\mat{N}^{-1/2}\Delta\vec{x}_k$}.  We denote
the $i$-components of $\vec{c}_k$ by
\mbox{$c_{ki}=[\vec{c}_{k}]_i$}. For a fixed $i$, we adaptively smooth
the 1D frequency distribution of these points $\{c_{ki}\}$ to produce
the histogram $Q_{\rm ica}^{(i)}(x;c_{ki})$ and use its value at
\mbox{$x=c_i$} as an estimator of
$P_c^{(i)}(c_i|\vec{p})$. Specifically, to obtain $Q_{\rm
  ica}^{(i)}(x;c_{ki})$ we apply a $10$th neighbour adaptive
smoothing: for a given value $x$ we find the distance
\mbox{$d_{10,i}(x):=|x-c_{10,i}|$} to the 10th nearest neighbour
sampling point \mbox{$c_{10,i}\in\{c_{ki}\}$} for fixed $i$ and then
compute
\begin{equation}
  Q_{\rm ica}^{(i)}(x;c_{ki})=\frac{10}{d_{10,i}(x)}\;.
\end{equation}
In doing so, we bin for every index $i$ the 1D distribution of values
$\{c_{ki}\}$ with an adaptive bin width $d_{10,i}(x)$ that depends on
the local density of the points $\{c_{ki}\}$ at $x$
\citep{0132.38905}. We find that this technique is more robust
compared to the Gaussian kernel method with fixed kernel size in
\citet{2009A&A...504..689H}, if the PDF-sampling has a few extreme
outliers in the tail of the distribution.

Finally, our approximation of the logarithm of $P_{\Delta
  d}(\Delta\vec{d}|\vec{p})$ is given by
\begin{eqnarray}
  \label{eq:loglike}
  \ln{P_{\Delta d}(\Delta\vec{d}|\vec{p})}&\approx&
  \ln{Q_{\rm ica}(\Delta\vec{d};\Delta\vec{x}_k)}\\
  &:=&\nonumber
  \eta+\sum_{i=1}^{N_{\rm d}}
  \ln{Q_{\rm ica}^{(i)}\Big([\mat{M}\mat{N}^{-1/2}\Delta\vec{d}]_i;c_{ki}\Big)}
\end{eqnarray}
where $\e^\eta$ is the normalisation of $Q_{\rm
  ica}(\Delta\vec{d};\Delta\vec{x}_k)$.  The normalisation $\eta$ can
be ignored in our case because for this study it is identical for
every position in the parameter space.

\subsection{Data  compression}
\label{sec:KL}

\begin{figure}
  \begin{center}
    \psfig{file=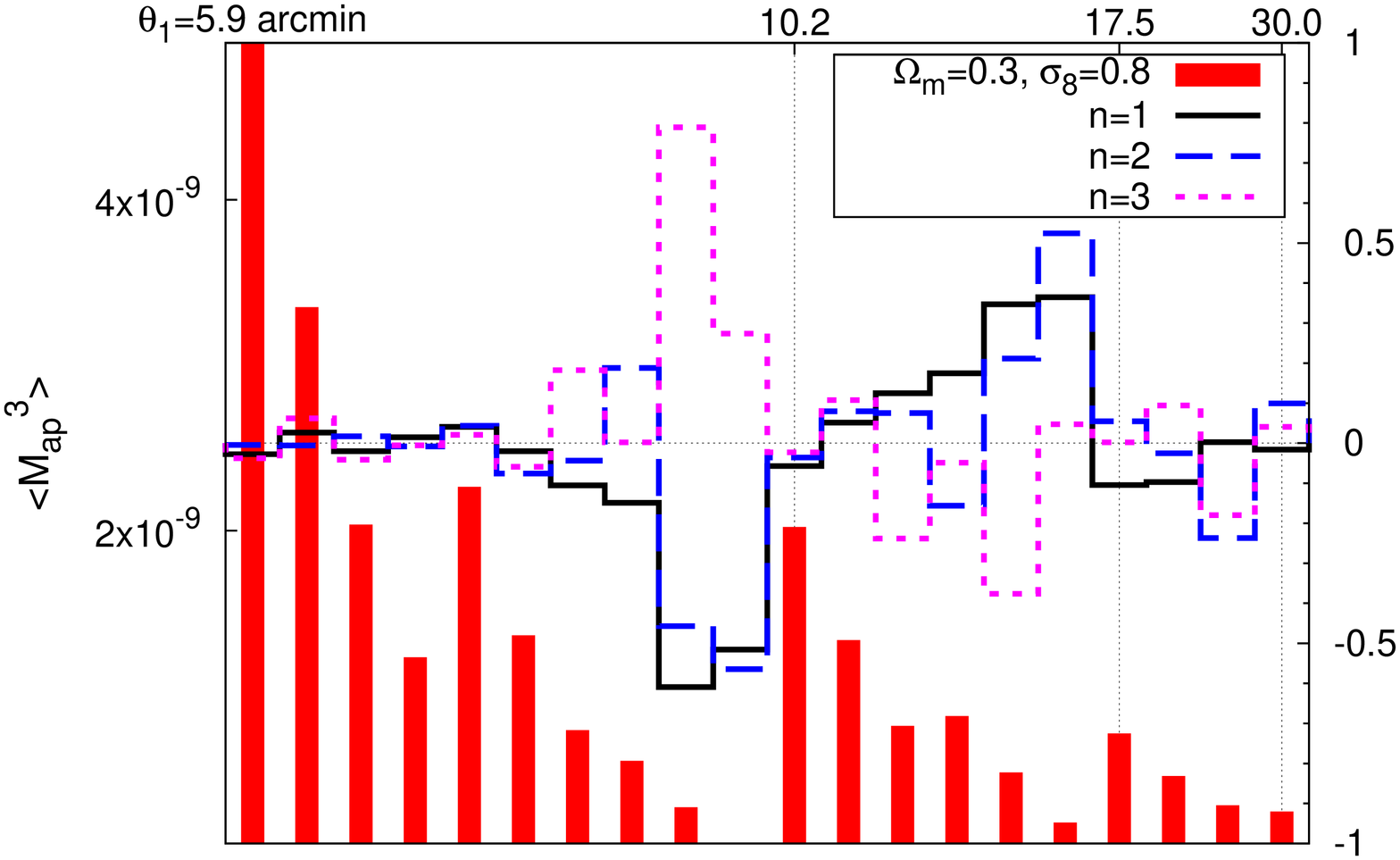,width=85mm,angle=-90}
    \psfig{file=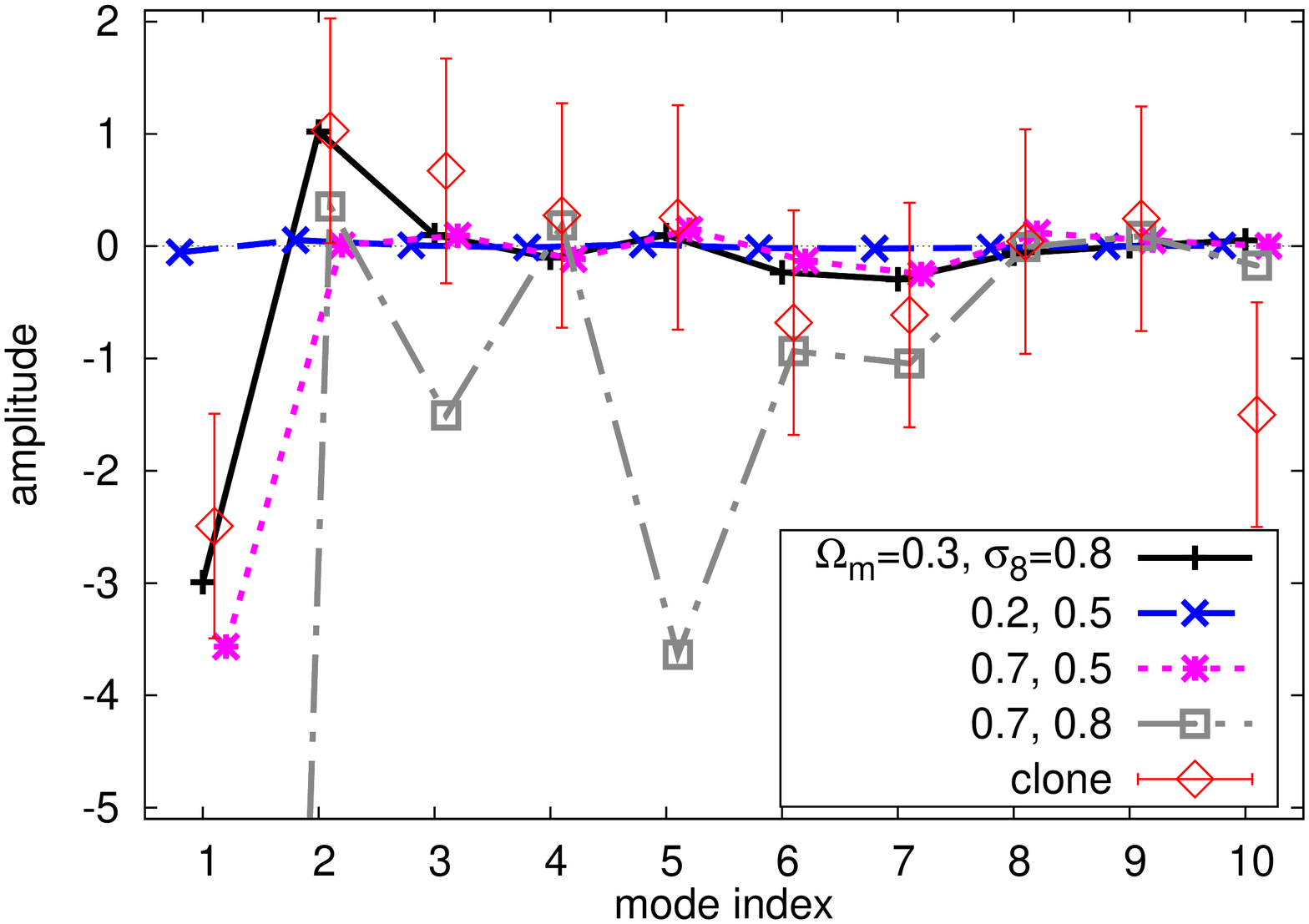,width=85mm,angle=-90}
  \end{center}
  \caption{\label{fig:KLexample} Illustration of the
    Karhunen-Lo\`{e}ve data compression. All models assume
    $h=0.7$. \emph{Top panel:} The red sticks show simulated data
    $\ave{M_{\rm ap}^3}$ with $N_{\rm d}=20$ elements sorted by
    increasing $\theta_1\le\theta_2\le\theta_3$;
    $\theta_i=5.9,10.2,17.5,30.0$ arcmin.  The KL modes are averages
    of all data points with relative weights as indicated by the lines
    for the mode indices $n=1,2,3$ (right hand $y$-axis).
    \emph{Bottom panel:} Amplitudes of KL modes as function of the
    mode index for four different noise-free model vectors (lines);
    the open square data point at \mbox{$n=1$} is off the chart at an
    amplitude of approximately -30. The open diamonds show a one noisy
    {\tt clone} data vector in comparison.}
\end{figure}

The ICA interpolation technique is prone to a bias that
under-estimates the size of the credible intervals of the model
parameters \citep{2009A&A...504..689H}. This bias is large if
\mbox{$N_{\rm d}/N_{\Delta d}\sim1$} and small for \mbox{$N_{\rm
    d}/N_{\Delta d} \ll 1$}; here \mbox{$N_{\Delta d}=200$} is our
number of sampling points of the likelihood and \mbox{$N_{\rm d}=20$}
is the size of the uncompressed data vector. A similar problem affects
traditional Gaussian likelihood analyses when the inverse covariance
has been estimated from simulations
\citep{2007A&A...464..399H,2013MNRAS.432.1928T}.  In order to reduce
this bias, we compress the size $N_{\rm d}$ of the data vector
$\vec{d}$, and we apply the same compression to the realisations
$\Delta\vec{x}_k$ in the ICA interpolation. This data compression also
has the benefit to produce a smoother posterior of $\vec{p}$ as the
sampling noise in the interpolated likelihood is reduced.

Our data compression is based on a Karhunen-Lo\`{e}ve (KL) transform
\citep{1997ApJ...480...22T,2006MNRAS.366..983K,2014AS}. It identifies
$N_{\rm d}$ signal-to-noise (SN) modes of $\vec{d}$ and rejects modes
with a low SN.  This procedure allows us to find a linear projection
\mbox{$\vec{d}^\prime=\mat{C}\vec{d}$} that reduces the number of
elements of $\vec{d}$ to \mbox{$N_{\rm d^\prime}<N_{\rm d}$}.  The
projection matrix $\mat{C}$ is chosen such that most of the
cosmological information on $\vec{p}$ is still available in
$\vec{d}^\prime$. Importantly, the KL transform -- or any projection
matrix $\mat{C}$ for that matter -- at most decreases the constraints
on $\vec{p}$ but does not bias the result provided the data can be fit
by the model. The KL transform affects the cosmological analysis and
the ICA decomposition only in that far that we now use as data vector
$\vec{d}^\prime$ instead of $\vec{d}$, the compressed model vector
$\vec{m}^\prime(\vec{p})=\mat{C}\vec{m}(\vec{p})$, and the Monte-Carlo
realisations $\Delta\vec{x}_k^\prime=\mat{C}\Delta\vec{x}_k$.

For a data compression matrix $\mat{C}$, we would like to identify
linear combinations of the components of $\vec{d}$ that vary strongly
when changing the parameters $\vec{p}$ (signal variance). On the other
hand, in the presence of measurement noise and theory errors, we would
also like to identify combinations that are most significant compared
to noise. The KL transform finds a compromise of both. This means the
KL transform identifies modes $\vec{e}_i$ in $\vec{d}$-space that have
large ratios of signal variance and noise variance. The compression
matrix $\mat{C}=(\vec{e}_1,\ldots,\vec{e}_{N_{\rm d^\prime}})^{\rm T}$
comprises the $N_{\rm d^\prime}$ KL modes with the highest SN ratio.
To construct $\mat{C}$, we proceed in two steps.

First, we compute the signal covariance $\mat{S}$ of model vectors
$\vec{m}(\vec{p})$ over a volume $V$ in parameter space that is
defined by our prior information $P_p(\vec{p})$ on the cosmological
parameters,
\begin{equation}
  \mat{S}
  =\int\frac{\d\Omega_{\rm m}\d\sigma_8\d h\,P_p(\vec{p})}{V}
  \big(\vec{m}(\vec{p})-\ave{\vec{m}}\big)
  \big(\vec{m}(\vec{p})-\ave{\vec{m}}\big)^{\rm T}\;,
\end{equation}
where \mbox{$V=\int\d\Omega_{\rm m}\d\sigma_8\d h P_p(\vec{p})$}, and
\begin{equation}
  \ave{\vec{m}}
  =
  \int\frac{\d\Omega_{\rm m}\d\sigma_8\d h\,
  P_p(\vec{p})}{V}\vec{m}(\vec{p})
\end{equation}
is the mean of $\vec{m}(\vec{p})$. For the computation of $\mat{S}$ by
integration, we adopt a flat prior $P_p(\vec{p})$ for
\mbox{$\Omega_{\rm m}\in[0.1,1.0]$}, \mbox{$\sigma_8\in[0.4,1.0]$},
and $h\in[0.6,0.8]$. The previous integrals could in principle be
estimated by a Monte-Carlo process for which we (i) randomly and
repeatedly draw parameters $\vec{p}$ from the prior $P_p(\vec{p})$,
(ii) compute $\vec{m}(\vec{p})$, and (iii) determine the mean
$\ave{\vec{m}}$ and covariance $\mat{S}$ of all realisations of
$\vec{m}(\vec{p})$. The Eigenvectors of $\mat{S}$ with the largest
Eigenvalue determine modes of $\vec{d}$ that are most sensitive with
regard to $\vec{p}$.

Second, we compute the noise covariance, which is simply the
covariance matrix $\mat{N}$ of $\Delta\vec{x}$ for a chosen fiducial
cosmology (Sect. \ref{sec:ica}). The data compression is thus optimal
for the fiducial model.  The KL modes are then the Eigenvectors
$\vec{e}_i$ of the generalised Eigenproblem
\begin{equation}
  \label{eq:KL}
  \mat{S}\vec{e}_i=\lambda_i\mat{N}\vec{e}_i\,, 
\end{equation}
where $\lambda_i$ are the Eigenvalues of $\vec{e}_i$; the SN ratio of
the mode $\vec{e}_i$ is $\sqrt{\lambda_i}$.

In the top panel of Fig. \ref{fig:KLexample}, we show simulated data
$\ave{M_{\rm ap}^3}$ for our \mbox{$N_{\rm d}=20$} combinations of
aperture radii (red bars). The KL modes are linear combinations of the
aperture radii with relative weights defined by the components of
$\vec{e}_i$.  These weights are displayed as lines for the first three
modes (right hand $y$-axis). Importantly, the weights can have
positive and negative signs. The lines show that our KL modes are
mostly sensitive to large aperture scale radii with
$\theta_i\ge10.2\,\rm arcmin$. The bottom panel shows as lines four
compressed model vectors $\mat{C}\vec{m}(\vec{p})$ that do not contain
noise in comparison to one noisy {\tt clone} data vector.  The error
bars reflect the $1\sigma$ uncertainty for noise as in our CFHTLenS
data. Except for models with large values of $\Omega_{\rm m}$ and
$\sigma_8$ (grey dashed-dotted line) all models are essentially zero
for KL mode indices larger than three.  This means their cosmological
information is mostly inside the first couple of modes. Note that
errors between different KL modes are uncorrelated and of equal
variance by construction, i.e., $\ave{\Delta d^\prime_i\Delta
  d^\prime_j}=\delta^{\rm K}_{ij}$. Nevertheless, in the case of a
non-Gaussian statistics, there may be higher-order correlations such
as \mbox{$\ave{\Delta d^\prime_i\Delta d^\prime_j\Delta
    d^\prime_k}\ne0$} present in the data.

\subsection{Bias correction}

After a data compression with \mbox{$N_{\rm d^\prime}\le10$} and the
following correction of the likelihood, we expect the bias in the size
of the credible intervals of our sampled posterior to be negligible
for the following reason. \citet{2009A&A...504..689H} report that this
bias is similar to the one known for a Gaussian likelihood for which
the data covariance is estimated from realisations of the data. In the
Gaussian case, to de-bias the likelihood the inverse of the estimated
covariance has to be rescaled by a factor $f_{\rm corr}=(N_{\Delta
  d}-N_{\rm d^\prime}-2)/(N_{\Delta d}-1)$ or, equivalently, ${\cal
  L}\mapsto{\cal L}^{f_{\rm corr}}$. We apply the latter correction to
our sampled likelihood $Q_{\rm
  ica}(\Delta\vec{d}^\prime;\Delta\vec{x}_k^\prime)$ in
Eq. \Ref{eq:loglike}.  However, this correction is small since
\mbox{$f_{\rm corr}\ge0.94$} for \mbox{$N_{\rm d^\prime}\le10$}.

\subsection{Sampling algorithm of posterior}
\label{sec:like}

We compute the posterior $P_{\rm p}(\Omega_{\rm m},\sigma_8|\vec{d})$
on a $100\times100$ grid that covers the $(\Omega_{\rm
  m},\sigma_8)$-plane. To compute the posterior value for a
  grid pixel we proceed in four steps:
\begin{enumerate}
\item Generation of a set of $N_{\Delta d}$ realisations $\vec{d}_k$
  to sample the error $P_{\Delta d}(\Delta\vec{d}|\vec{p})$ for a
  given $\vec{p}$: We obtain the $\vec{d}_k$ by measuring $\ave{M_{\rm
      ap}^3}$ on the {\it noisy-sample} of the {\tt clone}, hence both
  the intrinsic galaxy shapes and the cosmic variance are
  simulated. In the simulation, the 1D variance of intrinsic source
  ellipticities is set to \mbox{$\sigma_\epsilon=0.28$} to account for
  both the true intrinsic variance and the measurement error of
  ellipticities; both are assumed to be Gaussian. From the {\tt clone}
  patches, we compute a simulated $\ave{M^3_{\rm ap}}$ for an
  effective CFHTLenS survey area of roughly 100 square degree by
  combining the {\tt clone} measurements of seven randomly selected
  sight lines. Each {\tt clone} line-of-sight, out of 184 in total,
  covers $12.84\,{\rm deg^2}$. Repeating this procedure
  \mbox{$N_{\Delta d}=200$} times gives us the sample
  \mbox{$\Delta\vec{x}_k=\vec{d}_k-\vec{\bar{d}}$} where
  $\vec{\bar{d}}$ expresses the average of all realisations
  $\vec{d}_k$.
\item Determination of the data compression matrix $\mat{C}$ for a
  given covariance $\mat{N}$ of $\{\Delta\vec{x}_k\}$, models
  $\vec{m}(\vec{p})$, and the model prior $P_p(\vec{p})$: see
  Sect. \ref{sec:KL} for details. From this we compute the compressed
  vectors \mbox{$\vec{d}^\prime=\mat{C}\vec{d}$} and
  \mbox{$\Delta\vec{x}_k^\prime=\mat{C}\Delta\vec{x}_k$}.  We compute
  the compression matrix $\mat{C}$ once for a fiducial cosmology and
  use it for all $\vec{p}$.
\item ICA factorisation of the compressed $P_{\Delta
    d^\prime}(\Delta\vec{d}^\prime|\vec{p})$ into a product of
  interpolated, 1D histograms $Q_{\rm ica}^{(i)}(x;c_{ki})$ by means
  of an ICA of the sample $c_{ki}=[\Delta\vec{x}_k^\prime]_i$: see
  Sect. \ref{sec:ica} for details.
\item Marginalisation over the multiplicative factor $f$ and the
  Hubble parameter $h$ to obtain the posterior value $P_p(\Omega_{\rm
    m},\sigma_8|\vec{d}^\prime)$: We perform the marginalisation by another
  Monte-Carlo integration that draws \mbox{$N_{\rm fh}=500$} values
  $f_i\curvearrowleft P_f(f)$ and the same amount of values
  $h_i\curvearrowleft N(0.7,\sigma_{\rm h})$ from a Gaussian prior for
  the Hubble parameter. With these values we compute
  \begin{equation}
    \label{eq:likelihood2}
    P_p(\Omega_{\rm m},\sigma_8|\vec{d}^\prime)\approx
    \frac{1}{N_{\rm fh}}\sum_{i=1}^{N_{\rm fh}}
    \left[Q_{\rm ica}(\vec{d}^\prime-\vec{m}^\prime_{{\rm f},i}(\vec{p}_i);\Delta\vec{x}_{k}^\prime)\right]^{f_{\rm corr}}\;,
  \end{equation}
  where \mbox{$\vec{m}_{{\rm f},i}^\prime(\vec{p}_i)=\vec{m}^\prime(\vec{p}_i)(1+0.2f_i)$}
  is the rescaled model $\vec{m}^\prime(\vec{p}_i)$ for
  $\vec{p}_i=(\Omega_{\rm m},\sigma_8,h_i)$. The approximation $Q_{\rm
    ica}(\Delta\vec{d}^\prime;\Delta\vec{x}_k^\prime)$ depends on the factors
  $Q_{\rm ica}^{(i)}(x;c_{ki})$, see Eq. \Ref{eq:loglike}.
\end{enumerate}

Ideally, step (i) is repeated for any new value of $\vec{p}$ based on
a simulation for a fiducial cosmology with parameters $\vec{p}$.  To
keep the simulation effort viable, however, we assume that the scatter
of $\vec{d}_k$ is as in the {\tt clone} for any $\vec{p}$ in the
entire parameter space explored. Therefore step (i) only has to be
performed once. For a Gaussian likelihood, this assumption would be
equivalent to using the same likelihood covariance for all $\vec{p}$,
which is indeed a common assumption as e.g. in F14 (for a discussion
see \citealt{2009A&A...502..721E} and
\citealt{2013MNRAS.430.2200K}). In a future application of our
technique, subject to less prohibitive computational constraints, a
computation of the sampled likelihood for different $\vec{p}$ is
conceivable.

Finally, it is likely that we moderately underestimate the scatter of
$\Delta\vec{x}_k$ in step (i) due to cosmic variance. The 184
realisations of the {\tt clone} allow only for 26 independent
realisations of a $100\,\rm deg^2$ survey; the cosmic variance between
the $\Delta\vec{x}_k$ is partially correlated.

\begin{figure}
  \begin{center}
    \psfig{file=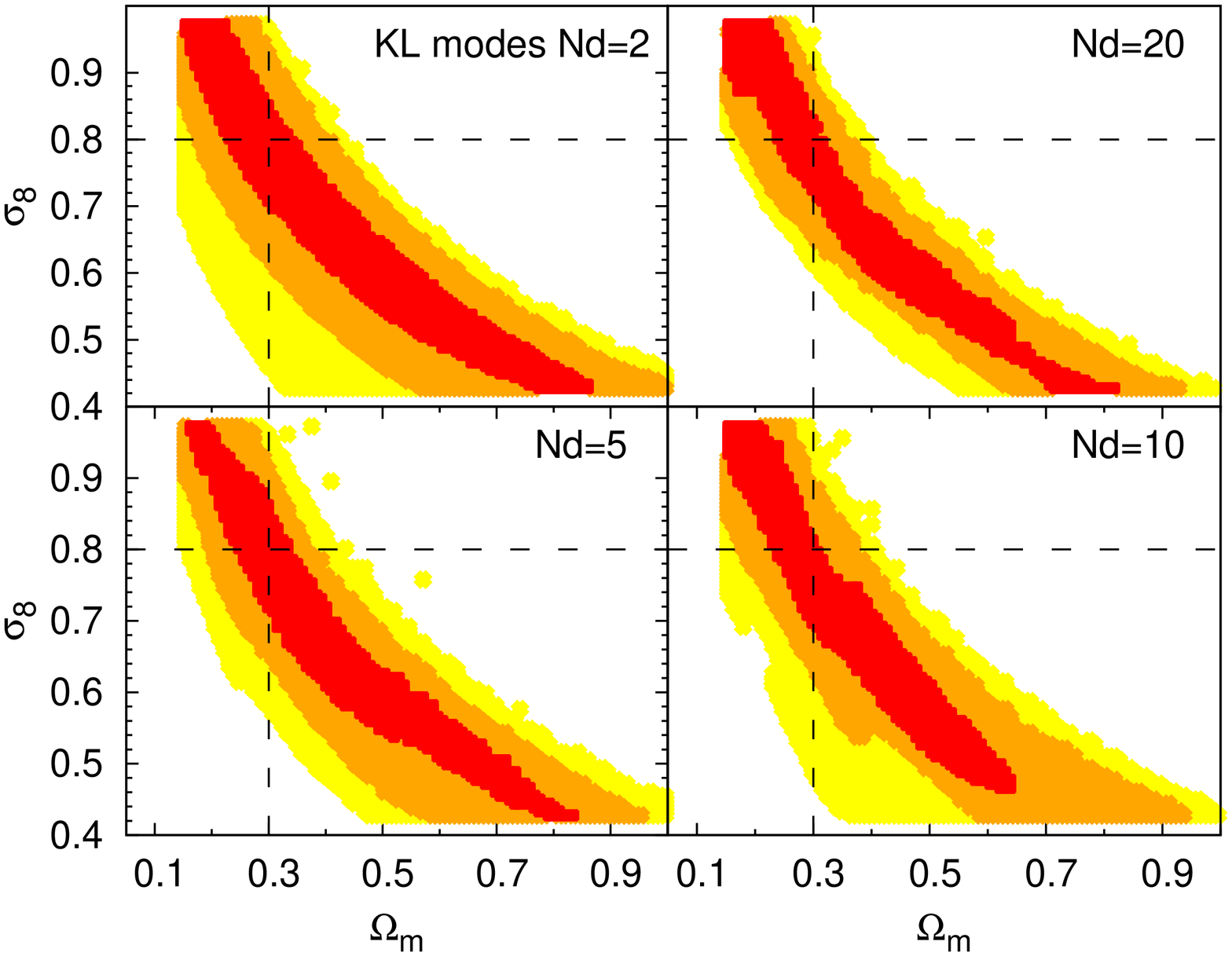,width=90mm,angle=-90}
    \psfig{file=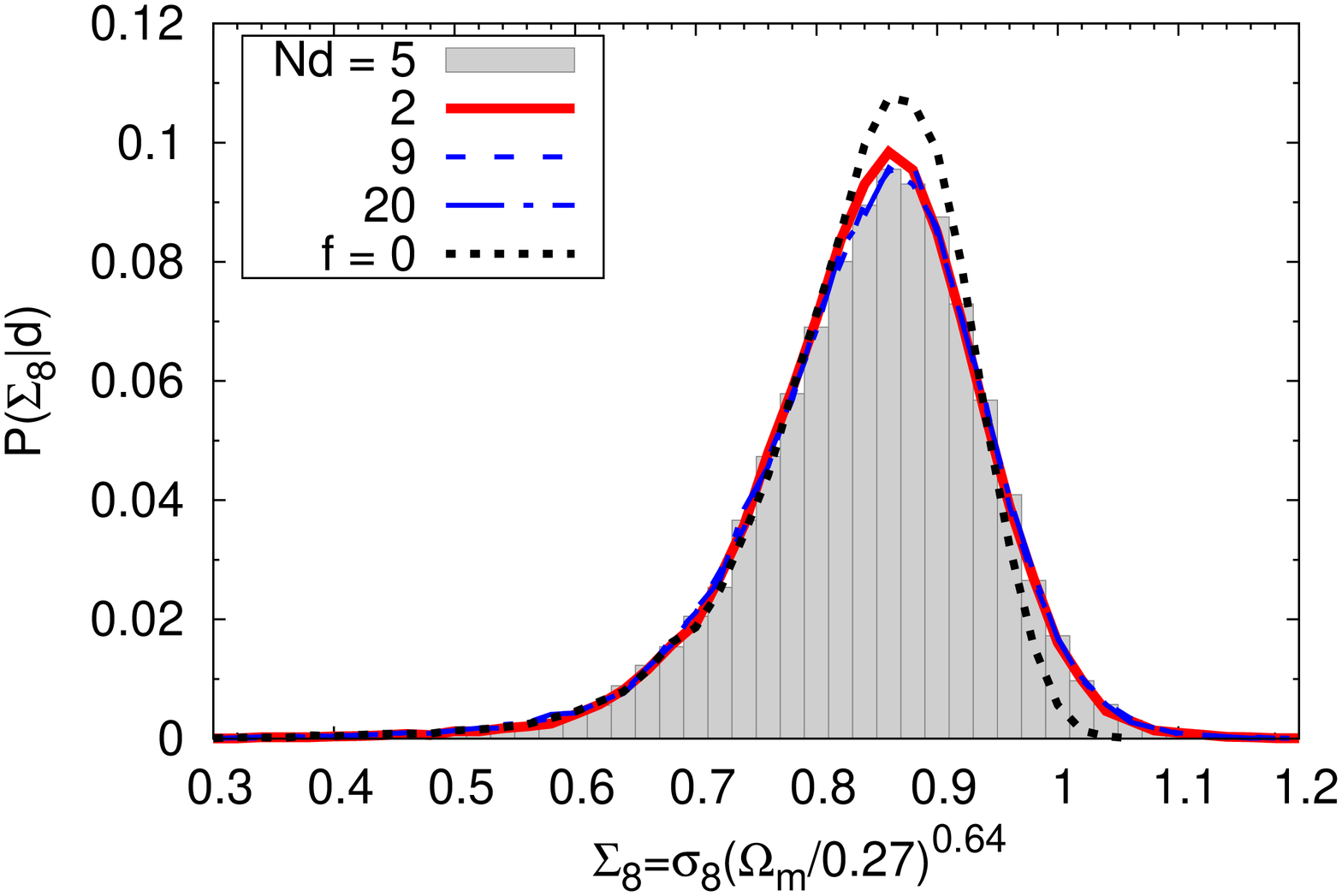,width=83mm,angle=-90}
    \psfig{file=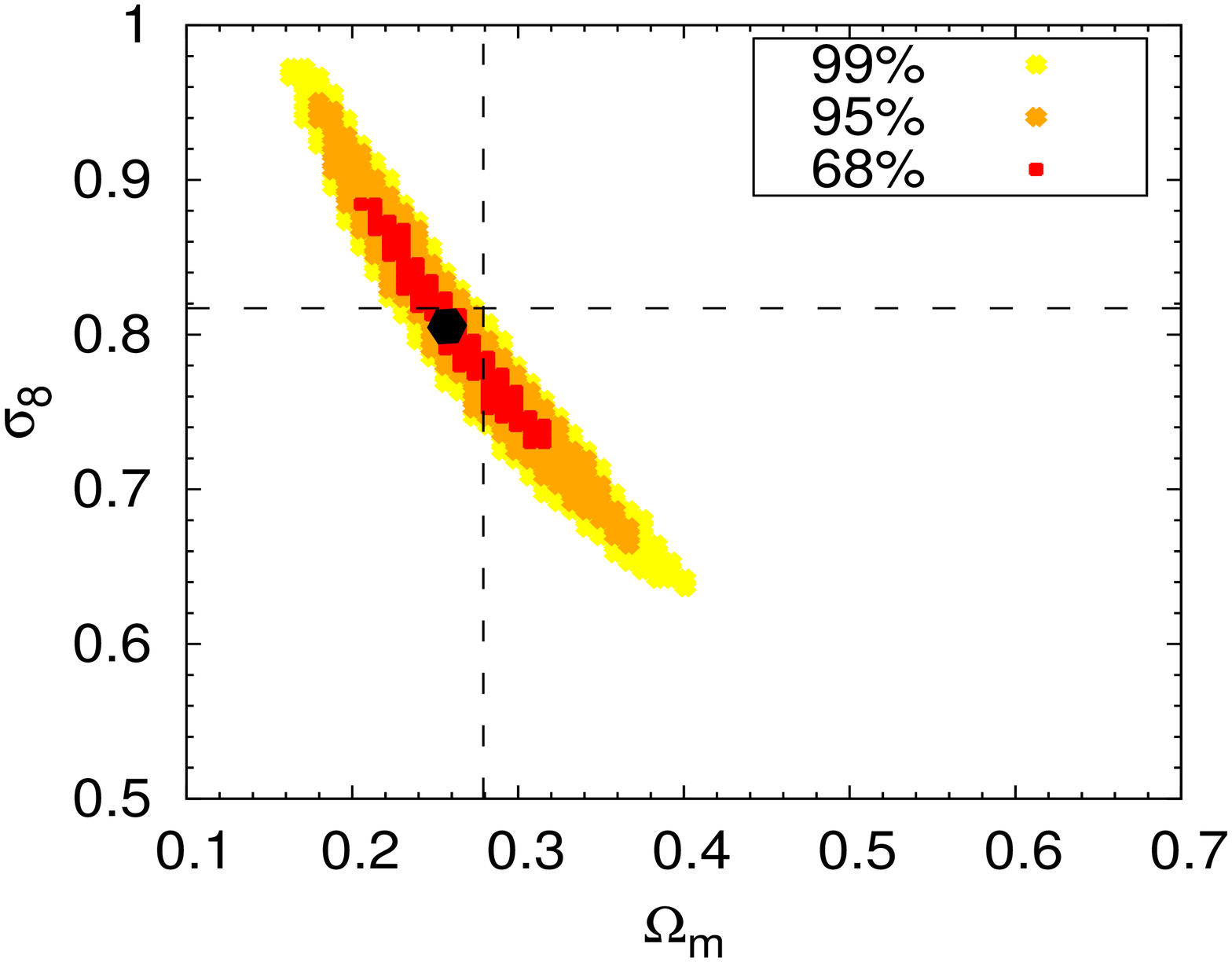,width=83mm,angle=0}
    \vspace{-0.5cm}
  \end{center}
  \caption{\label{fig:testrun} Results of our code tests running on
    mock data. \emph{Top panel:} A series of cosmological analyses
    using different degrees of data compression. All four runs use
    same simulated data vector with \mbox{$\Omega_{\rm m}=0.3$},
    \mbox{$\sigma_8=0.8$}, and \mbox{$h=0.7$} but a different number
    of KL modes as indicated. The coloured regions highlight the
    $68\%$, $95\%$, and $99\%$ credible regions for a flat
    cosmology. Insets with more KL modes \mbox{$N_{\rm d^\prime}$} are
    subject to more numerical noise. \emph{Middle panel:} The
    posterior of \mbox{$\Sigma_8=\sigma_8(\Omega_{\rm
        m}/0.27)^{0.64}$} for the test data vector for different
    degrees of compression and a Gaussian likelihood. The dotted black
    solid line uses \mbox{$N_{\rm d^\prime}=5$} and no error in the
    model amplitude (\mbox{$f=0$}). \emph{Bottom panel:} Results of a
    verification run of the analysis code. The credible regions
    reflect the combined constraints from 26 simulated
    \emph{noise-free} CFHTLenS measurements in the \texttt{clone}
    simulation. The posterior is offset with respect to the fiducial
    model ($\Omega_{\rm m}=0.279$, $\sigma_8=0.817$; dashed black
    lines).}
\end{figure}

\subsection{Tests of the posterior construction}

In this section, we test the robustness and accuracy of our code
implementation of Sect. \ref{sec:like} before its application to the
CFHTLenS data. 

We start by looking at different degrees of data compression in the
top panel of Fig. \ref{fig:testrun}. For this purpose, we use a
noise-free model vector $\vec{m}(\vec{p})$ with the parameters
$\vec{p}=(\Omega_{\rm m},\sigma_8,h)=(0.3,0.8,0.7)$ as input to the
cosmological analysis. We plot three credible regions of the resulting
posterior $P_p(\Omega_{\rm m},\sigma_8|\vec{d})$.  The maximum of all
posteriors coincides with the parameters of the input vector (not
shown). The figure supports our earlier expectation that only few KL
modes are required for the analysis: there is no clear improvement for
\mbox{$N_{\rm d^\prime}\ge3$}. The differences between the panels are
likely due to numerical noise in the sampled likelihood for the
following reason.  In Fig. \ref{fig:testrun} we plot the posterior of
the combined quantity \mbox{$\Sigma_8=\sigma_8(\Omega_{\rm
    m}/0.27)^{0.64}$} for the previous model data vector but devising
now a Gaussian likelihood. This likelihood model is not subject to
sampling noise as its analytical form is completely determined. As can
be seen, the posterior $P_p(\Sigma_8|\vec{d}^\prime)$ is virtually
unchanged for \mbox{$N_{\rm d^\prime}\ge3$}. Therefore, the first two
modes contain all cosmological information of the model data. The same
is presumably true for the non-Gaussian case so that all variation in
the top panel can be attributed to numerical noise.

As can be seen in Fig. \ref{fig:KLexample}, (unlikely) models with
values high in both $\Omega_{\rm m}$ and $\sigma_8$ exhibit some
signal until the mode \mbox{$n=7$}. Therefore, we set \mbox{$N_{\rm
    d^\prime}=5$} in our final CFHTLenS analysis as compromise between
sampling noise and cosmological information. Moreover, the middle
panel shows that the inclusion of a $20\%$ systematic error in the
model amplitude has a small impact on the posterior information (solid
black line). It slightly stretches the uncertainty toward larger
values of $\Sigma_8$.

For the bottom panel of Fig. \ref{fig:testrun}, we test the accuracy
of our analytical matter bispectrum and that of the likelihood
analysis by means of the {\tt clone}. We run our analysis code for 26
independent simulated CFHTLenS-like data vectors created using {\it
  noise-free} realisations; we combine the 184 patches into 26 groups
of seven data vectors.  The figure shows three credible regions of the
combined posterior of all 26 analyses. We attain the combined
posterior by adding the 26 individual logarithmic posteriors on a
grid.  Ideally, the resulting posterior should be consistent with the
cosmological parameters in the \texttt{clone} that we indicate by the
intersection of the two dashed lines.  The contours, however, indicate
with about $3\sigma$ a model bias due to which we do not perfectly
recover the \texttt{clone} cosmology: the centre of the contours,
indicated by the black point, is offset in $\sigma_8$ by a few percent
and in $\Omega_{\rm m}$ by roughly 10 percent. However, this offset is
still small compared to the CFHTLenS noise levels, which can be seen
by comparing offset in the bottom panel to the size of the credible
regions in the top panel. To assure that the bias is unrelated to
either the data compression or the ICA interpolation of the
likelihood, we also determine the maximum of the posterior for the
average, uncompressed \texttt{clone} data vector for a simple Gaussian
likelihood.  We find biased values consistent with the non-Gaussian
results of compressed data. Our interpretation is that the bias
originates from a small mismatch between the analytic model and the
{\tt clone} average that is still visible at 5 arcmin (EEE mode in
Fig. \ref{fig:3pt_clone}).

In conclusion, there is a systematic bias in our analytic model for
the third-order aperture statistics as revealed by our comparison to
N-body simulated data. Since this bias is small relative to the levels
of shape noise and cosmic variance in CFHTLenS we can ignore it for
the scope of this paper. For the analysis of next generation surveys,
on the other hand, more accurate models of $\ave{M_{\rm ap}^3}$ will
be needed.

\section{Results}
\label{sec:results}

\begin{figure}
  \begin{center}
    \psfig{file=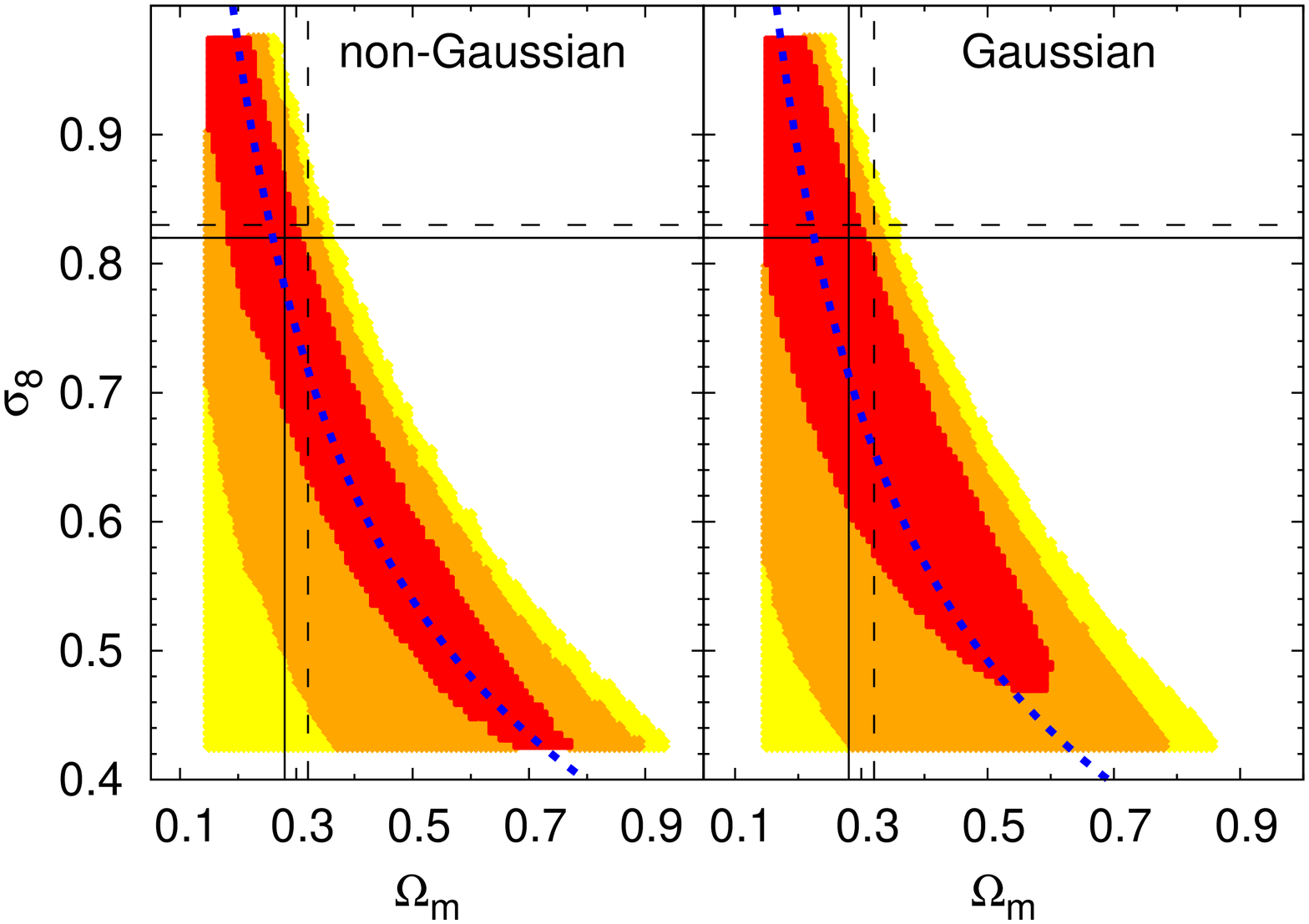,width=85mm,angle=-90}
    \psfig{file=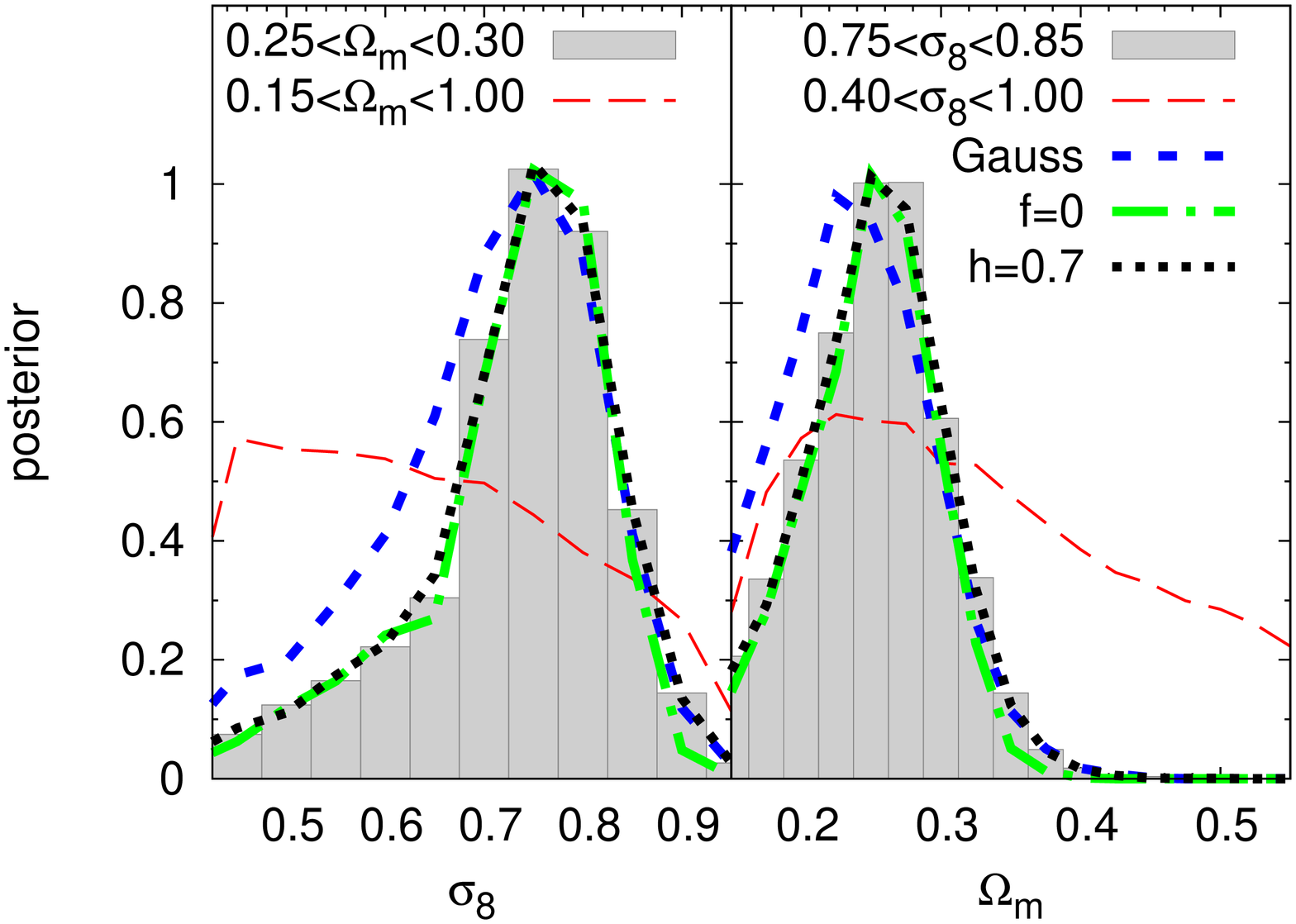,width=85mm,angle=-90}
    \vspace{-0.5cm}
  \end{center}
  \caption{\label{fig:cosmoresult1} \emph{Top panel:} Constraints in
    the \mbox{$\sigma_8-\Omega_{\rm m}$} plane from the CFHTLenS
    $\ave{M_{\rm ap}^3}$ for a flat $\Lambda\rm CDM$ cosmology with
    5\% uncertainty on $h=0.7$ and a 20\% uncertainty in the model
    amplitude (credibility red: 68\%, orange: 95\%, yellow: 99\%). The
    dashed lines indicate the Planck best-fit parameters, the solid
    lines WMAP9 parameters. The left inset depicts the posterior for a
    non-Gaussian likelihood, whereas the right inset is based on a
    Gaussian model. The dotted line on the left (right) corresponds to
    the best-fitting value of \mbox{$\Sigma_8=\sigma_8(\Omega_{\rm
        m}/0.27)^{0.64}=0.8 (0.73)$}. \emph{Bottom panel:}
    Marginalised cosmological results for $\sigma_8$ (left) and
    $\Omega_{\rm m}$ (right) alone with different top-hat priors on
    the parameters (as indicated).  The thick blue dashed line
    indicates constraints from the Gaussian likelihood in comparison
    to the non-Gaussian likelihood (grey area). The thick dotted black
    line are non-Gaussian constraints for a fixed \mbox{$h=0.7$},
    whereas the thick dotted-dashed green lines \mbox{$f=0$} do not
    include the $20\%$ multiplicative error in the model
    amplitude. Except for the red dashed lines, the posteriors are
    normalised to their value at the maximum.}
\end{figure}

We present our constraints of $\sigma_8$ and $\Omega_{\rm m}$ for the
CFHTLenS data for a flat $\Lambda\rm CDM$ model in
Fig. \ref{fig:cosmoresult1}. Therein we include only measurements of
the EEE mode of $\ave{M^3_{\rm ap}(\theta_1,\theta_2,\theta_3)}$ for
$\theta_i=5.9,10.2,17.5,30.0$ arcmin. In order to avoid confirmation
bias in the generation of this result, we blindly analysed four data
sets simultaneously of which three were noisy mock data and one was
the CFHTLenS data vector with the results shown here. The CFHTLenS
data vector was revealed only after the cosmological
analysis. Furthermore, we performed the analysis after the systematics
checks of the data and after our final decision on the usable range of
angular scales.  In particular, we did not use the CFHTLenS data
during the development and testing phase of the analysis code in any
way (Sect. \ref{sec:likelihood}).

In the top panel of Fig. \ref{fig:cosmoresult1}, we show the 68\%,
95\%, and 99\% credible regions for the joint constraints of the
matter density parameter $\Omega_{\rm m}$ and the amplitude $\sigma_8$
of the matter density fluctuations. We performed the analysis twice:
once with the non-Gaussian model of the likelihood for the left inset
(non-Gaussian), and once with a Gaussian likelihood for the right
inset (Gaussian). The Gaussian likelihood is based on the noise
covariance matrix $\mat{N}$ in Sect. \ref{sec:ica}. Both posteriors
use the same degree of data compression (five KL modes). For reference
we have indicated the slightly differing best-fitting values of WMAP9
and Planck as solid and dashed lines, respectively
\citep{2013ApJS..208...19H,2013arXiv1303.5076P}. In summary, there is
little difference in the posteriors, mostly visible between the 68\%
credible regions. Both constraints are highly degenerate and basically
only exclude simultaneously large values of $\Omega_{\rm m}$ and
$\sigma_8$. 

In order to break the degeneracy of the parameters for the bottom
panel of Fig. \ref{fig:cosmoresult1}, we assume an additional prior on
either $\Omega_{\rm m}$ (left) or $\sigma_8$ (right). The top-hat
priors are centred around \mbox{$\Omega_{\rm m}=0.275$} and
\mbox{$\sigma_8=0.8$}, close to the best-fitting values of the WMAP9
results. The widths of the top-hats are \mbox{$\Delta\Omega_{\rm
    m}=0.05$} and \mbox{$\Delta\sigma_8=0.1$}. The resulting
marginalised posteriors of the non-Gaussian data model are shown as
grey bars.  With these strong priors we infer \mbox{$\Omega_{\rm
    m}=0.27^{+0.05}_{-0.05}$} and
\mbox{$\sigma_8=0.77^{+0.07}_{-0.11}$} (68\% credibility).
Furthermore, without the imposed top-hat priors the marginalised
posterior of either parameter yields only weak constraints (thin
dashed red lines), which are essentially just the upper limits
\mbox{$\Omega_{\rm m}\le0.45(0.67)$} and
\mbox{$\sigma_8\le0.75(0.93)$} for a 68\% (95\%) credibility. These
upper limits depend on the adopted broad priors of
\mbox{$0.15\le\Omega_{\rm m}\le1$} and \mbox{$0.4\le\sigma_8\le1$} due
to the strong degeneracy of both parameters. Therefore, it is more
relevant to combine $(\Omega_{\rm m},\sigma_8)$ into
\mbox{$\Sigma_8=\sigma_8(\Omega_{\rm m}/0.27)^{0.64}$} as done in
F14. We plot the posterior of $\Sigma_8$ for the non-Gaussian data
model in the bottom panel of Fig. \Ref{fig:cosmoresult1} and obtain
\mbox{$\Sigma_8=0.79^{+0.08}_{-0.11}$} (grey area).

\begin{figure}
  \begin{center}
    \psfig{file=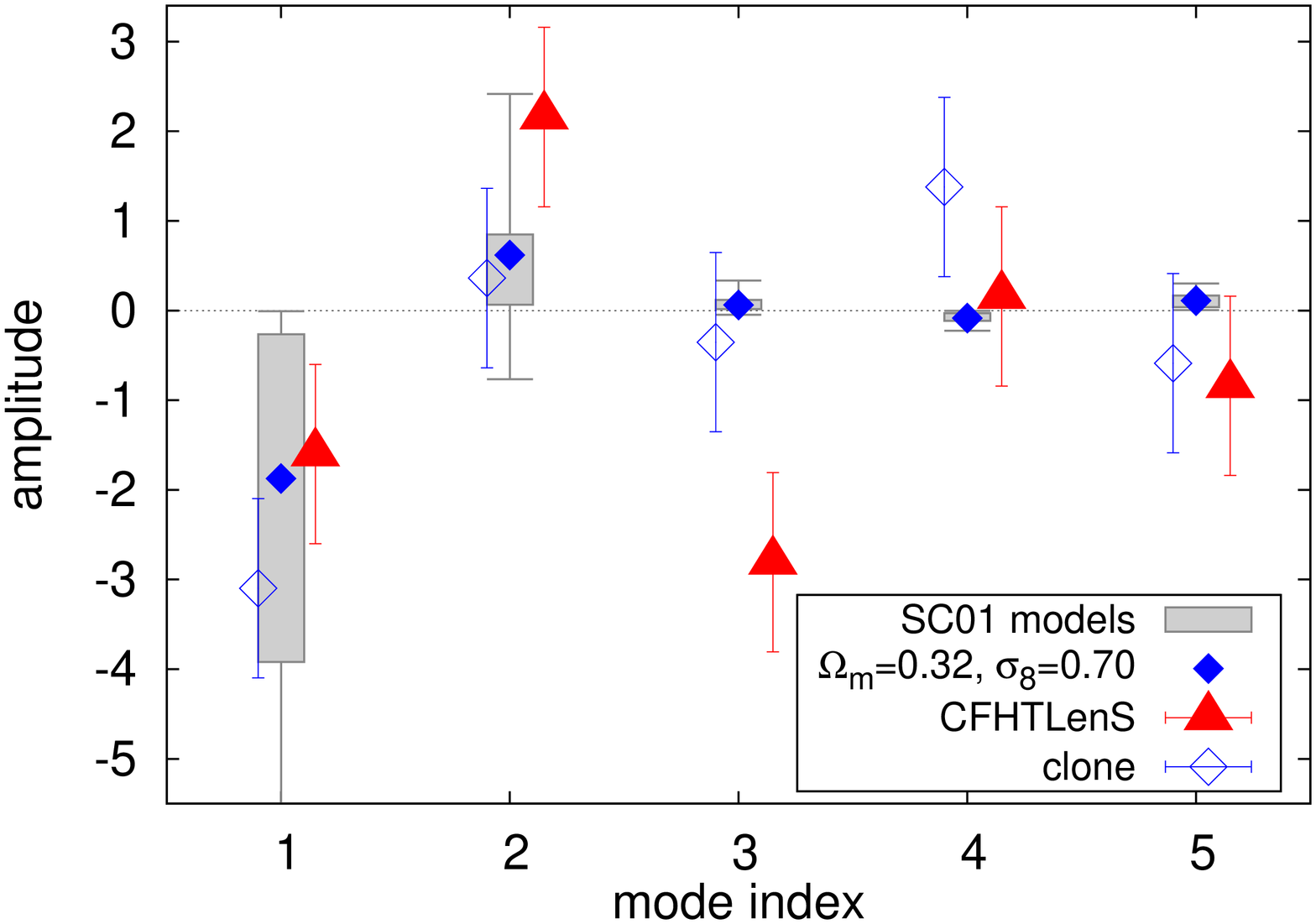,width=90mm,angle=-90}
    \psfig{file=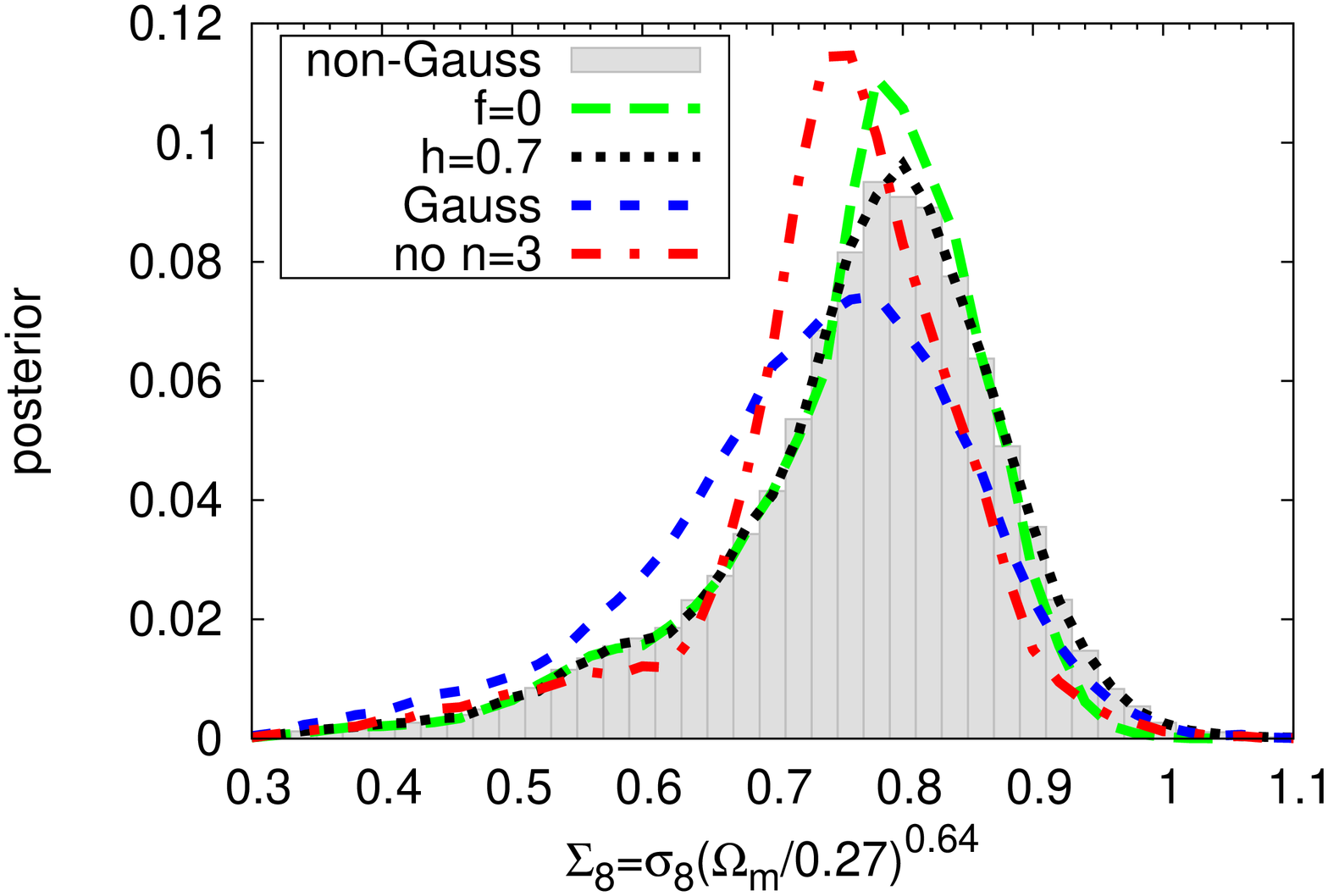,width=85mm,angle=-90}
    \vspace{-0.8cm}
  \end{center}
  \caption{\label{fig:cosmoresult2} \emph{Top panel:} The first five
    KL modes of the compressed CFHTLenS data vector (red triangles) in
    comparison to both the best-fitting model (filled blue diamonds)
    and the scatter of amplitudes of our models for a parameter range
    (grey whisker boxes). For the latter, the filled boxes denote the
    amplitude spread of 68\% of all models around the median for
    \mbox{$\Sigma_8=\sigma_8(\Omega_{\rm m}/0.27)^{0.64}\le1.0$} and
    \mbox{$h=0.7$}, whereas the whiskers bracket the entire amplitude
    range. There is a strong tension between models and data for
    \mbox{$n=3$}.  The error bars do not include the 20\%
    multiplicative error in the model amplitude. The $\chi^2$ per
    degree of freedom of CFHTLenS is $4.2$ or $2.9$ when the
    multiplicative error is included. The open diamonds denoted
    \emph{clone} correspond to one random realisation of a noisy {\tt
      clone} measurement.  \emph{Bottom panel:} Posterior of
    $\Sigma_8$ for the non-Gaussian data model (\emph{non-Gauss}) for
    various strong priors (lines \mbox{\emph{$f=0$}} and
    \mbox{\emph{$h=0.7$}}) compared to the posterior in the Gaussian
    model (line \emph{Gauss}). The red dotted-dashed line shows the
    posterior based on the KL modes \mbox{$n=1,2,4,5$} only.}
\end{figure}

The constraining power of the non-Gaussian model for low values of
\mbox{$\sigma_8\lesssim0.7$}, \mbox{$\Omega_{\rm m}\lesssim0.2$}, and
\mbox{$\Sigma_8\lesssim0.7$} is slightly better compared to the
Gaussian data model, as can be seen in the bottom panels of
Fig. \ref{fig:cosmoresult1} and Fig. \ref{fig:cosmoresult2} (compare
the grey histogram to the blue dashed line).  In addition to that, the
black dotted lines shows our non-Gaussian constraints for a fixed
\mbox{$h=0.7$} without the marginalisation of the Hubble parameter
that all other results are subject to. The difference to the posterior
in grey is small. Likewise, the impact of the $20\%$ error in the
predicted model amplitude is also small, as indicated by the
dotted-dashed green lines ($f=0$).

Figure \ref{fig:cosmoresult2} displays the goodness of the SC01 model
with respect to the CFHTLenS data. We plot the first five uncorrelated
KL modes that are used in our analysis (red triangles).  For
comparison we also plot the KL modes of one random noisy {\tt clone}
data vector into the plot (open diamonds).  The blue filled diamonds
correspond to the best-fitting model of CFHTLenS with
\mbox{$\Omega_{\rm m}=0.32$} and \mbox{$\sigma_8=0.7$}.  We observe a
clear discrepancy between the best fit and the CFHTLenS data at
\mbox{$n=3$} and some weak discrepancy at \mbox{$n=2$}. The $\chi^2$
of the residuals of the best fit is $4.2$ per degree of freedom or
\mbox{$4.2/(1+20\%)^2=2.9$} if we account for a 20\% systematic error
in the predicted amplitude; the systematic error can be included by
increasing the errors of modes in the plot by 20\%.  In order to
illustrate that reasonable models are unable to fit the CFHTLenS data
at $n=3$, we have added the grey boxes to this figure. They depict the
scatter of 68\% of the model amplitudes for parameters
\mbox{$\Sigma_8\le1.0$} and $(\Omega_{\rm
  m},\sigma_8)\in[0.1,1]\times[0.4,1]$; current data constraints
confine $\Sigma_8$ to about $0.76\pm0.06$ with 68\% confidence
(F14). Additionally, the whiskers bracket the total amplitude range
covered by the models. Clearly, our \mbox{$n=3$} mode is well outside
the whisker region, underlining no model reproduces our measurement
here. Note that mixed higher-order moments between the errors of the
KL modes may exist even though their second-order correlations vanish.

Finally, the red dotted-dashed line in the bottom Fig.
\ref{fig:cosmoresult2} corresponds to the posterior of $\Sigma_8$
leaving out the KL mode \mbox{$n=3$}. Statistically the posterior is
consistent with the constraint for \emph{non-Gauss} although the mode
of the distribution shifts to slightly smaller values of $\Sigma_8$
and tightens a bit, i.e., \mbox{$\Sigma_8=0.77\pm0.08$}.

\section{Discussion}
\label{sec:discussion}

We conclude from our results that the model of a Gaussian likelihood
is sufficient for the cosmological analysis of the CFHTLenS
$\ave{M_{\rm ap}^3}$ data, or data of comparable surveys, despite the
evidence for a non-Gaussian distribution of errors.  To arrive at this
conclusion we measured the third moment $\ave{M^3_{\rm ap}}$ of the
aperture mass for aperture scales between 5 and 30 arcmin. For this we
excluded scales below 5 arcmin owing to indications of systematics in
the aperture statistics. Our choice of 5 arcmin as smallest usable
scale in the analysis had been made prior to the cosmological analysis
of the data to avoid confirmation bias. Our maximum scale is
determined by the geometry of the survey.  For this angular range, the
{\tt clone} simulation of our data in Fig. \ref{fig:distribution} and
in Fig. \ref{fig:quartiles} indicates a moderate non-Gaussian
distribution of the measurement errors between 10 and 30 arcmin. A
Gaussian likelihood hence may produce biased parameter estimates in a
cosmological analysis. To test the validity of a Gaussian likelihood,
we performed a comparative cosmological analysis with a non-Gaussian
likelihood that is constructed from the distribution of $\ave{M^3_{\rm
    ap}}$ estimates in our simulation. For this task, we devised a
novel technique that involves a data compression to reduce the
aperture statistics to a few essential uncorrelated modes that are
shown in Fig. \ref{fig:cosmoresult2} (higher-order correlations may be
non-vanishing). Our statistical analysis factors in both measurement
noise and uncertainties from the theory side without the need to
specify the analytic form of the error PDF. This practical algorithm
is an important contribution of this paper for future analyses because
its applicability is not restricted to third-order shear
statistics. The outcome of our statistical analysis for the two
cosmological parameters $(\Omega_{\rm m},\sigma_8)$ is displayed in
Fig. \ref{fig:cosmoresult1} and juxtaposed with a second analysis that
uses a Gaussian likelihood. The posteriors of the parameters exhibit
only little differences between the Gaussian and non-Gaussian
likelihood -- at least for a flat $\Lambda\rm CDM$ model which is
strongly favoured by recent CMB constraints
\citep{2013arXiv1303.5076P,2013ApJS..208...19H}. The main difference
between both models of the likelihood is illustrated by the bottom
panels of Fig. \ref{fig:cosmoresult1} and Fig. \ref{fig:cosmoresult2}:
the non-Gaussian model excludes more strongly low values of
$\sigma_8$, the lower 68\% bound shifts up from $0.63$ in the Gaussian
case to $0.66$ in the non-Gaussian; the lower limit of $\Sigma_8$
shifts from $0.60$ to $0.68$; the improvement of the lower limit of
$\Omega_{\rm m}$ is below $0.01$.  A qualitative similar behaviour for
$\Omega_{\rm m}$ and $\sigma_8$ was found with the non-Gaussian model
of \cite{2009A&A...504..689H} for the second-order statistics of
cosmic shear.  The observed changes are small compared to the width of
the posteriors though. Consequently our results support the recent
decision of F14 to apply a Gaussian model of the likelihood to angular
scales smaller than 15 arcmin as reasonable approximation, although
their cosmological constraints can probably be tightened by a
non-Gaussian data model.

For future surveys with increased survey area and higher source number
densities, we expect that non-Gaussian features will become more
prominent in the data; a Gaussian model may hence then no longer be
adequate.  As seen in the top panels of Fig. \ref{fig:distribution},
on small angular scales around 5 arcmin the signal is dominated by
shape noise of the sources, which tends to Gaussianise the
distribution of $\ave{M^3_{\rm ap}}$ for statistically independent
intrinsic shapes.  On larger scales of about 30 arcmin, on the other
hand, cosmic variance is dominating, which on our sub-degree scales is
still non-Gaussian due to the non-linear clustering of matter
(Fig. \ref{fig:quartiles}). For a fixed angular scale and for an
increasingly higher number of source triplets in the estimator, the
amplitude of cosmic variance grows greater relative to the shot noise
variance. Hence future lensing surveys will exhibit stronger
non-Gaussianities on the angular scales that are considered
here. However, for a quantitative assessment of the impact of these
non-Gaussianities on parameter constraints with \mbox{$\sim10^3$}
square degree surveys we need more independent simulated data vectors
than the {\tt clone} can provide. The {\tt clone} allows only for
about two independent realisations of a $\ave{M^3_{\rm ap}}$
measurement in this case. This is far too few to sample the data
likelihood. Still, considering that we already see an inaccuracy with
a Gaussian data model in CFHTLenS, we anticipate that a Gaussian model
will potentially bias the cosmological analysis in future surveys.

Our posterior for $(\Omega_{\rm m},\sigma_8)$ from $\ave{M^3_{\rm
    ap}}$ only is highly degenerate, and it is $68\%$ consistent with
recent constraints from the CMB and the Gaussian CFHTLenS analysis of
F14. The degeneracy is shown in Fig. \ref{fig:cosmoresult1}. The joint
constraints on two parameters naturally have to be degenerate along a
one-dimensional line, as the cosmological information in the data
essentially consists only of one significant KL mode, see
Fig. \ref{fig:cosmoresult2}. From the joint, degenerate constraints we
infer \mbox{$\Sigma_8=\sigma_8(\Omega_{\rm
    m}/0.27)^{0.64}=0.79^{+0.08}_{-0.11}$}, in accordance with
F14. Note that F14 used only cosmological scales between 5.5 and 15
arcmin as well as a more accurate model for the matter bispectrum
which explains minor difference between our posterior contours. For a
comparison to WMAP9 results, we break the parameter degeneracy by
imposing a narrow prior on either parameter $\Omega_{\rm m}$ or
$\sigma_8$ that is consistent with WMAP9, i.e., \mbox{$\Omega_{\rm
    m}\in[0.25,0.30]$} or \mbox{$\sigma_8\in[0.75,0.85]$}. For a prior
on one parameter, the constraint of the other parameter is then also
consistent with the WMAP9 best fit (\mbox{$\sigma_8=0.82$} and
\mbox{$\Omega_{\rm m}=0.28$}): we find \mbox{$\Omega_{\rm
    m}=0.27^{+0.05}_{-0.05}$} and
\mbox{$\sigma_8=0.77^{+0.07}_{-0.11}$}.  The recent results from
\citet{2013arXiv1303.5076P}, on the other hand, favour somewhat larger
best-fit values \mbox{$\sigma_8=0.83$} and \mbox{$\Omega_{\rm
    m}=0.31$} due to a smaller \mbox{$h=0.67\pm0.014$}. These values
are also consistent with our CFHTLenS findings although there appears
to be a $1\sigma$ tension between the values of $\Omega_{\rm m}$. This
tension is, however, comparable to the inaccuracy in our statistical
methodology or model as seen in our verification run (bottom panel of
Fig. \ref{fig:testrun}); $\Omega_{\rm m}$ and $\sigma_8$ are somewhat
underestimated in the analysis. In conclusion, our $\ave{M^3_{\rm
    ap}}$ cannot distinguish a WMAP9 from a Planck cosmology; both are
equally consistent with our CFHTLenS results.

In spite of the broad consistency of $(\Omega_{\rm m},\sigma_8)$ with
the standard cosmology model, the CFHTLenS $\ave{M^3_{\rm ap}}$ data
have features that cannot be explained by any of our SC01 models or
more accurate models of the matter bispectrum in dark-matter-only
universes (flat $\Lambda\rm CDM$). A dark-matter-only model
consequently is not good enough to describe our observed third-order
correlations in the cosmic shear field, or there are remaining
systematics in CFHTLenS on scales greater than $\sim10$ arcmin that
are relevant for the third-order statistics.  The only moderately good
model fit to the data becomes easily obvious when we inspect the
compressed CFHTLenS data in Fig. \ref{fig:cosmoresult2}; error bars
are uncorrelated in this representation, and they have equal variance;
the best-fit has \mbox{$\chi^2=4.2$} per degree of freedom without
20\% error in the model amplitude and \mbox{$\chi^2=2.9$} including
the amplitude error; the number of degrees of freedom is 3.  All
models with a generous cut of \mbox{$\Sigma_8\le1.0$} are essentially
zero for KL modes \mbox{$n\ge3$}. The CFHTLenS data at \mbox{$n=3$},
on the other hand, is $3\sigma$ away from any of these models. Only
models that are simultaneously large in \mbox{$\Omega_{\rm m}\sim0.7$}
and \mbox{$\sigma_8\sim0.8$} get closer to the \mbox{$n=3$} data point
but quickly move away from \mbox{$n=1$} and \mbox{$n=5$} at the same
time and cannot explain our observation either; see e.g. the
dashed-dotted line in the bottom panel of
Fig. \ref{fig:KLexample}. The discrepancy is also visible in
Fig. \ref{fig:3pt_multi} where we plot the best-fitting model (filled
squares) in comparison to the CFHTLenS data (open diamonds): the
CFHTLenS data points are systematically above the model data points
for \mbox{$\theta_2=5.91$} arcmin (third column) in order to be
consistent for larger values of $\theta_2$ (columns 4-5).  Certainly,
this disagreement with theory, reflected by the \mbox{$n=3$} KL mode,
could be a shortfall of SC01. However, no analytic model can be better
than the bispectrum power in cosmological simulations that are used to
either test or calibrate bispectrum models in the non-linear regime
\citep{2012A&A...541A.162V,2012GILMAR,ScCo01}. Therefore, improved
modelling would at best reproduce the {\tt clone} data points which
are plotted in Fig. \ref{fig:KLexample}, open diamonds in the lower
panel, and Fig. \ref{fig:cosmoresult2}, filled triangles in the upper
panel.  Clearly, these data points do not exhibit the \mbox{$n=3$}
feature seen in CFHTLenS; we find the same for the {\tt clone} data
points in the other 183 line-of-sights. Since both the {\tt clone} and
the CFHTLenS two-point statistics are consistent with the standard
cosmological model \citep{2014MNRAS.441.2725F,2014MNRAS.442.1326K}, we
therefore conclude that also a bispectrum model more advanced than
SC01 or a dark-matter-only standard $\Lambda\rm CDM$ in general cannot
explain our \mbox{$n=3$} mode of CFHTLenS.  The $n=3$ mode is most
sensitive at
$(\theta_1,\theta_2,\theta_3)=(5^\prime\!.9,17^\prime\!.5,30^\prime)$
and $(10^\prime\!.2, 17^\prime\!.5,17^\prime\!.5)$ (top panel of
Fig. \ref{fig:KLexample}). We hence broadly locate the discrepancy
between \mbox{$\theta\approx10-20$} arcmin, or correspondingly
\mbox{$\ell\approx248-496$}.

Despite the evidence of some shear systematics in the data, we are
unable to conclusively identify either intrinsic alignments (IA) or
residual PSF systematics as origin of the model discrepancy. However,
there is evidence for IA playing only a minor role in this
context. After the application of the systematics test in Appendix
\ref{sec:systematics}, we additionally rejected nine CFHTLenS fields
from the H12 sample. This mildly affects the third-order aperture
statistics between 10-20 arcmin (left panel of
Fig. \ref{fig:3pt_data_selected}): the \emph{pass fields} have a
higher signal for the equilateral $\ave{M^3_{\rm ap}}$. We hence
suspect that PSF systematics are at least in part responsible for a
signal deficiency around $15$ arcmin.  Furthermore, after the removal
of the early-type galaxies from our shear catalogue, we found little
difference in the EEE signal but rather an increase of the BBB signal
on small scales (Fig. \ref{fig:3pt_data}). This behaviour is the
opposite of what is expected for IA contaminated data: most of the IA
signal is associated with early type galaxies
\citep{2006MNRAS.367..611M,2011A&A...527A..26J,2011MNRAS.410..844M}. Therefore
it is unlikely that the signal drop near $15$ arcmin as well as the
remaining EEB, EBB, and BBB signals below 5 arcmin can be explained
with IA. This supports theoretical models of IA that predict a
insignificant contribution of IA correlations to the EEE signal above
$\sim5$ arcmin for CFHTLenS
\citep{2014MNRAS.445.2918M,2014A&A...561A..53V,Seetal08}.  In
conclusion, further research is required to remove the remaining EEB,
EBB, BBB systematics and to decide whether the tension between data
and theoretical models persist. Finally, F14 do not report a
significant B-mode on scales below 5 arcmin for the second-order
aperture statistics. But they agree with our finding of a EEB, EBB,
and BBB signal on these scales for the third-order statistics. This
suggests that the here reported PSF systematics become only relevant
for higher-order cosmic shear statistics.

As additional online material we provide a Monte-Carlo sample of
$(\Omega_{\rm m},\sigma_8)$ based on the posterior in the top left
panel of Fig. \ref{fig:cosmoresult1} and a set of 200 realisations of
$\ave{M^3_{\rm ap}}$ data vectors that we produced from our
\emph{noisy} {\tt clone} simulations.

\section*{Acknowledgements}

This work is based on observations obtained with MegaPrime/MegaCam, a
joint project of CFHT and CEA/IRFU, at the Canada-France-Hawaii
Telescope (CFHT) which is operated by the National Research Council
(NRC) of Canada, the Institut National des Sciences de l'Univers of
the Centre National de la Recherche Scientifique (CNRS) of France, and
the University of Hawaii. This research used the facilities of the
Canadian Astronomy Data Centre operated by the National Research
Council of Canada with the support of the Canadian Space Agency.  We
thank the CFHT staff for successfully conducting the CFHTLS
observations and in particular Jean-Charles Cuillandre and Eugene
Magnier for the continuous improvement of the instrument calibration
and the Elixir detrended data that we used. We also thank TERAPIX for
the quality assessment and validation of individual exposures during
the CFHTLS data acquisition period, and Emmanuel Bertin for developing
some of the software used in this study. CFHTLenS data processing was
made possible thanks to significant computing support from the NSERC
Research Tools and Instruments grant program, and to HPC specialist
Ovidiu Toader. The early stages of the CFHTLenS project was made
possible thanks to the support of the European Commission's Marie
Curie Research Training Network DUEL (MRTN-CT-2006-036133) which
directly supported six members of the CFHTLenS team (LF, HHo, PS, BR,
CB, MV) between 2007 and 2011 in addition to providing travel support
and expenses for team meetings.  This material is based in part upon
work supported in part by the National Science Foundation Grant No.
1066293 and the hospitality of the Aspen Center for Physics.

The N-body simulations used in this analysis were performed on the TCS
supercomputer at the SciNet HPC Consortium. SciNet is funded by: the
Canada Foundation for Innovation under the auspices of Compute Canada;
the Government of Ontario; Ontario Research Fund - Research
Excellence; and the University of Toronto.

ES acknowledges support from the Netherlands Organisation for
Scientific Research (NWO) grant number 639.042.814 and support from
the European Research Council under the EC FP7 grant number 279396.
LVW acknowledges support from the Natural Sciences and Engineering
Research Council of Canada (NSERC) and the Canadian Institute for
Advanced Research (CIfAR, Cosmology and Gravity program). HHo
acknowledges support from Marie Curie IRG grant 230924, the
Netherlands Organisation for Scientific Research (NWO) grant number
639.042.814 and from the European Research Council under the EC FP7
grant number 279396. JHD acknowledges the support from the NSERC and a
CITA National Fellowship. CH acknowledges support from the European
Research Council under the EC FP7 grant number 240185.  HHi is
supported by the DFG Emmy Noether grant Hi 1495/2-1. TE is supported
by the Deutsche Forschungsgemeinschaft (DFG) through project ER
327/3-1 and the Transregional Collaborative Research Centre TR 33 --
``The Dark Universe''. PS also receives support by the DFG
  through the project SI 1769/1-1.

Author Contributions: All authors contributed to the development and
writing of this paper. The authorship list reflects the lead authors
of this paper (P. Simon, E. Semboloni, L. van Waerbeke, H. Hoekstra)
followed by two alphabetical groups. The first alphabetical group
includes key contributors to the science analysis and interpretation
in this paper, the founding core team and those whose long-term
significant effort produced the final CFHTLenS data product. The
second group covers members of the CFHTLenS team who made a
significant contribution to either the project, this paper, or
both. The CFHTLenS collaboration was co-led by CH and LvW and the
CFHTLenS Cosmology Working Group was led by TK.

\bibliographystyle{mn2e} 
\bibliography{CFHTLenS_3pt}

\begin{thebibliography}{90}
\expandafter\ifx\csname natexlab\endcsname\relax\def\natexlab#1{#1}\fi

\bibitem[{{Asgari} \& {Schneider}(2014)}]{2014AS}
{Asgari} M., {Schneider} P., 2014, ArXiv:1409.0863

\bibitem[{{Bartelmann} \& {Schneider}(2001)}]{BaSc01}
{Bartelmann} M., {Schneider} P., 2001, \physrep, 340, 291

\bibitem[{{Benitez}(2000)}]{2000ApJ...536..571B}
{Benitez} N., 2000, \apj, 536, 571

\bibitem[{{Benjamin} {et~al}\mbox{.}(2007){Benjamin}, {Heymans}, {Semboloni},
  {van Waerbeke}, {Hoekstra}, {Erben}, {Gladders}, {Hetterscheidt}, {Mellier},
  \& {Yee}}]{Benjamin07}
{Benjamin} J. {et~al.}, 2007, \mnras, 381, 702

\bibitem[{{Benjamin} {et~al}\mbox{.}(2013){Benjamin}, {van Waerbeke},
  {Heymans}, {Kilbinger}, {Erben}, {Hildebrandt}, {Hoekstra}, {Kitching},
  {Mellier}, {Miller}, {Rowe}, {Schrabback}, {Simpson}, {Coupon}, {Fu},
  {Harnois-D{\'e}raps}, {Hudson}, {Kuijken}, {Semboloni}, {Vafaei}, \&
  {Velander}}]{2013MNRAS.431.1547B}
{Benjamin} J. {et~al.}, 2013, \mnras, 431, 1547

\bibitem[{{Berg{\'e}} {et~al}\mbox{.}(2010){Berg{\'e}}, {Amara}, \&
  {R{\'e}fr{\'e}gier}}]{Beetal09}
{Berg{\'e}} J., {Amara} A., {R{\'e}fr{\'e}gier} A., 2010, \apj, 712, 992

\bibitem[{{Bernardeau} {et~al}\mbox{.}(1997){Bernardeau}, {van Waerbeke}, \&
  {Mellier}}]{Beetal97}
{Bernardeau} F., {van Waerbeke} L., {Mellier} Y., 1997, \aap, 322, 1

\bibitem[{{Bernardeau} {et~al}\mbox{.}(2003){Bernardeau}, {van Waerbeke}, \&
  {Mellier}}]{2003A&A...397..405B}
{Bernardeau} F., {van Waerbeke} L., {Mellier} Y., 2003, \aap, 397, 405

\bibitem[{{Calabretta} \& {Greisen}(2000)}]{CaGi00}
{Calabretta} M., {Greisen} E.~W., 2000, in Astronomical Society of the Pacific
  Conference Series, Vol. 216, Astronomical Data Analysis Software and Systems
  IX, {Manset} N., {Veillet} C., {Crabtree} D., eds., p. 571

\bibitem[{{Cooray} \& {Hu}(2002)}]{Co&Hu02}
{Cooray} A., {Hu} W., 2002, \apj, 574, 19

\bibitem[{{Crittenden} {et~al}\mbox{.}(2002){Crittenden}, {Natarajan}, {Pen},
  \& {Theuns}}]{2002ApJ...568...20C}
{Crittenden} R.~G., {Natarajan} P., {Pen} U.-L., {Theuns} T., 2002, \apj, 568,
  20

\bibitem[{{Dodelson}(2003)}]{2003moco.book.....D}
{Dodelson} S., 2003, {Modern cosmology}. Academic Press, Amsterdam
  (Netherlands)

\bibitem[{{Eifler} {et~al}\mbox{.}(2009){Eifler}, {Schneider}, \&
  {Hartlap}}]{2009A&A...502..721E}
{Eifler} T., {Schneider} P., {Hartlap} J., 2009, \aap, 502, 721

\bibitem[{{Eisenstein} \& {Hu}(1998)}]{EiHu98}
{Eisenstein} D.~J., {Hu} W., 1998, \apj, 496, 605

\bibitem[{{Erben} {et~al}\mbox{.}(2013){Erben}, {Hildebrandt}, {Miller}, {van
  Waerbeke}, {Heymans}, {Hoekstra}, {Kitching}, {Mellier}, {Benjamin}, {Blake},
  {Bonnett}, {Cordes}, {Coupon}, {Fu}, {Gavazzi}, {Gillis}, {Grocutt}, {Gwyn},
  {Holhjem}, {Hudson}, {Kilbinger}, {Kuijken}, {Milkeraitis}, {Rowe},
  {Schrabback}, {Semboloni}, {Simon}, {Smit}, {Toader}, {Vafaei}, {van Uitert},
  \& {Velander}}]{Erbenetal12}
{Erben} T. {et~al.}, 2013, \mnras, 433, 2545

\bibitem[{{Fu} {et~al}\mbox{.}(2014){Fu}, {Kilbinger}, {Erben}, {Heymans},
  {Hildebrandt}, {Hoekstra}, {Kitching}, {Mellier}, {Miller}, {Semboloni},
  {Simon}, {van Waerbeke}, {Coupon}, {Harnois-D{\'e}raps}, {Hudson}, {Kuijken},
  {Rowe}, {Schrabback}, {Vafaei}, \& {Velander}}]{2014MNRAS.441.2725F}
{Fu} L. {et~al.}, 2014, \mnras, 441, 2725

\bibitem[{{Fu} {et~al}\mbox{.}(2008){Fu}, {Semboloni}, {Hoekstra}, {Kilbinger},
  {van Waerbeke}, {Tereno}, {Mellier}, {Heymans}, {Coupon}, {Benabed},
  {Benjamin}, {Bertin}, {Dor{\'e}}, {Hudson}, {Ilbert}, {Maoli}, {Marmo},
  {McCracken}, \& {M{\'e}nard}}]{Fuetal08}
{Fu} L. {et~al.}, 2008, \aap, 479, 9

\bibitem[{{Gil-Marin} {et~al}\mbox{.}(2012){Gil-Marin}, {Wagner}, {Fragkoudi},
  {Jimenez}, \& {Verde}}]{2012GILMAR}
{Gil-Marin} H., {Wagner} C., {Fragkoudi} F., {Jimenez} R., {Verde} L., 2012,
  JCAP, 2, 47

\bibitem[{{Gillis} {et~al}\mbox{.}(2013){Gillis}, {Hudson}, {Erben}, {Heymans},
  {Hildebrandt}, {Hoekstra}, {Kitching}, {Mellier}, {Miller}, {van Waerbeke},
  {Bonnett}, {Coupon}, {Fu}, {Hilbert}, {Rowe}, {Schrabback}, {Semboloni}, {van
  Uitert}, \& {Velander}}]{2013MNRAS.431.1439G}
{Gillis} B.~R. {et~al.}, 2013, \mnras, 431, 1439

\bibitem[{{Hamana} {et~al}\mbox{.}(2002){Hamana}, {Colombi}, {Thion},
  {Devriendt}, {Mellier}, \& {Bernardeau}}]{Hamanaetal02}
{Hamana} T., {Colombi} S.~T., {Thion} A., {Devriendt} J.~E.~G.~T., {Mellier}
  Y., {Bernardeau} F., 2002, \mnras, 330, 365

\bibitem[{{Hamilton} {et~al}\mbox{.}(2006){Hamilton}, {Rimes}, \&
  {Scoccimarro}}]{2006MNRAS.371.1188H}
{Hamilton} A.~J.~S., {Rimes} C.~D., {Scoccimarro} R., 2006, \mnras, 371, 1188

\bibitem[{{Harnois-D{\'e}raps} {et~al}\mbox{.}(2012){Harnois-D{\'e}raps},
  {Vafaei}, \& {van Waerbeke}}]{HDetal12}
{Harnois-D{\'e}raps} J., {Vafaei} S., {van Waerbeke} L., 2012, \mnras, 426,
  1262

\bibitem[{{Harnois-D{\'e}raps} \& {van Waerbeke}(2014)}]{2014arXiv1406.0543H}
{Harnois-D{\'e}raps} J., {van Waerbeke} L., 2014, ArXiv:1406.0543

\bibitem[{{Harnois-D{\'e}raps} {et~al}\mbox{.}(2014){Harnois-D{\'e}raps}, {van
  Waerbeke}, {Viola}, \& {Heymans}}]{2014arXiv1407.4301H}
{Harnois-D{\'e}raps} J., {van Waerbeke} L., {Viola} M., {Heymans} C., 2014,
  ArXiv: 1407.4301

\bibitem[{{Hartlap} {et~al}\mbox{.}(2009){Hartlap}, {Schrabback}, {Simon}, \&
  {Schneider}}]{2009A&A...504..689H}
{Hartlap} J., {Schrabback} T., {Simon} P., {Schneider} P., 2009, \aap, 504, 689

\bibitem[{{Hartlap} {et~al}\mbox{.}(2007){Hartlap}, {Simon}, \&
  {Schneider}}]{2007A&A...464..399H}
{Hartlap} J., {Simon} P., {Schneider} P., 2007, \aap, 464, 399

\bibitem[{{Heavens} {et~al}\mbox{.}(2000){Heavens}, {Refregier}, \&
  {Heymans}}]{2000MNRAS.319..649H}
{Heavens} A., {Refregier} A., {Heymans} C., 2000, \mnras, 319, 649

\bibitem[{{Heitmann} {et~al}\mbox{.}(2014){Heitmann}, {Lawrence}, {Kwan},
  {Habib}, \& {Higdon}}]{2014ApJ...780..111H}
{Heitmann} K., {Lawrence} E., {Kwan} J., {Habib} S., {Higdon} D., 2014, \apj,
  780, 111

\bibitem[{{Heymans} {et~al}\mbox{.}(2013){Heymans}, {Grocutt}, {Heavens},
  {Kilbinger}, {Kitching}, {Simpson}, {Benjamin}, {Erben}, {Hildebrandt},
  {Hoekstra}, {Mellier}, {Miller}, {van Waerbeke}, {Brown}, {Coupon}, {Fu},
  {Harnois-D{\'e}raps}, {Hudson}, {Kuijken}, {Rowe}, {Schrabback}, {Semboloni},
  {Vafaei}, \& {Velander}}]{Heetal13}
{Heymans} C. {et~al.}, 2013, \mnras, 432, 2433

\bibitem[{{Heymans} {et~al}\mbox{.}(2012){Heymans}, {van Waerbeke}, {Miller},
  {Erben}, {Hildebrandt}, {Hoekstra}, {Kitching}, {Mellier}, {Simon},
  {Bonnett}, {Coupon}, {Fu}, {Harnois D{\'e}raps}, {Hudson}, {Kilbinger},
  {Kuijken}, {Rowe}, {Schrabback}, {Semboloni}, {van Uitert}, {Vafaei}, \&
  {Velander}}]{Heymans12}
{Heymans} C. {et~al.}, 2012, \mnras, 427, 146

\bibitem[{{Hildebrandt} {et~al}\mbox{.}(2012){Hildebrandt}, {Erben}, {Kuijken},
  {van Waerbeke}, {Heymans}, {Coupon}, {Benjamin}, {Bonnett}, {Fu}, {Hoekstra},
  {Kitching}, {Mellier}, {Miller}, {Velander}, {Hudson}, {Rowe}, {Schrabback},
  {Semboloni}, \& {Ben{\'{\i}}tez}}]{Hildebrandtetal12}
{Hildebrandt} H. {et~al.}, 2012, \mnras, 421, 2355

\bibitem[{{Hinshaw} {et~al}\mbox{.}(2013){Hinshaw}, {Larson}, {Komatsu},
  {Spergel}, {Bennett}, {Dunkley}, {Nolta}, {Halpern}, {Hill}, {Odegard},
  {Page}, {Smith}, {Weiland}, {Gold}, {Jarosik}, {Kogut}, {Limon}, {Meyer},
  {Tucker}, {Wollack}, \& {Wright}}]{2013ApJS..208...19H}
{Hinshaw} G. {et~al.}, 2013, \apjs, 208, 19

\bibitem[{{Hirata} \& {Seljak}(2003)}]{HiSe03}
{Hirata} C.~M., {Seljak} U., 2003, \prd, 68, 083002

\bibitem[{{Hoekstra} {et~al}\mbox{.}(2006){Hoekstra}, {Mellier}, {van
  Waerbeke}, {Semboloni}, {Fu}, {Hudson}, {Parker}, {Tereno}, \&
  {Benabed}}]{Hoekstraetal06}
{Hoekstra} H. {et~al.}, 2006, \apj, 647, 116

\bibitem[{Hyvarinen(1999)}]{761722}
Hyvarinen A., 1999, Neural Networks, IEEE Transactions on, 10, 626

\bibitem[{{Jarvis} {et~al}\mbox{.}(2004){Jarvis}, {Bernstein}, \&
  {Jain}}]{JaBeJa04}
{Jarvis} M., {Bernstein} G., {Jain} B., 2004, \mnras, 352, 338

\bibitem[{{Joachimi} {et~al}\mbox{.}(2011){Joachimi}, {Mandelbaum}, {Abdalla},
  \& {Bridle}}]{2011A&A...527A..26J}
{Joachimi} B., {Mandelbaum} R., {Abdalla} F.~B., {Bridle} S.~L., 2011, \aap,
  527, A26

\bibitem[{{Kayo} {et~al}\mbox{.}(2013){Kayo}, {Takada}, \& {Jain}}]{Kayoetal13}
{Kayo} I., {Takada} M., {Jain} B., 2013, \mnras, 429, 344

\bibitem[{{Keitel} \& {Schneider}(2011)}]{2011A&A...534A..76K}
{Keitel} D., {Schneider} P., 2011, \aap, 534, A76

\bibitem[{{Kilbinger}(2014)}]{2014arXiv1411.0115K}
{Kilbinger} M., 2014, ArXiv: 1411.0115

\bibitem[{{Kilbinger} {et~al}\mbox{.}(2013){Kilbinger}, {Fu}, {Heymans},
  {Simpson}, {Benjamin}, {Erben}, {Harnois-D{\'e}raps}, {Hoekstra},
  {Hildebrandt}, {Kitching}, {Mellier}, {Miller}, {van Waerbeke}, {Benabed},
  {Bonnett}, {Coupon}, {Hudson}, {Kuijken}, {Rowe}, {Schrabback}, {Semboloni},
  {Vafaei}, \& {Velander}}]{2013MNRAS.430.2200K}
{Kilbinger} M. {et~al.}, 2013, \mnras, 430, 2200

\bibitem[{{Kilbinger} \& {Munshi}(2006)}]{2006MNRAS.366..983K}
{Kilbinger} M., {Munshi} D., 2006, \mnras, 366, 983

\bibitem[{{Kilbinger} \& {Schneider}(2005)}]{KiSc05}
{Kilbinger} M., {Schneider} P., 2005, \aap, 442, 69

\bibitem[{{Kitching} {et~al}\mbox{.}(2012){Kitching}, {Balan}, {Bridle},
  {Cantale}, {Courbin}, {Eifler}, {Gentile}, {Gill}, {Harmeling}, {Heymans},
  {Hirsch}, {Honscheid}, {Kacprzak}, {Kirkby}, {Margala}, {Massey}, {Melchior},
  {Nurbaeva}, {Patton}, {Rhodes}, {Rowe}, {Taylor}, {Tewes}, {Viola},
  {Witherick}, {Voigt}, {Young}, \& {Zuntz}}]{2012MNRAS.423.3163K}
{Kitching} T.~D. {et~al.}, 2012, \mnras, 423, 3163

\bibitem[{{Kitching} {et~al}\mbox{.}(2014){Kitching}, {Heavens}, {Alsing},
  {Erben}, {Heymans}, {Hildebrandt}, {Hoekstra}, {Jaffe}, {Kiessling},
  {Mellier}, {Miller}, {van Waerbeke}, {Benjamin}, {Coupon}, {Fu}, {Hudson},
  {Kilbinger}, {Kuijken}, {Rowe}, {Schrabback}, {Semboloni}, \&
  {Velander}}]{2014MNRAS.442.1326K}
{Kitching} T.~D. {et~al.}, 2014, \mnras, 442, 1326

\bibitem[{{Kitching} {et~al}\mbox{.}(2008){Kitching}, {Miller}, {Heymans}, {van
  Waerbeke}, \& {Heavens}}]{2008MNRAS.390..149K}
{Kitching} T.~D., {Miller} L., {Heymans} C.~E., {van Waerbeke} L., {Heavens}
  A.~F., 2008, \mnras, 390, 149

\bibitem[{{Komatsu} {et~al}\mbox{.}(2009){Komatsu}, {Dunkley}, {Nolta},
  {Bennett}, {Gold}, {Hinshaw}, {Jarosik}, {Larson}, {Limon}, {Page},
  {Spergel}, {Halpern}, {Hill}, {Kogut}, {Meyer}, {Tucker}, {Weiland},
  {Wollack}, \& {Wright}}]{WMAP5}
{Komatsu} E. {et~al.}, 2009, \apjs, 180, 330

\bibitem[{{Laureijs} {et~al}\mbox{.}(2011){Laureijs}, {Amiaux}, {Arduini},
  {Augu{\`e}res}, {Brinchmann}, {Cole}, {Cropper}, {Dabin}, {Duvet}, {Ealet},
  \& et~al.}]{redbook}
{Laureijs} R. {et~al.}, 2011, ArXiv:1110.3193

\bibitem[{{Li} {et~al}\mbox{.}(2014){Li}, {Hu}, \&
  {Takada}}]{2014PhRvD..89h3519L}
{Li} Y., {Hu} W., {Takada} M., 2014, \prd, 89, 083519

\bibitem[{Loftsgaarden \& Quesenberry(1965)}]{0132.38905}
Loftsgaarden D., Quesenberry C., 1965, Ann. Math. Stat., 36, 1049

\bibitem[{{Mandelbaum} {et~al}\mbox{.}(2011){Mandelbaum}, {Blake}, {Bridle},
  {Abdalla}, {Brough}, {Colless}, {Couch}, {Croom}, {Davis}, {Drinkwater},
  {Forster}, {Glazebrook}, {Jelliffe}, {Jurek}, {Li}, {Madore}, {Martin},
  {Pimbblet}, {Poole}, {Pracy}, {Sharp}, {Wisnioski}, {Woods}, \&
  {Wyder}}]{2011MNRAS.410..844M}
{Mandelbaum} R. {et~al.}, 2011, \mnras, 410, 844

\bibitem[{{Mandelbaum} {et~al}\mbox{.}(2006){Mandelbaum}, {Hirata}, {Ishak},
  {Seljak}, \& {Brinkmann}}]{2006MNRAS.367..611M}
{Mandelbaum} R., {Hirata} C.~M., {Ishak} M., {Seljak} U., {Brinkmann} J., 2006,
  \mnras, 367, 611

\bibitem[{{Merkel} \& {Sch{\"a}fer}(2014)}]{2014MNRAS.445.2918M}
{Merkel} P.~M., {Sch{\"a}fer} B.~M., 2014, \mnras, 445, 2918

\bibitem[{{Miller} {et~al}\mbox{.}(2013){Miller}, {Heymans}, {Kitching}, {van
  Waerbeke}, {Erben}, {Hildebrandt}, {Hoekstra}, {Mellier}, {Rowe}, {Coupon},
  {Dietrich}, {Fu}, {Harnois-D{\'e}raps}, {Hudson}, {Kilbinger}, {Kuijken},
  {Schrabback}, {Semboloni}, {Vafaei}, \& {Velander}}]{Milleretal12}
{Miller} L. {et~al.}, 2013, \mnras, 429, 2858

\bibitem[{{Miller} {et~al}\mbox{.}(2007){Miller}, {Kitching}, {Heymans},
  {Heavens}, \& {van Waerbeke}}]{Milleretal07}
{Miller} L., {Kitching} T.~D., {Heymans} C., {Heavens} A.~F., {van Waerbeke}
  L., 2007, \mnras, 382, 315

\bibitem[{{Pen} {et~al}\mbox{.}(2003){Pen}, {Zhang}, {van Waerbeke}, {Mellier},
  {Zhang}, \& {Dubinski}}]{Peetal03}
{Pen} U.-L., {Zhang} T., {van Waerbeke} L., {Mellier} Y., {Zhang} P.,
  {Dubinski} J., 2003, \apj, 592, 664

\bibitem[{{Planck Collaboration} {et~al}\mbox{.}(2014){Planck Collaboration},
  {Ade}, {Aghanim}, {Armitage-Caplan}, {Arnaud}, {Ashdown}, {Atrio-Barandela},
  {Aumont}, {Baccigalupi}, {Banday}, \& et~al.}]{2013arXiv1303.5076P}
{Planck Collaboration} {et~al.}, 2014, \aap, 571, A16

\bibitem[{{Press} {et~al}\mbox{.}(1992){Press}, {Teukolsky}, \&
  {Vetterling}}]{1992nrca.book.....P}
{Press} W.~H., {Teukolsky} S.~A., {Vetterling} W.~T.~e., 1992, {Numerical
  recipes in C. The art of scientific computing}. Cambridge: University Press,
  1992, 2nd ed.

\bibitem[{{Rimes} \& {Hamilton}(2006)}]{2006MNRAS.371.1205R}
{Rimes} C.~D., {Hamilton} A.~J.~S., 2006, \mnras, 371, 1205

\bibitem[{{Sato} {et~al}\mbox{.}(2011){Sato}, {Ichiki}, \&
  {Takeuchi}}]{2011PhRvD..83b3501S}
{Sato} M., {Ichiki} K., {Takeuchi} T.~T., 2011, \prd, 83, 023501

\bibitem[{{Schneider}(2003)}]{2003A&A...408..829S}
{Schneider} P., 2003, \aap, 408, 829

\bibitem[{{Schneider}(2006)}]{2006glsw.conf..269S}
{Schneider} P., 2006, in Saas-Fee Advanced Course 33: Gravitational Lensing:
  Strong, Weak and Micro, {G.~Meylan, P.~Jetzer, P.~North, P.~Schneider,
  C.~S.~Kochanek, \& J.~Wambsganss}, ed., pp. 269--451

\bibitem[{{Schneider} {et~al}\mbox{.}(2010){Schneider}, {Eifler}, \&
  {Krause}}]{ScEiKr10}
{Schneider} P., {Eifler} T., {Krause} E., 2010, \aap, 520, A116

\bibitem[{{Schneider} {et~al}\mbox{.}(2005){Schneider}, {Kilbinger}, \&
  {Lombardi}}]{ScKiLo05}
{Schneider} P., {Kilbinger} M., {Lombardi} M., 2005, \aap, 431, 9

\bibitem[{{Schneider} \& {Lombardi}(2003)}]{ScLo03}
{Schneider} P., {Lombardi} M., 2003, \aap, 397, 809

\bibitem[{{Schneider} {et~al}\mbox{.}(1998){Schneider}, {van Waerbeke}, {Jain},
  \& {Kruse}}]{Scetal98}
{Schneider} P., {van Waerbeke} L., {Jain} B., {Kruse} G., 1998, \mnras, 296,
  873

\bibitem[{{Scoccimarro} \& {Couchman}(2001)}]{ScCo01}
{Scoccimarro} R., {Couchman} H.~M.~P., 2001, \mnras, 325, 1312

\bibitem[{{Semboloni} {et~al}\mbox{.}(2008){Semboloni}, {Heymans}, {van
  Waerbeke}, \& {Schneider}}]{Seetal08}
{Semboloni} E., {Heymans} C., {van Waerbeke} L., {Schneider} P., 2008, \mnras,
  388, 991

\bibitem[{{Semboloni} {et~al}\mbox{.}(2012){Semboloni}, {Hoekstra}, \&
  {Schaye}}]{Seetal12}
{Semboloni} E., {Hoekstra} H., {Schaye} J., 2012, \mnras, 417, 2020

\bibitem[{{Semboloni} {et~al}\mbox{.}(2006){Semboloni}, {Mellier}, {van
  Waerbeke}, {Hoekstra}, {Tereno}, {Benabed}, {Gwyn}, {Fu}, {Hudson}, {Maoli},
  \& {Parker}}]{Sembolonietal06}
{Semboloni} E. {et~al.}, 2006, \aap, 452, 51

\bibitem[{{Semboloni} {et~al}\mbox{.}(2011){Semboloni}, {Schrabback}, {van
  Waerbeke}, {Vafaei}, {Hartlap}, \& {Hilbert}}]{Seetal11a}
{Semboloni} E., {Schrabback} T., {van Waerbeke} L., {Vafaei} S., {Hartlap} J.,
  {Hilbert} S., 2011, \mnras, 410, 143

\bibitem[{{Simon} {et~al}\mbox{.}(2013){Simon}, {Erben}, {Schneider},
  {Heymans}, {Hildebrandt}, {Hoekstra}, {Kitching}, {Mellier}, {Miller}, {van
  Waerbeke}, {Bonnett}, {Coupon}, {Fu}, {Hudson}, {Kuijken}, {Rowe},
  {Schrabback}, {Semboloni}, \& {Velander}}]{2013MNRAS.430.2476S}
{Simon} P. {et~al.}, 2013, \mnras, 430, 2476

\bibitem[{{Simpson} {et~al}\mbox{.}(2013){Simpson}, {Heymans}, {Parkinson},
  {Blake}, {Kilbinger}, {Benjamin}, {Erben}, {Hildebrandt}, {Hoekstra},
  {Kitching}, {Mellier}, {Miller}, {van Waerbeke}, {Coupon}, {Fu},
  {Harnois-D{\'e}raps}, {Hudson}, {Kuijken}, {Rowe}, {Schrabback}, {Semboloni},
  {Vafaei}, \& {Velander}}]{2013MNRAS.429.2249S}
{Simpson} F. {et~al.}, 2013, \mnras, 429, 2249

\bibitem[{{Smith} {et~al}\mbox{.}(2003){Smith}, {Peacock}, {Jenkins}, {White},
  {Frenk}, {Pearce}, {Thomas}, {Efstathiou}, \& {Couchman}}]{Smetal03}
{Smith} R.~E. {et~al.}, 2003, \mnras, 341, 1311

\bibitem[{{Takada} \& {Jain}(2003)}]{TaJa03b}
{Takada} M., {Jain} B., 2003, \mnras, 344, 857

\bibitem[{{Takahashi} {et~al}\mbox{.}(2012){Takahashi}, {Sato}, {Nishimichi},
  {Taruya}, \& {Oguri}}]{2012ApJ...761..152T}
{Takahashi} R., {Sato} M., {Nishimichi} T., {Taruya} A., {Oguri} M., 2012,
  \apj, 761, 152

\bibitem[{{Taylor} {et~al}\mbox{.}(2013){Taylor}, {Joachimi}, \&
  {Kitching}}]{2013MNRAS.432.1928T}
{Taylor} A., {Joachimi} B., {Kitching} T., 2013, \mnras, 432, 1928

\bibitem[{{Tegmark} {et~al}\mbox{.}(1997){Tegmark}, {Taylor}, \&
  {Heavens}}]{1997ApJ...480...22T}
{Tegmark} M., {Taylor} A.~N., {Heavens} A.~F., 1997, \apj, 480, 22

\bibitem[{{Vafaei} {et~al}\mbox{.}(2010){Vafaei}, {Lu}, {van Waerbeke},
  {Semboloni}, {Heymans}, \& {Pen}}]{Vaetal10}
{Vafaei} S., {Lu} T., {van Waerbeke} L., {Semboloni} E., {Heymans} C., {Pen}
  U.-L., 2010, Astroparticle Physics, 32, 340

\bibitem[{{Valageas}(2014)}]{2014A&A...561A..53V}
{Valageas} P., 2014, \aap, 561, A53

\bibitem[{{Valageas} \& {Nishimichi}(2011{\natexlab{a}})}]{VaNi11a}
{Valageas} P., {Nishimichi} T., 2011{\natexlab{a}}, \aap, 527, A87

\bibitem[{{Valageas} \& {Nishimichi}(2011{\natexlab{b}})}]{VaNi11b}
{Valageas} P., {Nishimichi} T., 2011{\natexlab{b}}, \aap, 532, A4

\bibitem[{{Valageas} {et~al}\mbox{.}(2012){Valageas}, {Sato}, \&
  {Nishimichi}}]{2012A&A...541A.162V}
{Valageas} P., {Sato} M., {Nishimichi} T., 2012, \aap, 541, A162

\bibitem[{{van Waerbeke}(1998)}]{Wae98}
{van Waerbeke} L., 1998, \aap, 334, 1

\bibitem[{{van Waerbeke} {et~al}\mbox{.}(2013){van Waerbeke}, {Benjamin},
  {Erben}, {Heymans}, {Hildebrandt}, {Hoekstra}, {Kitching}, {Mellier},
  {Miller}, {Coupon}, {Harnois-D{\'e}raps}, {Fu}, {Hudson}, {Kilbinger},
  {Kuijken}, {Rowe}, {Schrabback}, {Semboloni}, {Vafaei}, {van Uitert}, \&
  {Velander}}]{2013MNRAS.433.3373V}
{van Waerbeke} L. {et~al.}, 2013, \mnras, 433, 3373

\bibitem[{{van Waerbeke} {et~al}\mbox{.}(1999){van Waerbeke}, {Bernardeau}, \&
  {Mellier}}]{1999A&A...342...15V}
{van Waerbeke} L., {Bernardeau} F., {Mellier} Y., 1999, \aap, 342, 15

\bibitem[{{van Waerbeke} {et~al}\mbox{.}(2001){van Waerbeke}, {Hamana},
  {Scoccimarro}, {Colombi}, \& {Bernardeau}}]{Waetal01}
{van Waerbeke} L., {Hamana} T., {Scoccimarro} R., {Colombi} S., {Bernardeau}
  F., 2001, \mnras, 322, 918

\bibitem[{{Velander} {et~al}\mbox{.}(2014){Velander}, {van Uitert}, {Hoekstra},
  {Coupon}, {Erben}, {Heymans}, {Hildebrandt}, {Kitching}, {Mellier}, {Miller},
  {van Waerbeke}, {Bonnett}, {Fu}, {Giodini}, {Hudson}, {Kuijken}, {Rowe},
  {Schrabback}, \& {Semboloni}}]{2014MNRAS.437.2111V}
{Velander} M. {et~al.}, 2014, \mnras, 437, 2111

\bibitem[{{Wilking} \& {Schneider}(2013)}]{2013A&A...556A..70W}
{Wilking} P., {Schneider} P., 2013, \aap, 556, A70

\bibitem[{{Zhang} \& {Pen}(2005)}]{ZhPe05}
{Zhang} L.~L., {Pen} U.-L., 2005, New Astronomy, 10, 569

\end{thebibliography}

\appendix

\section{Refined test for shear systematics}
\label{sec:systematics}

The level of residual systematics in the CFHTLenS lensing catalogue is
quantified by H12 using the two-point shear statistics. It is
therefore not guaranteed that residual systematics for higher-order
statistics would lead to the same field selection. For this paper, we
therefore assess the level of systematics affecting three-point shear
statistics by using a refinement of the H12 methodology. We apply the
refined test to the 129 CFHTLenS pointings that have already passed
the systematics criteria of H12. Out of these 129 we reject another
nine pointings for our cosmological analysis of the three-point shear
statistics. The details of the refinement and its results for CFHTLenS
follow below.

\subsection{Description of the method}

We adopt the residual systematics PSF model of H12: any residual
systematics is a linear combination of the stellar anisotropy
$\vec{\epsilon}^\star_a$ of all exposures $a$. Therefore, the observed
galaxy ellipticity, $\vec{\epsilon}^{\rm obs}$, originating from the
lack of a perfect PSF correction, is given by:
\begin{equation}
  \label{eq:syst} 
  \epsilon^{\rm
    obs}_i=\epsilon^{\rm int}_i+\gamma_i+\eta_i + \sum_{a=1} ^{n_{\rm exp}}
  \alpha_a \epsilon^\star_{a, i}\,, ~~~ i=1,2\,. 
\end{equation}
This thus expresses, for each ellipticity component $i$, the observed
(PSF-corrected) ellipticity $\epsilon^{\rm obs}_i$ as sum of (i) the
intrinsic ellipticity $\epsilon^{\rm int}_i$, (ii) the cosmic shear
$\gamma_i$, and (iii) the random noise $\eta_i$. In addition, (iv) the
last term expresses the PSF residual systematic error at the location
of the galaxy as a linear combination of the original PSF ellipticity
$\epsilon^\star_a$ directly measured from the stars in each exposure $a$ out
of $n_{\rm exp}$ exposures; the $\alpha_a$ are the coefficients of the
linear combination.

Moreover, following the conclusion from H12 we assume that the
zero-lag correlations of the ellipticities already contain all the
relevant information about the residual PSF correlations for the
three-point correlation functions. For the PSF residuals described by
\Ref{eq:syst}, the average measured zero-lag on a given pointing is
\begin{equation}
  \label{eq:resdef} 
  \ave{\epsilon^{\rm obs}_i\epsilon^{\rm obs}_j\epsilon^{\rm obs}_k}=
  \ave{\gamma_i \gamma_j \gamma_k} + \ave{\Delta(\epsilon^{\rm obs}_i\epsilon^{\rm
      obs}_j\epsilon^{\rm obs}_k)}\,,
\end{equation}
where $i,j,k=1,2$ indicate the projections of $\vec{\epsilon}^{\rm obs}$ and
$\gg$, and 
\begin{eqnarray}
  \label{eq:residual}
  \lefteqn{\ave{\Delta(\epsilon^{\rm obs}_i\epsilon^{\rm obs}_j\epsilon^{\rm obs}_k)}}
  \\
  \nonumber
  &=&
  \sum_{a,b,c=1 }^{n_{\rm exp}} \alpha_a \alpha_b\alpha_c \ave{\epsilon^\star_{a,i}\epsilon^\star_{b,j}
    \epsilon^\star_{c,k}}
  \\
  \nonumber
  &+&3\Big [ \ave{\gamma_j \gamma_k}+ \ave{ \eta_j \eta_k}+ \ave{
    \epsilon_j^{\rm int}\epsilon_k^{\rm int}}\Big ]  \sum_{a=1}^{n_{\rm
      exp}}\alpha_a \ave{\epsilon^\star_{a,i}} \nonumber\,.  
\end{eqnarray}
The indices $(a,b,c)$ indicate the various exposures, while the
average $\ave{\ldots}$ is over all zero-lag triplets.  Similarly, the
PSF-galaxy cross-correlations are:
\begin{eqnarray}
  \label{eq:cross1}
  \lefteqn{\ave{\epsilon^{\rm obs}_i\epsilon^{\rm obs}_j\epsilon^{\star}_{a, k
      }}}\\
  \nonumber&=&
  \Big[\ave{\gamma_i \gamma_j}  + \ave{\eta_i \eta_j} + \ave{\epsilon^{\rm
      int}\epsilon^{\rm int}}\Big] \ave{\epsilon^\star_{a,k}}\\
  \nonumber&+&
  \sum _{b,c=1}^{n_{\rm exp}} \alpha_a \alpha_b\alpha_c \ave{\epsilon^\star_{a, k}\epsilon^\star_{b,i}
    \epsilon^\star_{c, j}}
\end{eqnarray}
and
\begin{equation}
  \label{eq:cross2}
  \ave{\epsilon^{\rm obs}_i\epsilon^\star_{a,j}\epsilon^\star_{b,k}}=
  \sum _{c=1} ^{n_{\rm exp}} \alpha_{c} \ave{\epsilon^\star_{c,i}\epsilon^\star_{a,j}\epsilon^\star_{b,
      k}}\,. 
\end{equation}
We estimate the zero lag triplets by interpolating the stellar
anisotropy at the position of the source galaxy. For the derivation of
the equations \Ref{eq:residual}--\Ref{eq:cross2}, we assume that the
PSF is uncorrelated with the intrinsic ellipticity, the shear and the
random noise. In the absence of any systematics, equations
\Ref{eq:residual}--\Ref{eq:cross2} have to vanish, hence a non-zero
signal for either average can be used as indicator for
systematics. However, in the presence of systematics the expectation
values of both \Ref{eq:residual} and \Ref{eq:cross1} do directly
depend on cosmology through the terms $\ave{\gamma_i\gamma_j\gamma_k}$
and $\ave{\gamma_i\gamma_j}$, and they depend on the details of IA
through $\ave{\epsilon^{\rm int}_j\epsilon^{\rm int}_k}$. Therefore,
for an evaluation of the significance of a non-zero signal we have to
assume a fiducial model for the shear and IA correlations. Conversely,
the expectation value of \Ref{eq:cross2} is free of these assumption
so that we focus on \Ref{eq:cross2} as systematics indicator in the
following. Nevertheless some cosmology dependence enters for this
indicator too because the variance of this indicator
\begin{eqnarray}
  \nonumber
  \ave{(\epsilon^{\rm
      obs}_i\epsilon^\ast_{a,j}\epsilon^\ast_{b,k})^2}&=&
  \ave{(\epsilon^{\rm
      int}_i+\gamma_i+\eta_i)^2}\ave{(\epsilon^\ast_{a,j}\epsilon^\ast_{b,k})^2}\\
  &+&
  \ave{\left(\sum_{c=1}^{n_{\rm exp}}\alpha_c\epsilon^\ast_{c,i}\epsilon^\ast_{a,j}\epsilon^\ast_{b,k}\right)^2}
\end{eqnarray}
contains second-order correlations between $\gamma_i$ and $\epsilon^{\rm
  int}_j$ on the right hand side; the intrinsic variance for the null
hypothesis, i.e., no systematics, has to be known to test for a
presence of systematics. Contrary to \Ref{eq:residual} and
\Ref{eq:cross1} however, we expect a weak impact of the cosmology on
the test results, though, as on small scales (zero lag) the variance
$\ave{(\epsilon^{\rm int}_i+\gamma_i+\eta_i)^2}\approx\ave{(\epsilon^{\rm
    int}_i)^2}$ is likely dominated by shape noise. 
  
\subsection{Null hypothesis}

\begin{figure}
  \psfig{figure=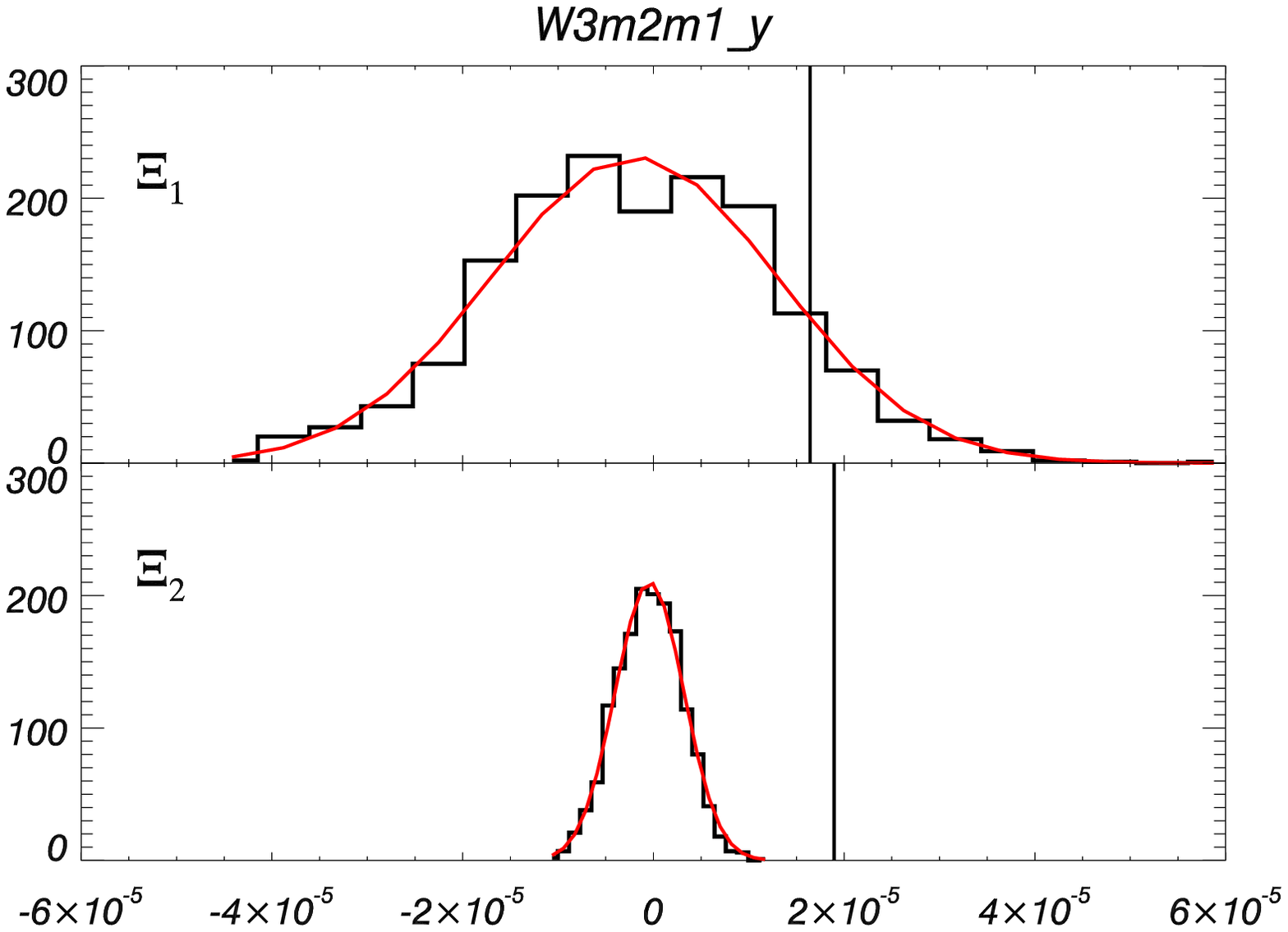,width=0.5\textwidth}
  \caption{\label{fig:cross} \emph{Top panel}: comparison of the
    systematics indicators $\Xi_i$ of W3m2m1\_y (vertical black line)
    to the simulated distribution of the null hypothesis
    (histogram). The red solid line is the best-fitting Gaussian to
    the histogram. The $\Xi_1$ is consistent with no
    systematics. \emph{Bottom panel}: the same as the top panel but
    for $\Xi_2$. Here the measured value is inconsistent with the null
    hypothesis on a high confidence level; the pointing is hence
    rejected for the cosmological analysis of $\ave{M^3_{\rm ap}}$.}
\end{figure}

In the following, we describe the set of simulations we use for the
measurement of the PDF of the systematics indicator
$\ave{\epsilon^{\rm obs}_i\epsilon^\ast_{a,j}\epsilon^\ast_{a,k}}$ for
the null hypothesis. This distribution is the key element in the
evaluation of the residual systematics for the 3-points function
measurement. Using the same set of simulations described in H12, there
are 160 realisations per pointing available, each with a shear signal
assigned from a projected mass map extracted from independent
lines-of-sights. These maps originate from the same N-body simulations
that have been used for the {\tt clone}. This means for the null
hypothesis:
\begin{itemize}
  \item a WMAP5 cosmology;
  \item no intrinsic alignments of the sources;
  \item no correlations of intrinsic ellipticities and shear;
  \item no $B$-modes of the shear field;
  \item distribution of intrinsic shapes and source positions as in
    CFHTLenS.
\end{itemize}
A shear value is assigned to each galaxy, depending on its 3D position
in the sky.  Shape noise is added to the shear as an ellipticity
component obtained by randomising the orientation of the ellipticity
of the CFHTLenS galaxies. This procedure guarantees that the simulated
catalogues have similar shape noise and intrinsic ellipticity
characteristics as the real CFHTLenS pointing.  Note that the
simulated shape noise is marginally larger than in the real data
because the shear is also randomised when computing the noise, it is
not removed; this is only a one percent effect and hence negligible
for the test.  For each pointing, we create 1600 realisations by
generating for each of the 160 lines-of-sight ten different noise
realisations.  The PSF ellipticity at each galaxy location,
$\vec{\epsilon}^\star_a$ is derived from the \lensfit~PSF model
\citep{Milleretal12}. For each pointing, the PDF of
$\ave{\epsilon^{\rm obs}_i\epsilon^\star_{a,j}\epsilon^\star_{a,k}}$
is then constructed from the distribution of \Ref{eq:cross2} in the
1600 realisations.

The probability of measuring a given value of 
\begin{equation}
  \label{eq:chidef} 
  \Xi_i=\sum_{a=1}^{\rm n_{\rm
      exp}}\ave{\epsilon^{\rm obs}_{i}\epsilon^\star_{a,i}\epsilon^\star_{a,i}}\,,~~ i=1,2 
\end{equation}
for a CFHTLenS pointing given its specific PSF and noise properties
can now be quantified against the assumptions of the null
hypothesis. We obtain the zero lag correlation $\ave{\epsilon^{\rm
    obs}_i \epsilon^\star_{a,j}\epsilon^\star_{b,k}}$ by interpolating
$\epsilon^\star_{a,j}$ and $\epsilon^\star_{a,k}$ at the position of
the galaxy with $\vec{\epsilon}^{\rm obs}$. For simplicity, we
restrict the analysis to the correlations $\ave{\epsilon^{\rm
    obs}_{i}\epsilon^\star_{a,i}\epsilon^\star_{a,i}}$ measured in the
same exposure $a$, and then sum over all the exposures of a given
pointing; cross-correlations between exposures are ignored.  This
strategy provides the strongest signal.  We then judge the systematics
significance of a value of $\Xi_i$ in CFHTLenS by the correlation
excess with respect to our simulated PDF. See Fig. \ref{fig:cross},
based on the pointing W3m2m1\_y, as an example. Here we plot the null
hypothesis as histogram (black solid lines) that we successfully fit
by a Gaussian distribution (red solid lines). The vertical solid lines
are the corresponding actual values in CFHTLenS for comparison. We
reject this pointing due to its large excess of $\Xi_2$.
    
\subsection{Application to the CFHTLenS data}

We evaluate the cross-correlation defined in \Ref{eq:cross2} for each
pair of exposures $(a,b)$ and for different projections $(i,j,k)$ of
the vectors $(\vec{\epsilon}^\star$, $\vec{\epsilon}^{\rm obs})$.
From this, we estimate the statistics $\Xi_i$ for every CFHTLenS
pointing and test the values against the null hypothesis. For this
test, we assume that a Gaussian distribution is a good approximation
of the null distribution of $\Xi_i$. This is a valid assumption for
the two-points cross-correlations (see H12), and we have checked that
it is also a valid assumption for histograms of $\Xi_i$, which we fit
by a Gaussian of mean $\nu_i$ and variance $\sigma_i$ for every
pointing. As expected, the averages $\nu_i$ are always consistent with
zero.

\begin{table}
  \begin{center}
    \begin{tabular}{@{}l@{}|@{}c@{}|@{}c@{}|@{}c@{}}    
      & $\Xi_1$ & $\Xi_2$ & null (Gaussian)\\
      \hline
      $|\Xi_i| < \sigma_i$  &\, 90 (70\%) &\, 69 (53\%) &\, $\sim 80(68.2\%)$\\
      $|\Xi_i| < 2 \sigma_i$   &\, 126 (98\%) &\,  109 (84\%)&\,  $\sim 123(95.4\%)$\\
      $|\Xi_i| < 3 \sigma_i$  &\, 129 (100\%) &\,  124 (96\%) &\, $\sim 129(99.6\%)$\\
    \end{tabular}
  \end{center}
  \caption{\label{tab:table1} Number of pointings with $|\Xi_i|$ below
    $\sigma_i$ (first line), $2\sigma_i$ (second line), and
    $3\sigma_i$ (third line); $\sigma_i$ is the  standard deviation of
    the null hypothesis (no  systematics). We indicate in parentheses
    the  fractional value corresponding to this number. We show
    the results both for $\Xi_1$ (first column) and $\Xi_2$ (second
    column) as well as the expectation for a null signal (third
    column). The total number of 129 pointings used here complies with the
    systematics criteria of H12.}
\end{table}

As a null $\Xi_i$ is supposed to obey Gaussian statistics, $31.80\%$
of the pointings should have \mbox{$|\Xi_i|>\sigma_i$}, $4.6\%$
\mbox{$|\Xi_i|>2\sigma_i$}, and $0.04\%$
\mbox{$|\Xi_i|>3\sigma_i$}. The Table \ref{tab:table1} compares
systematics indicators of 129 CFHTLenS pointings to the null
hypothesis with respect to the $\sigma_i$, $2\sigma_i$, and
$3\sigma_i$ thresholds. These 129 pointings have been pre-selected by
the criteria defined in H12. Within $3\sigma_i$ the statistics of
$|\Xi_1|$ is consistent with the null hypothesis, whereas $|\Xi_2|$
reveals too many outliers with
\mbox{$|\Xi_2|>3\sigma_i$}. 

For the final cosmological analysis, we decided to reject pointings
that are within the Gaussian $1\%$ tail of the null hypothesis. The
false-positive rate of our test is consequently $1\%$. Based on this
cut, we reject the following pointings: ${\rm W1m3p3\_i}$, ${\rm
  W1p3p1\_y}$, ${\rm W2m1m0\_i}$, ${\rm W3m3m0\_i}$, ${\rm
  W3m2m1\_y}$, ${\rm W3p1m1\_i}$, ${\rm W4m3p1\_i}$, ${\rm
  W4m3m0\_i}$, and ${\rm W4m3p3\_y}$. All these fields are rejected
owing to too high values of $|\Xi_2|$ alone. The signal with and
without the rejected fields can be seen in the left panel of
Fig. \ref{fig:3pt_data_selected}.

\end{document}